\documentclass[twocolumn,prd,amsmath,amssymb,showpacs,nofootinbib,superscriptaddress]{revtex4-1}
\usepackage{graphicx}
\usepackage{rotating}
\usepackage{paralist}
\newcommand{\be}{\begin{equation}}
\newcommand{\ee}{\end{equation}}
\def \ea{e-ASTROGAM }
\def \eap{e-ASTROGAM}

\def \mtt{ }
\def \mat{ }

\def\gsim{\lower.5ex\hbox{$\; \buildrel > \over \sim \;$}}
\begin{document}

\setcounter{tocdepth}{4}

\title{The e-ASTROGAM mission\\
(exploring the extreme Universe with gamma rays in the MeV~--~GeV  range)}


\author{Alessandro~De~Angelis}
\thanks{Corresponding Author}
\affiliation{Istituto Nazionale di Fisica Nucleare, Sezione di Padova, I-35131 Padova, Italy}
\affiliation{Istituto Nazionale di Astrofisica, Padova, Italy}
\affiliation{Dipartimento di Matematica, Informatica e Fisica, Universit\`a di Udine, I-33100 Udine, Italy}
\affiliation{Laboratorio de Instrumenta\c{c}ao e Particulas and Instituto Superior Tecnico, Lisboa, Portugal}
\author{Vincent~Tatischeff}	 
\thanks{Corresponding Author}
%\affiliation{CSNSM,  F-91405 Orsay Campus, France}
%\affiliation{IN2P3-CNRS/Univ. Paris-Sud, F-91405 Orsay Campus, France}
%\affiliation{Universit\'e Paris-Saclay, F-91405 Orsay Campus, France}
\affiliation{CSNSM, CNRS and University of Paris Sud, F-91405, Orsay, France}
\author{Marco~Tavani}	
\affiliation{INAF/IAPS, via del Fosso del Cavaliere 100, I-00133, Roma, Italy}
\affiliation{University of Roma ``Tor Vergata'', I-00133, Roma, Italy}
\affiliation{Istituto Nazionale di Fisica Nucleare, Sezione di Roma ``Tor Vergata", I-00133 Roma, Italy}	
\author{Uwe~Oberlack}
\affiliation{Institute of Physics and PRISMA Excellence Cluster, Johannes Gutenberg University Mainz, D-55099 Mainz, Germany}		 \author{Isabelle~A.~Grenier}	
\affiliation{AIM Paris-Saclay, CEA/IRFU, CNRS, Univ Paris Diderot, F-91191 Gif-sur-Yvette, France}	
\author{Lorraine~Hanlon}	
\affiliation{School of Physics, University College Dublin, Ireland}	 
\author{Roland~Walter} 
\affiliation{University of Geneva, Chemin d'Ecogia 16, CH-1290 Versoix, Switzerland}		 
\author{Andrea~Argan}
\affiliation{INAF Headquarters, Viale del Parco Mellini, 84, I-00136, Roma, Italy}		 
\author{Peter~von~Ballmoos}  
\affiliation{IRAP Toulouse, 9 av. du Colonel-Roche - BP 44 346, F-31028 Toulouse Cedex 4, France}
\author{Andrea~Bulgarelli}	
\affiliation{INAF/IASF Bologna, Via  Gobetti 101, I-40129 Bologna, Italy}	 
\author{Immacolata~Donnarumma}  
\affiliation{INAF/IAPS, via del Fosso del Cavaliere 100, I-00133, Roma, Italy}
\affiliation{Now at Agenzia Spaziale Italiana, Roma, Italy}
\author{Margarita~Hernanz}
\affiliation{ICE (CSIC-IEEC), Campus  UAB, Carrer Can Magrans s/n, E-08193 Cerdanyola del Valles, Barcelona, Spain}
\author{Irfan~Kuvvetli}
\affiliation{DTU Space, National Space Institute, Technical University of Denmark, Kgs. Lyngby, Denmark}
\author{Mark~Pearce} 
\affiliation{KTH Royal Institute of Technology, Dept. of Physics, 10691 Stockholm, Sweden}
\author{Andrzej~Zdziarski}
\affiliation{Nicolaus Copernicus Astronomical Center, Polish Academy of Sciences, Bartycka 18, PL-00-716 Warszawa, Poland}
\author{Alessio~Aboudan} 	
\affiliation{Dept. of Physics and Astronomy University of Padova and INAF, via Marzolo 8, I-35131 Padova, Italy}
\author{Marco~Ajello}
\affiliation{Department of Physics and Astronomy, Clemson University, Clemson, SC 29634, USA} 
\author{Giovanni~Ambrosi}
\affiliation{INFN Perugia, Perugia, Italy}	
\author{Denis~Bernard}
\affiliation{Institute of Physics and PRISMA Excellence Cluster, Johannes Gutenberg University Mainz, 55099 Mainz, Germany} 	
\author{Elisa~Bernardini}
\affiliation{DESY, Platanen Allee 6, D-15738 Zeuthen, Germany}	
\author{Valter~Bonvicini}
\affiliation{INFN Trieste, via A. Valerio, I-34127 Trieste, Italy}
\author{Andrea~Brogna}	  
\affiliation{Institute of Physics and PRISMA Excellence Cluster, Johannes Gutenberg University Mainz, 55099 Mainz, Germany}
\author{Marica~Branchesi}
\affiliation{Universit\`a degli Studi di Urbino, DiSPeA, I-61029 Urbino, Italy}
\affiliation{Istituto Nazionale di Fisica Nucleare, Sezione di Firenze, Italy}	
\author{Carl Budtz-J{\o}rgensen}
\affiliation{DTU Space, National Space Institute, Technical University of Denmark, Kgs. Lyngby, Denmark}
\author{Andrei~Bykov}
\affiliation{Ioffe Institute, St.Petersburg 194021, Russia}
\author{Riccardo~Campana}
\affiliation{INAF/IASF Bologna, Via  Gobetti 101, I-40129 Bologna, Italy}
\author{Martina~Cardillo}
\affiliation{INAF/IAPS, via del Fosso del Cavaliere 100, I-00133, Roma, Italy}
\author{Paolo~Coppi}
\affiliation{Department of Astronomy, Yale University, P.O. Box 208101, New Haven, CT 06520-8101, USA}		 
\author{Domitilla~De~Martino}	
\affiliation{NAF - Osservatorio Astronomico di Capodimonte, Salita Moiariello 16, I-80131 Napoli, Italy}
\author{Roland~Diehl}
\affiliation{Max Planck Institut fuer extraterrestrische Physik, Giessenbachstr.1, D-85748 Garching, Germany;  Excellence Cluster Universe, Germany}	 
\author{Michele~Doro}	
\affiliation{Istituto Nazionale di Fisica Nucleare, Sezione di Padova, I-35131 Padova, Italy}	
\affiliation{Dipartimento di Fisica e Astronomia ``G. Galilei'', Universit\`a di Padova, I-35131 Padova, Italy}		
\author{Valentina~Fioretti}	
\affiliation{INAF/IASF Bologna, Via  Gobetti 101, I-40129 Bologna, Italy} 
\author{Stefan~Funk}
\affiliation{Erlangen Centre for Astroparticle Physics, D-91058 Erlangen, Germany}			
\author{Gabriele~Ghisellini	}
\affiliation{INAF - Osservatorio di Brera, via E. Bianchi 46, I-23807 Merate, Italy}		
\author{J.~Eric~Grove}		
\affiliation{U.S. Naval Research Laboratory,  4555 Overlook Ave SW, Washington, DC 20375,  USA}
\author{Clarisse~Hamadache}
\affiliation{CSNSM,  F-91405 Orsay Campus, France}
\affiliation{IN2P3-CNRS/Univ. Paris-Sud, F-91405 Orsay Campus, France}
\affiliation{Universit\'e Paris-Saclay, F-91405 Orsay Campus, France}
\author{Dieter~H.~Hartmann}
\affiliation{Department of Physics and Astronomy, Clemson University, Clemson, SC 29634, USA} 		 
\author{Masaaki~Hayashida}	
\affiliation{Institute for Cosmic Ray Research, the University of Tokyo, Kashiwa, Chiba, 277-8582, Japan	}
\author{Jordi~Isern}	
\affiliation{ICE (CSIC-IEEC), Campus  UAB, Carrer Can Magrans s/n, E-08193 Cerdanyola del Valles, Barcelona, Spain}
\author{Gottfried~Kanbach}
\affiliation{Max-Planck-Institut fur Extraterrestrische Physik, Postfach 1312, 85741 Garching, Germany}
\author{J\"urgen~Kiener}	
\affiliation{CSNSM,  F-91405 Orsay Campus, France}
\affiliation{IN2P3-CNRS/Univ. Paris-Sud, F-91405 Orsay Campus, France}
\affiliation{Universit\'e Paris-Saclay, F-91405 Orsay Campus, France}	
\author{J\"urgen~Kn\"odlseder}
\affiliation{IRAP Toulouse, 9 av. du Colonel-Roche - BP 44 346, 31028 Toulouse Cedex 4, France}
\author{Claudio~Labanti} 
\affiliation{INAF/IASF Bologna, Via  Gobetti 101, I-40129 Bologna, Italy}
\author{Philippe~Laurent}	
\affiliation{PC, Univ Paris Diderot, CNRS/IN2P3, CEA/Irfu, Observatoire de Paris, 10 rue Alice Domont et L\'eonie Duquet, F-75205 Paris Cedex 13, France}\author{Olivier~Limousin} 	
\affiliation{CEA/Saclay IRFU/Department of Astrophysics, Bat. 709, F-91191, Gif-Sur-Yvette, France}
\author{Francesco~Longo}
\affiliation{Istituto Nazionale di Fisica Nucleare, Sezione di Trieste, I-34127 Trieste, Italy}
\affiliation{Dipartimento di Fisica, Universit\`a di Trieste, I-34127 Trieste, Italy} 
\author{Karl~Mannheim}
\affiliation{Universitaet Wuerzburg, Campus Hubland Nord, Lehrstuhl fuer Astronomie, Emil-Fischer-Strasse 31, D-97074 Wuerzburg, Germany}		
\author{Martino~Marisaldi} 
\affiliation{University of Bergen, Norway} 
\affiliation{INAF/IASF Bologna, Via  Gobetti 101, I-40129 Bologna, Italy}
\author{Manel~Martinez}
\affiliation{IFAE-BIST, Edifici Cn. Universitat Autonoma de Barcelona, E-08193 Bellaterra, Spain}
\author{Mario~N.~Mazziotta}
\affiliation{Istituto Nazionale di Fisica Nucleare, Sezione di Bari, I-70126 Bari, Italy}	
\author{Julie McEnery}
\affiliation{NASA Goddard Space Flight Center, Greenbelt, MD 20771, USA}			
\author{Sandro~Mereghetti}
\affiliation{INAF/IASF, Via Bassini 15, I-20133 Milano, Italy}
\author{Gabriele~Minervini	}
\affiliation{INAF/IAPS, via del Fosso del Cavaliere 100, I-00133, Roma, Italy}	 
\author{Alexander~Moiseev}
\affiliation{CRESST/NASA/GSFC and University of Maryland, College Park, USA}
\author{Aldo Morselli}
\affiliation{Istituto Nazionale di Fisica Nucleare, Sezione di Roma ``Tor Vergata", I-00133 Roma, Italy}	
\author{Kazuhiro~Nakazawa}	
\affiliation{Department of Physics, the University of Tokyo, 7-3-1 Hongo, Bunkyo-ku, Tokyo 113-0033}
\author{Piotr~Orleanski}
\affiliation{Space Research Center of Polish Academy of Sciences, Bartycka 18a, PL-00-716 Warszawa, Poland}
\author{Josep~M.~Paredes}	
\affiliation{Departament de Fisica Quantica i Astrofisica, ICCUB, Universitat de Barcelona, IEEC-UB, Marti i Franques 1, E-08028 Barcelona, Spain}
\author{Barbara~Patricelli}
\affiliation{Istituto Nazionale di Fisica Nucleare, Sezione di Pisa, I-56127 Pisa, Italy}
\affiliation{Scuola Normale Superiore, Piazza dei Cavalieri 7, I-56126 Pisa, Italy}
\author{Jean~Peyr\'e}	
\affiliation{CSNSM,  F-91405 Orsay Campus, France}
\affiliation{IN2P3-CNRS/Univ. Paris-Sud, F-91405 Orsay Campus, France}
\affiliation{Universit\'e Paris-Saclay, F-91405 Orsay Campus, France}		 
\author{Giovanni~Piano} 		
\affiliation{INAF/IAPS, via del Fosso del Cavaliere 100, I-00133, Roma, Italy} 
\author{Martin~Pohl}
\affiliation{Institute of Physics and Astronomy, University of Potsdam, 14476 Potsdam, Germany} 		 
\author{Harald Ramarijaona}	
\affiliation{CSNSM,  F-91405 Orsay Campus, France}
\affiliation{IN2P3-CNRS/Univ. Paris-Sud, F-91405 Orsay Campus, France}
\affiliation{Universit\'e Paris-Saclay, F-91405 Orsay Campus, France}	 
\author{Riccardo~Rando}
\affiliation{Istituto Nazionale di Fisica Nucleare, Sezione di Padova, I-35131 Padova, Italy}		
\affiliation{Dipartimento di Fisica e Astronomia ``G. Galilei'', Universit\`a di Padova, I-35131 Padova, Italy}
\author{Ignasi~Reichardt}  
\affiliation{Universitat Rovira i Virgili, Carrer de l`Escorxador, E-43003 Tarragona, Spain}
\author{Marco~Roncadelli} 	
\affiliation{INFN Pavia, via A. Bassi 6, I-27100 Pavia, Italy}
\affiliation{INAF Milano, Milano, Italy}
\author{Rui~Curado~da~Silva}  
\affiliation{LIP, Departamento de F'sica Universidade de Coimbra, P-3004-516 Coimbra, Portugal }
\author{Fabrizio~Tavecchio}	 
\affiliation{INAF - Osservatorio di Brera, via E. Bianchi 46, I-23807 Merate, Italy}
\author{David~J.~Thompson} 
\affiliation{NASA Goddard Space Flight Center, Greenbelt, MD, USA}
\author{Roberto~Turolla}	 
\affiliation{Dept. of Physics and Astronomy University of Padova, via Marzolo 8, I-35131 Padova, Italy}
\affiliation{University College London, United Kingdom}
\author{Alexei~Ulyanov}
\affiliation{School of Physics, University College Dublin, Belfield, Dublin 4, Ireland}
\author{Andrea~Vacchi}
\affiliation{University of Udine and INFN GC di Udine, via delle Scienze, I-33100 Udine, Italy}		
\author{Xin~Wu} 	
\affiliation{University of Geneva, Department of Nuclear and Particle Physics, 24 quai Ernest-Ansermet, 
CH-1211 Geneva 4, Switzerland}			 
\author{Andreas~Zoglauer}
\affiliation{University of California at Berkeley, Space Sciences Laboratory, 7 Gauss Way, Berkeley, CA 94720, USA}
\collaboration{On behalf of the e-ASTROGAM Collaboration}
\noaffiliation

%\institute{Alessandro De Angelis \at
%              INFN Padova, Via Marzolo 8, I-35141 Padova, Italy;\\
%              also at INAF Padova, Udine University, and LIP/IST Lisboa \\
%              Tel.: +39-049-967.7364\\
% \email{alessandro.deangelis@pd.infn.it}           %  \\
%%             \emph{Present address:} of F. Author  %  if needed
%           Vincent Tatischeff \at
%           CSNSM, CNRS and Univ Paris-Sud, F-91405 Orsay Cedex, France \\
%           Tel.: +33-(0)1-69 15 52 41\\
%           \email{vincent.tatischeff@csnsm.in2p3.fr}           %  \\
%}

%\date{Version 0.9 - November 3, 2016}
%\date{February 24, 2017; To be published in Experimental Astronomy.}

%\authorrunning{A. De Angelis {\em et al.}, on behalf of the e-ASTROGAM Collaboration}
%\authorrunning{The e-ASTROGAM Collaboration}

%\markleft{The e-ASTROGAM Collaboration}


%\linenumbers


\begin{abstract}
\ea (`enhanced ASTROGAM') is a breakthrough Observatory space mission, {with a detector composed by a Silicon tracker, a calorimeter, and an anticoincidence system,} dedicated to the study of the non-thermal Universe in the photon energy range from 0.3 MeV to 3 GeV {-- the lower energy limit can be pushed to energies as low as 150 keV, albeit with rapidly degrading angular resolution, for the tracker, and  to  30 keV for calorimetric detection}. The mission is based on an advanced space-proven detector technology, with unprecedented sensitivity, angular and energy resolution, combined with polarimetric capability. 
Thanks to its performance in the MeV-GeV domain,  substantially improving its predecessors, \ea will open a new window on the {non-thermal 
Universe,} making pioneering observations of the most powerful  Galactic and extragalactic  sources, elucidating the nature of their relativistic outflows and  their effects on the surroundings. 
With a line sensitivity in the MeV energy range one to two orders of magnitude better than previous generation instruments, \ea will determine the origin of key isotopes fundamental {for the understanding of supernova explosion and the chemical evolution of our Galaxy}. The mission will provide unique  data of significant interest to a broad astronomical community, complementary to powerful observatories such as LIGO-Virgo-GEO600-KAGRA, SKA, ALMA, E-ELT, TMT, LSST, JWST, Athena, CTA, IceCube, KM3NeT, and the promise of eLISA. 

\vskip 7mm

\end{abstract}

\maketitle

\tableofcontents

%\twocolumns

\section{Introduction}
\label{intro}

{\ea is a gamma-ray mission concept proposed as a response to the European Space Agency (ESA) Call for the fifth Medium-size mission (M5) of the {\it Cosmic Vision} Science Programme. The planned launch date is 2029. 

The main constituents of the \ea payload will be:
\begin{itemize}
\item A {\bf  Tracker} in which the cosmic $\gamma$-rays can undergo a  Compton scattering or a pair conversion, based on 56 planes of double-sided Si strip detectors, each plane with total area of $\sim$1 m$^2$;
\item A {\bf Calorimeter} to measure the energy of the secondary particles, made of an array of CsI (Tl) bars of 5$\times$5$\times$80 mm$^3$ each, with relative energy resolution of 4.5\% at 662 keV;
\item  An {\bf Anticoincidence system} (AC), composed of a standard plastic scintillator AC shielding and a Time of Flight,  to veto the 
charged particle background.
\end{itemize}

If selected,} \ea will operate in a maturing gravitational wave and multimessenger epoch, opening up entirely new and exciting synergies. The mission will provide unique and complementary data of significant interest to a broad astronomical community, in a decade of powerful observatories such as LIGO-Virgo-GEO600-KAGRA, SKA, ALMA, E-ELT, LSST, JWST, Athena, CTA and the promise of eLISA. 

The core mission science of \ea addresses three major topics of modern astrophysics.

\begin{itemize} \itemsep 0cm \topsep 0cm
\item \textbf{\em{Processes at the heart of the extreme Universe: prospects for the Astronomy of
the 2030s}}

Observations of relativistic jet and outflow sources (both in our Galaxy and
in active galactic nuclei, AGNs) in the X-ray and  GeV--TeV energy ranges have shown that the  MeV--GeV band holds the key to understanding the  transition from the low energy continuum to a spectral range shaped by very poorly understood particle acceleration processes. 
{\mtt   
\ea will:
(1) determine the composition (hadronic or leptonic) of the outflows and jets, which strongly influences the environment -- breakthrough polarimetric capability and spectroscopy 
providing the keys to unlocking this long-standing question;
(2) identify the physical acceleration processes in these outflows and jets (e.g. diffusive shocks, magnetic field reconnection, plasma effects), that may lead to dramatically different particle energy distributions; 
(3) clarify the role of the magnetic field in powering ultrarelativistic {jets in gamma-ray bursts (GRBs)}, through time-resolved polarimetry and spectroscopy.
In addition, measurements in the \ea energy band will have a big  impact on multimessenger astronomy in the 2030s. Joint detection of gravitational waves and gamma-ray transients would be ground-breaking.

\item \textit{\textbf{The origin and impact of high-energy particles on galaxy evolution, from cosmic rays to antimatter}}

{\mtt \ea   will  resolve the outstanding issue of the origin and propagation of low-energy cosmic rays affecting star formation. 
It  will measure cosmic-ray diffusion in interstellar clouds
and their impact on gas dynamics and state; it will provide crucial diagnostics about the wind outflows and their feedback on the Galactic 
environment (e.g., Fermi bubbles, Cygnus cocoon).}
\ea will have  optimal sensitivity and energy resolution to detect  line
emissions from 511 keV up to 10 MeV, and  a variety of 
issues will be resolved, in particular: (1)  origin
of the gamma-ray and positron excesses toward the Galactic
 inner regions;  (2)  determination of the astrophysical 
 sources of the local positron population from a very sensitive observation of pulsars and supernova remnants (SNRs). 
 As a consequence \ea will be able to discriminate the backgrounds to dark matter (DM) signals.

\item \textbf{\textit{Nucleosynthesis and the chemical enrichment
of our Galaxy}}

The \ea line sensitivity is more than an order of magnitude  better than previous instruments.  The deep exposure of the Galactic plane region will determine how different  isotopes are created in stars and distributed in the interstellar medium; it will also unveil the recent history of supernova explosions in the Milky Way.  
Furthermore, \ea will detect a significant number of Galactic novae and supernovae in nearby galaxies, thus addressing fundamental issues in the explosion mechanisms of both core-collapse and thermonuclear supernovae. The $\gamma$-ray data will provide a much better understanding of Type Ia supernovae and their evolution with look-back time and metallicity, which is a pre-requisite for their use as standard candles for precision cosmology.  

}
\end{itemize}

\begin{figure*}[ht]
\centering
\includegraphics[width=0.8\textwidth]{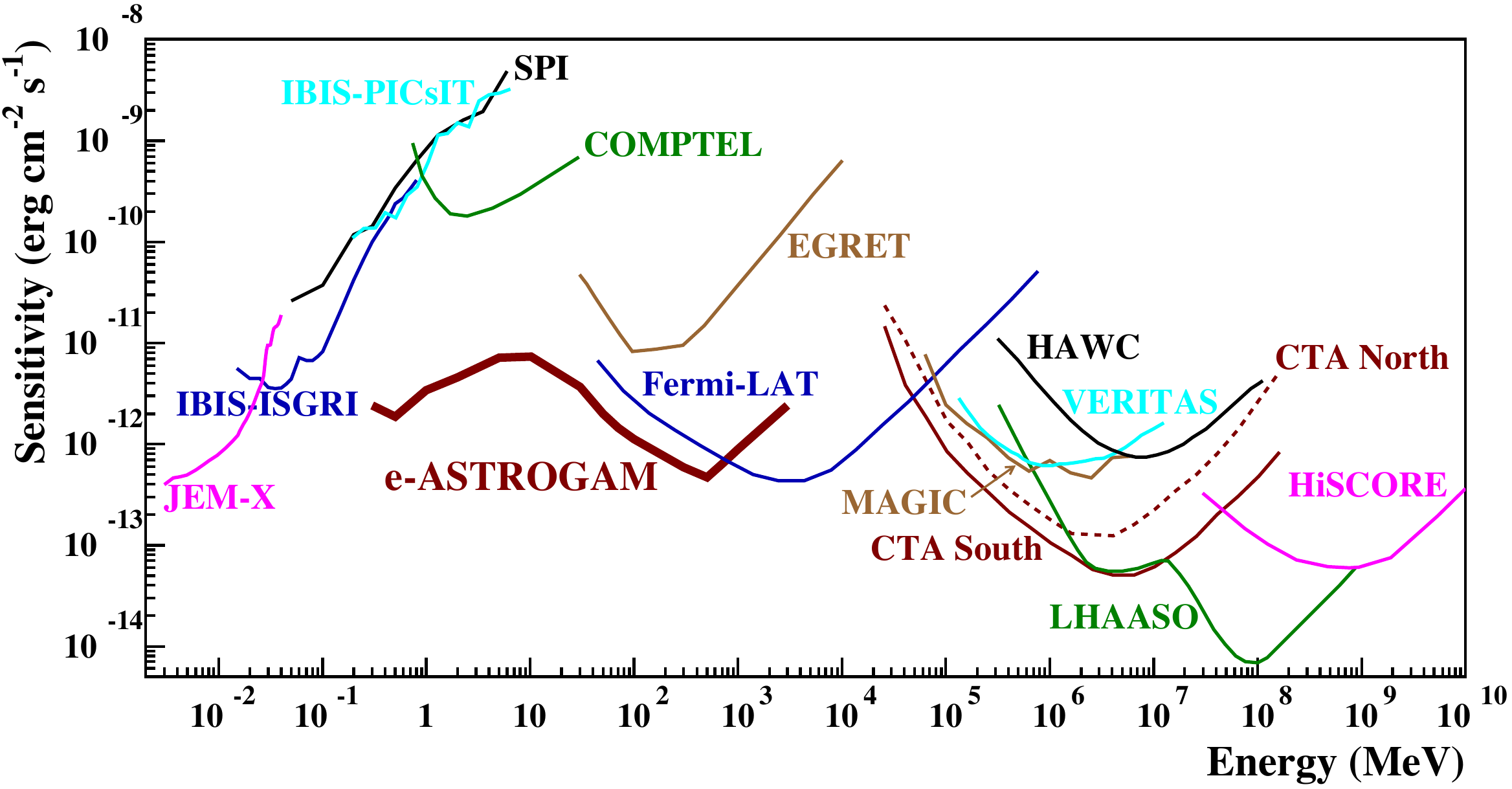}
\caption{{Point source continuum differential sensitivity of different X- and $\gamma$-ray instruments. The curves for {\it INTEGRAL}/JEM-X, IBIS (ISGRI and PICsIT), and SPI are for an effective observation time $T_{\rm obs}$ = 1 Ms. The COMPTEL and EGRET sensitivities are given for {the typical observation time accumulated during the $\sim$9 years of the {\it CGRO} mission (see Fig. 1 in \cite{tak13}). The {\it Fermi}/LAT sensitivity is for a high Galactic latitude source in 10 years of  observation in survey mode}. For MAGIC, VERITAS (sensitivity of H.E.S.S. is similar), and CTA, the sensitivities are given for $T_{\rm obs}$ = 50 hours. For HAWC $T_{\rm obs}$ = 5 yr, for LHAASO $T_{\rm obs}$ = 1~yr, and for HiSCORE $T_{\rm obs}$ = 1000 h. The e-ASTROGAM sensitivity is calculated at $3\sigma$ for an effective exposure of 1 year and for a source at high Galactic latitude.}
\label{fig:sensitivity}}
\end{figure*}

In addition to addressing its core scientific goals, \ea will 
achieve many serendipitous discoveries (the unknown unknowns) through its combination of wide field of view (FoV) and improved sensitivity, 
measuring  in 3 years the spectral energy distributions of thousands of Galactic and extragalactic sources,  
and providing new information on solar flares and terrestrial gamma-ray flashes (TGF). \ea will become a key contributor to multiwavelength time-domain astronomy. 
{\mtt The mission } % and therefore it
has outstanding discovery potential {\mtt as an Observatory facility that is 
open to a wide astronomical community.}

\ea  is designed to achieve:
\begin{itemize} \itemsep 0cm
\item Broad energy coverage (0.3 MeV to 3 GeV), with  one-two orders of magnitude improvement in continuum sensitivity in the range 0.3 MeV -- 100 MeV compared to previous instruments (the lower energy limit can be pushed to energies as low as 150 keV, albeit with rapidly degrading angular resolution, for the tracker, and to  30 keV for calorimetric detection);
\item Unprecedented performance for $\gamma$-ray lines, with, for example, a sensitivity for the 847~keV line from Type Ia SNe 70 times better than that of INTEGRAL/SPI;
\item Large FoV ($>$2.5 sr), ideal to detect transient sources and hundreds of GRBs;
\item Pioneering polarimetric capability for both steady and transient sources;
\item Optimized source identification capability afforded by the best angular resolution achievable by state-of-the-art detectors in this energy range (about 0.15 degrees at 1 GeV);
\item Sub-millisecond trigger and alert capability for GRBs and other cosmic and terrestrial transients;
\item Combination of Compton and pair-production detection techniques allowing model-independent control on the detector systematic uncertainties.
\end{itemize}

%\newpage

\section{Science Case}\label{sciencecase}

%Using Silicon detector technology that is backed by substantial space heritage and industry know-how, 

\begin{figure*}%{0.52\textwidth}
\centering
\includegraphics[width=0.6\textwidth]{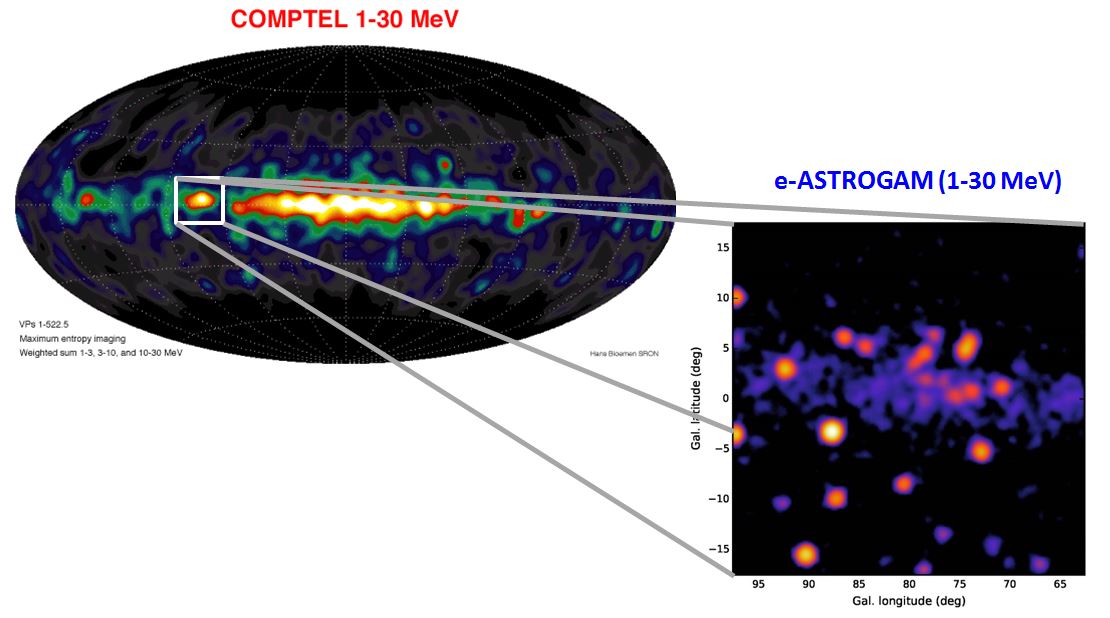}
\includegraphics[width=0.675\textwidth]{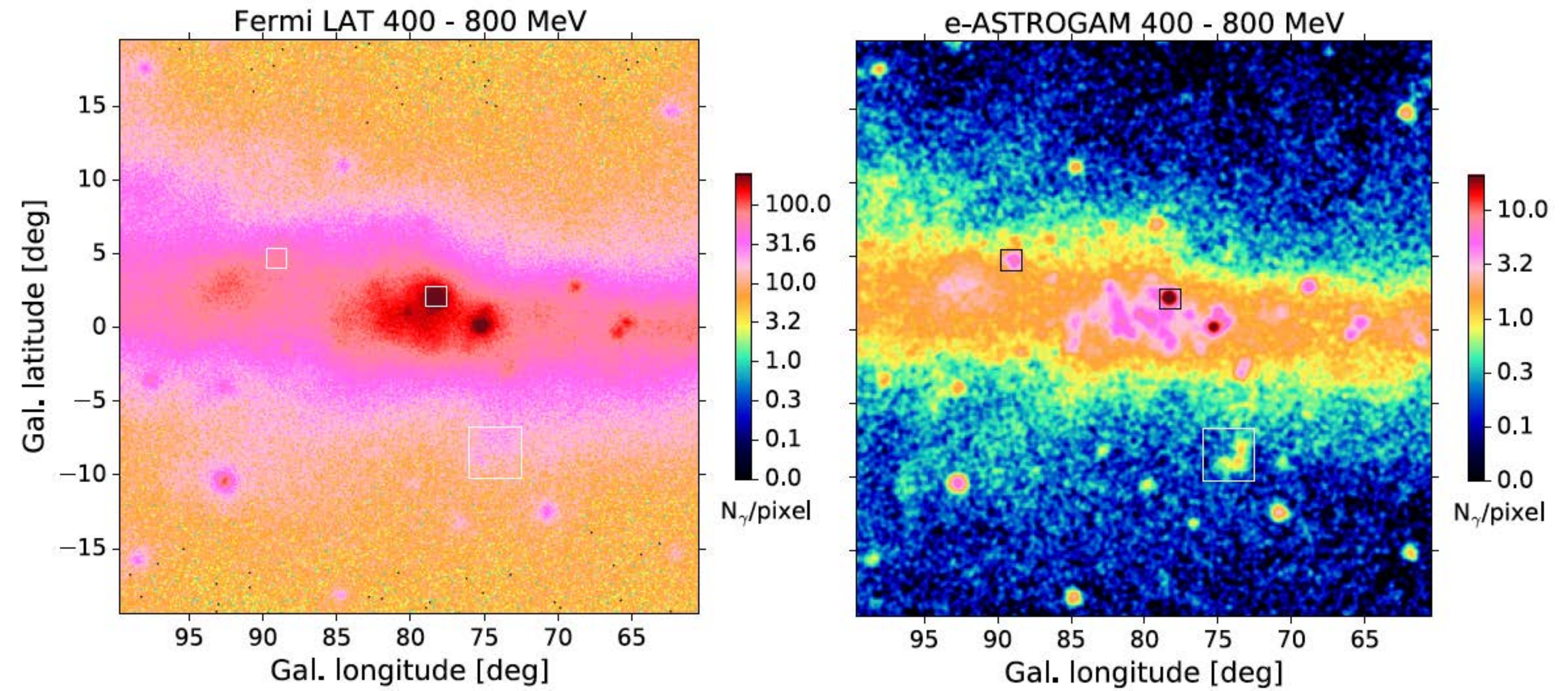}
\caption{An example of the capability of \ea to transform our 
 knowledge of the MeV-GeV sky. Upper panel: The upper left figure shows the
1-30 MeV sky as observed by COMPTEL in the 1990s; the lower right figure
  shows the simulated Cygnus region in the 1-30 MeV
energy region  from \eap. Lower panel: comparison between the view of the Cygnus region by Fermi in 8 years (left) and that by e-ASTROGAM in one year of effective exposure (right) between 400 MeV and 800 MeV.}
\label{fig:Jurgen}
\end{figure*}

\begin{figure}
\centering
\includegraphics[width=\columnwidth]{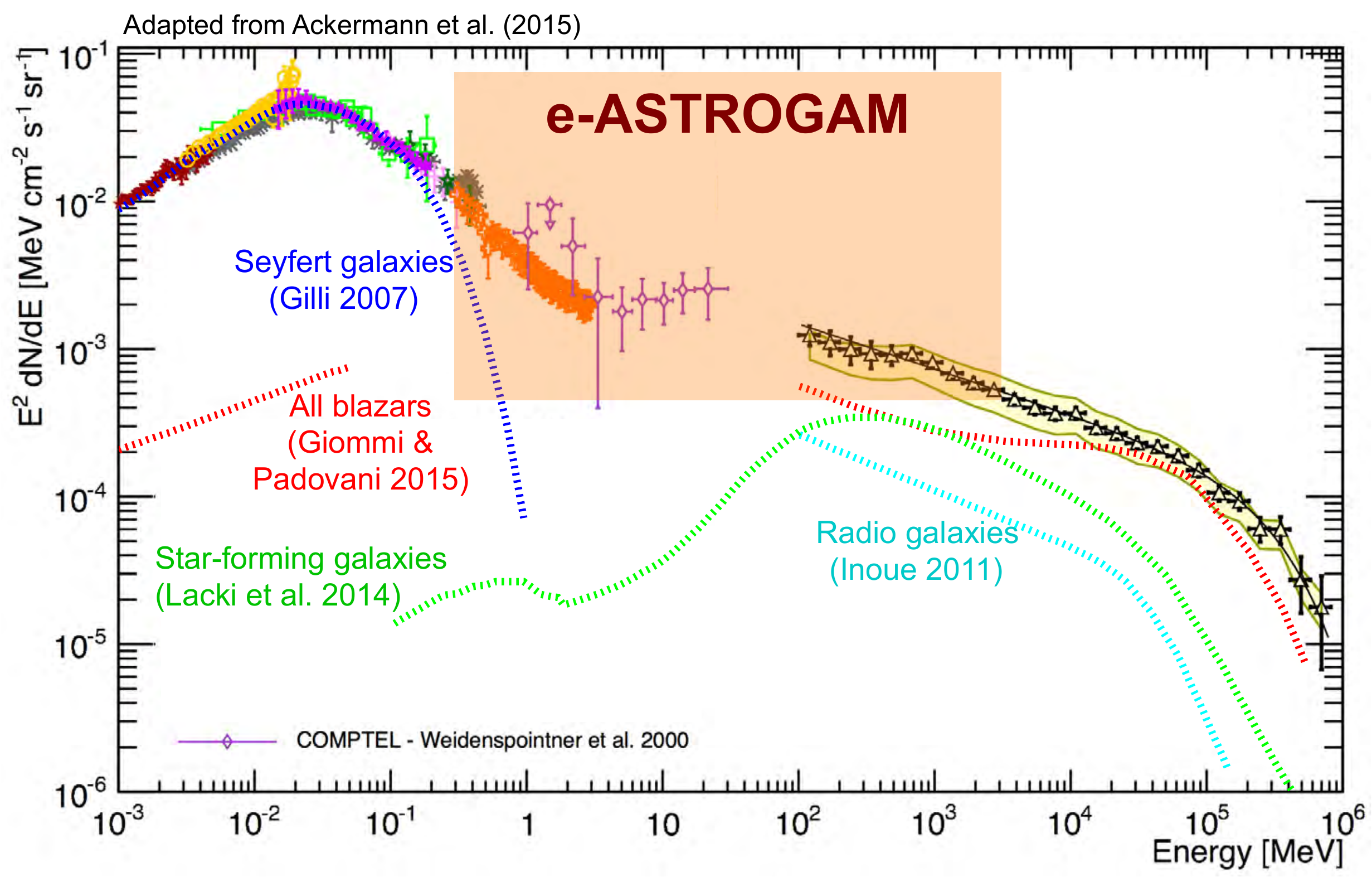}
\caption{Compilation of the measurements of the total extragalactic gamma-ray intensity between 1 keV and 820 GeV \cite{FermiEGB}, {with different components from current models}; the contribution from MeV blazars is largely unknown.
  The semi-transparent band indicates the energy region in which e-ASTROGAM will dramatically improve on present knowledge.\label{fig:egb}}
\end{figure}

\ea will open the MeV region for exploration, with an improvement of one-two orders of magnitude in sensitivity (Fig.~\ref{fig:sensitivity}) compared to the current state of the art, much of which was derived from the COMPTEL instrument more than two decades ago.  It will also achieve a spectacular improvement in terms of source localization accuracy (Fig.~\ref{fig:Jurgen}) and  energy resolution, and will allow to measure the contribution to the radiation of the Universe in an unknown range (Fig. \ref{fig:egb}). %At higher energies, reaching  beyond several tens of MeV, 
The sensitivity of \ea will reveal the transition from nuclear processes to those involving electro- and hydro-dynamical, magnetic and gravitational interactions.

An important characteristic of \ea is its ability to measure  polarization in the MeV range, which is afforded by Compton interactions in the detector.
Polarization encodes information about the geometry of magnetic fields and adds a new observational pillar, in addition to the temporal and spectral, through which fundamental processes governing the MeV emission can be determined. The addition of polarimetric information will be crucial for a variety of investigations, including accreting black-hole (BH) systems, magnetic field structures in jets, and the emission mechanisms of GRBs. Polarization will provide definitive insight into the presence of hadrons in extragalactic jets and the origin of ultra-high-energy cosmic rays (CR).

In the following sections, the core science questions \cite{physcase} to be addressed by \ea are presented. The requirements coming from the scientific objectives, and driving the  instrument design, are presented in Sect. \ref{sec:scirec}.

\subsection{Processes at the heart of the extreme Universe: prospects for the Astronomy of the
2030s}

The Universe accessible to \ea is dominated by strong particle acceleration. Ejection of plasma (jets or uncollimated outflows), ubiquitous in accreting systems, drives the transition from the keV energy range, typical of the accretion regime, to the  GeV-TeV range, through reprocessing of synchrotron radiation (e.g, inverse Compton, IC) or hadronic mechanisms. For some sources the MeV band naturally separates the acceleration and reprocessing energy ranges.
Other  systems, instead, radiate the bulk of their output in the MeV band. This is the most frequent case for AGNs at cosmological distances.

\ea will also study  extreme acceleration mechanisms  from 
compact objects such as neutron stars and (supermassive) black holes. Its polarimetric capabilities and its continuum sensitivity  will solve the problem of the nature of the highest energy radiation.

\begin{figure}
\vspace{-10pt}
\centering
\includegraphics[width=\columnwidth]{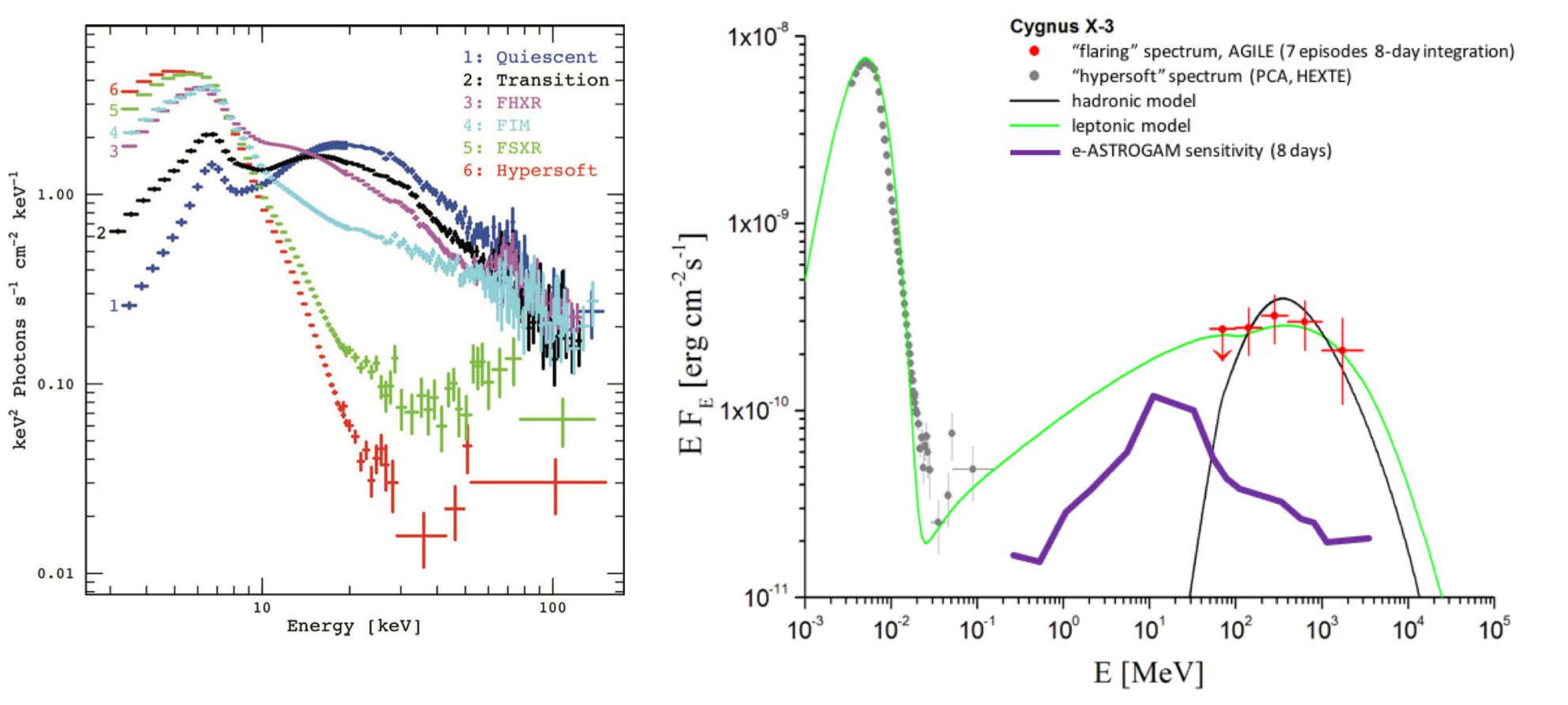} 
\caption{{Left panel: Different  states of Cyg X-3 in the soft X-ray/hard X-ray ranges (adapted from {\cite{Koljonen}}). Right panel: {spectral energy distribution (SED)} of Cyg X-3 
during the $\gamma$-ray flaring activity in 2011 (adapted from \cite{piano12}). The purple curve is the \ea 3$\sigma$ sensitivity for an 8-day observation,  from a simulation taking into account the expected background; the time matches the integration time of the AGILE $\gamma$-ray spectrum (red points)}. The green curve refers to a leptonic model and the black curve to a hadronic model.} 
\label{fig:CygX-3}
\vspace{-5pt}
\end{figure}

The transition to non-thermal processes involves, in particular, the emission of relativistic jets and winds.
{\mat In our Galaxy, this is relevant for compact
binaries and microquasars.
The interplay between accretion processes and
jet emission can best be studied in the MeV region, where disk Comptonization is expected to fade and other non-thermal components can originate from jet particles. \ea  observations of
Galactic compact objects and in particular of accreting BH
systems (such as Cygnus X-1 \cite{Zdziarski2014}, Cygnus X-3
(\cite{tavani2009,Abdocygx3}), V404 Cygni \cite{roques15}) will determine the nature of the
steady-state emission due to Comptonization and the transitions to highly non-thermal
radiation (Fig.~\ref{fig:CygX-3}). The main processes behind this
emission are Compton scattering by accelerated non-thermal
electrons and its attenuation/reprocessing by electron-positron pair production. The magnetic field in the BH vicinity can be quite strong, and have both random and ordered components; synchrotron emission by the   electrons in the
accretion flow may give rise to polarized MeV emission (e.g. 
\cite{Romero2014}), which can be  measured by
\eap, together with spectral transitions. Signatures of
$e^+e^-$ production and annihilation (e.g. \cite{Siegert2016}) can be  detected by \eap.}

%\end{wrapfigure}

\ea offers a unique way to study accelerated jets in
blazars on short and long timescales.
Among the unsolved questions are the origin of the photons
undergoing Comptonization, the location of the acceleration region in  jets (near to, or far from, the central black hole), and the {\mat presence of additional components of accelerated} electrons i.e. whether or not there is a mildly relativistic population of electrons. The latter question has important implications for the understanding of acceleration processes such as shock or magnetic
field reconnection.

\begin{figure} % [\sidecaptionrelwidth]
\centering
\includegraphics[width=\columnwidth]{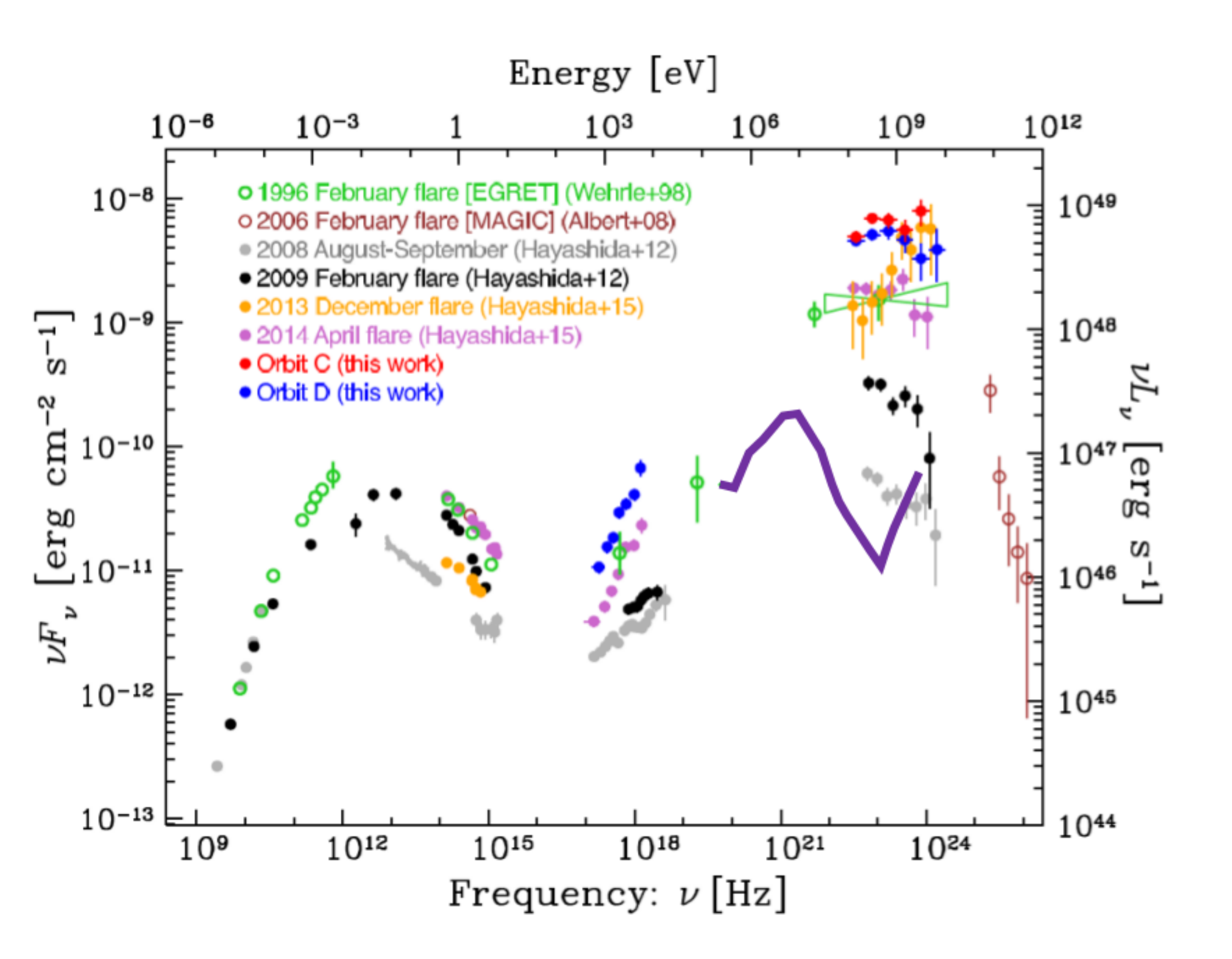} %IR: changed the sensitivity curve to purple, removed the (d) that remained top left
\caption{{SED from a collection of different spectral states of the FSRQ 3C 279 showing a dramatic gamma-ray flaring activity, including the  minute-timescale episode detected by Fermi in June 2015 \cite{Fermi3c279}. {The purple solid line is the 3$\sigma$ \ea sensitivity calculated for a 50\,ks exposure.}}} 
\label{fig:3c279} 
%\vspace{-20pt}
\end{figure}
%\end{wrapfigure}

Very fast variations of the gamma-ray emission {\mat have been
recently detected and are challenging current models;}
% poorly understood; a
an example is provided by the flat spectrum radio quasar (FSRQ) 3C 279
\cite{Fermi3c279}. {\mat During a flare in 2015} \cite{Fermi3c279}, variations of 
2-3 minutes {\mat were} detected in gamma-rays, i.e. well below
standard light-travel times usually assumed in theoretical models.
{These very rapid variability phenomena are the most compelling evidence of the occurrence of out-of-equilibrium particle acceleration most likely produced by magnetic field reconnection with remarkably high efficiency. On a smaller scale the magnetic field merging and the acceleration of relativistic electrons and ions have been typically observed in solar flares, where sub-second timescales and gamma-ray emissions into the 100 MeV range have been frequently detected. This process, called ``super-acceleration" (see \cite{tavani2013}), has recently been  invoked to explain the unexpected gamma-ray flare of the Crab Nebula (\cite{tavani2011,Abdocrab}), and leads to a very efficient mechanism for magnetic energy dissipation. It may apply to very rapid gamma-ray emission from compact objects with timescales and intensities incompatible with the current paradigms. }e-ASTROGAM will provide crucial information in the strongly variable spectral range 1-100 MeV (Fig. \ref{fig:3c279}), {constraining the electron population spectrum and the responsible particle acceleration mechanisms}.
An outstanding unsolved issue is the existence of gamma-rays
produced by hadronic processes.
The origin of % these MeV
 photons can be effectively probed both by {much improved spectral} measurements
 in the MeV-GeV band (detecting the ``pion bump''), and by polarimetric observations.
 Polarimetry is indeed a powerful tool to establish the
 nature of the emitters (hadrons vs. leptons) and,
 in case of leptonic (i.e., inverse Compton) emission,
 the nature of the soft photon target radiation
 \cite{ZhangBoettcher2013}. 
 
 An example is the very bright (about 50 times our sensitivity at 1 MeV) AGN 3C279  \cite{ZhangBoettcher2013}. In the 100 keV-10 MeV range, the polarization in leptonic emission models is low and rapidly decreasing with energy (from 40\% to 0), while for hadronic models it is high and increasing in energy ($>$60\% to 80\%).  Even in non optimal conditions (fields not well ordered, alignment to line of sight not optimal) the polarization signature would allow to identify unambiguously a hadronic scenario. The best targets for this study are blazars - both BL Lacs (suitable to study the transition from the synchrotron to the IC or hadronic-dominated component) and the powerful FSRQs (in which the transition from the highly polarized SSC radiation to the less polarized external Compton component can be revealed).

{Over 2/3 of the 3033 sources} from the 3rd {\it Fermi}-LAT Catalog of GeV-band sources (3FGL), have power-law
spectra (at energies larger than 100 MeV) steeper than $E^{-2}$,
implying that their peak energy output is below 100 MeV
(Fig.~\ref{fig:flare}).  The subset that is {\mat extragalactic hosts BHs with masses reaching $10^{10}$ $M_\odot$, and are often located at high redshift
($z\geq2-3$). They are therefore ideal tracers of the formation and history of super-massive BHs in the Universe
\cite{Ghisellini2010,Ghisellini2013}. In particular, the sources hosting the most massive BHs are elusive in the GeV band {\mat as} probed by {\it Fermi}-LAT. 
\begin{figure}
\begin{center}
\includegraphics[width=\linewidth]{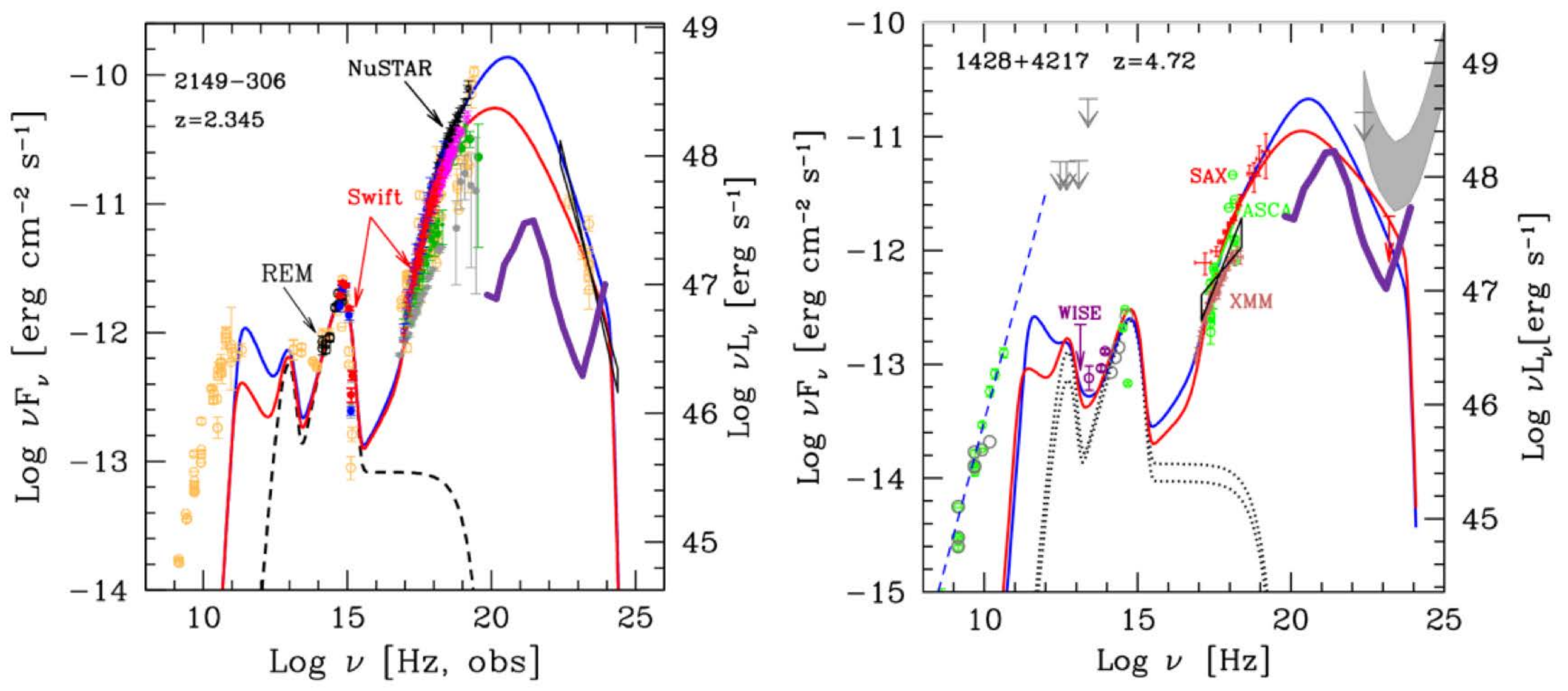}
\caption{SED of PKS\,2149-306 at $z=2.345$ (left panel) and the fitting model by \cite{tagliaferri2015} and GB 1428+4217 at $z = 4.72$ (right panel) and the fitting model by Ghisellini (private communication). Downward arrow is 2$\sigma$ Fermi upper limit over 7.5 years (see \cite{paliya2016}).
The sensitivity of e-ASTROGAM, calculated for an effective exposure of 1 year, is shown as a purple curve in the 0.3\,MeV -- 3\,GeV range. The bulk of the power is expected in the MeV band, around $10^{21}$~Hz.}
\label{fig:flare}
\vspace{-15pt}
\end{center}
\end{figure}

{Recent hard X-ray surveys \cite{ajello09,Ghisellini2010,ajello12} have been shown to be more effective in detecting higher redshift blazars compared to {GeV} $\gamma$-ray surveys.%, despite {{\it Fermi}-LAT's} sensitivity. 
 The main reason is that the SEDs of these sources peak in the MeV region  (hereafter, MeV blazars) and detection  becomes a difficult task for $\gamma$-ray instruments, even for  \textit{Fermi}-LAT. \ea will detect hundreds of these blazars, constraining their SED peaks very tightly (Fig. \ref{fig:flare}). 
These discoveries will revolutionize our understanding of: (i) how the two populations of AGNs (radio-quiet and radio-loud) evolve with redshift; (ii) the formation and growth of supermassive BHs; (iii) the connection between the jet and the central engine and (iv) the role of the jet in the feedback occurring in the host galaxies \cite{volon11}.}
%With an
%expected sample of \{id hundreds} of AGN detected, e-ASTROGAM  will
%understand how the two populations of AGN (radio-quiet and radio-loud) evolve in redshift, shedding light on the formation and growth of supermassive BHs and also the connection between the jet and the central engine. 
{\mat {\ea will very
substantially advance our knowledge of MeV-blazars up to redshift
$\sim 4.5$, with implications for blazar physics and cosmology --   the attenuation due to cosmological  absorption is negligible in the MeV region and the sensitivity has been shown by fits to the SED to be adequate (Fig. \ref{fig:flare})}}.
{{These} observations will be
invaluable and complementary 
 to data {from the future {\it ATHENA} mission} for the study of super-massive black holes.}
 
 By detecting the population of MeV-blazars up to redshift $\sim 4.5$, \ea will
 resolve the extragalactic gamma-ray background (EGB) in the MeV range ({Fig. \ref{fig:egb}). A possible
residual excess in the MeV range may have cosmological implications related to baryogenesis \cite{vonbal14}. Quasar-driven outflows could also account for the yet-not-understood part of the EGB \cite{loeb2016}.

\subsubsection{Gamma-Ray Bursts\label{sec:grb}}

GRBs are among the most intriguing and puzzling phenomena in astrophysics. They are believed to originate from coalescing NSs or BHs (short-GRBs, of duration  $\lesssim$2 s) or from the final collapses of massive stars (long-GRBs). Their radiative output is believed to originate from highly relativistic outflows.

e-ASTROGAM will allow unprecedented studies of both classes of GRBs, thanks to the combined interplay between the imaging Tracker, the Calorimeter, and the AC system, providing excellent sensitivity for spectral and timing studies combined with polarization capability. It will be possible to detect GRBs with durations from sub-millisecond to hundreds of seconds and study them in the energy range where their emission peaks. The e-ASTROGAM spectral performance will be very relevant, because of the role played by the MeV range in constraining theoretical models of particle acceleration. The total number of GRBs detectable by e-ASTROGAM is estimated to be $\sim$600 during the first 3 years (see Sect.~\ref{sec:polargrb}).

The e-ASTROGAM imaging Tracker can localize GRBs within 0.1$^\circ$-1$^\circ$ (depending on their intensity), and the information can be processed onboard for a fast communication. The  delay of the alerts is similar to $Fermi$-LAT, being the procedure (event rates onboard, plus simple localization based on onboard fast reconstruction) the same. This translates into an initial alert with an accuracy of 1$^\circ$-2$^\circ$ within (2-4) s, to be confirmed with similar accuracy within 30 s and then made public. Within 4h-8h the final alert accuracy of  0.1$^\circ$-1$^\circ$ can be reached. The alerts issued by e-ASTROGAM will be extremely valuable for  observatories such as CTA. The Calorimeter can act as an independent detector extending the energy range down to 30 keV: an on-board trigger logic spanning timescales from sub-ms up to seconds will be implemented.

For bright GRBs, e-ASTROGAM will  detect polarization in the MeV range. The Tracker can provide  information down to (150-200) keV, also for polarization measurements. We can estimate as 42 GRBs/year  the number of events with a detectable polarization fraction of 20\%; for a polarization fraction of 10\% the  number is about 16 GRBs/year. The polarization information, combined with spectroscopy in the MeV-GeV band, will provide a unique diagnostic to address the role of magnetic fields in the radiative output and dynamics of the most relativistic outflows in our Universe.

The e-ASTROGAM sensitivity to short and long GRBs will be very useful also for the detection of electromagnetic counterparts of impulsive gravitational wave events, as described in Sect.~\ref{sec:newastro}.

\subsubsection{ %Connection with different astrophysical messengers:
e-ASTROGAM and the new Astronomy}
\label{sec:newastro}

e-ASTROGAM fills the need for an MeV gamma--ray detector
operating at the same time as facilities such as SKA and CTA, as well as eLISA and neutrino detectors. It  guarantees the availability of complementary information to obtain a coherent picture of the transient sky and the sources of gravitational waves (GWs) and high-energy neutrinos. This will undoubtedly be an exciting new landscape for astronomy in the XXI century.

The first detections of GW signals from binary black hole (BH-BH) mergers, observed by  Advanced LIGO \cite{Abbott2016a}, marked the onset of the era of GW astronomy. The next breakthrough  will be the observation of their  electromagnetic counterparts, which will characterize the progenitor and its environment. Neutron star (NS)--BH  or 
%vt% BH-BH 
{NS-NS} mergers can eject relativistic outflows \cite{Nakar2007,Veres2014,Wang2016} or produce sub-relativistic omnidirectional high-energy emission  \cite{Takami2014}. In both cases emission up to the MeV energy range can be expected.
% The e-ASTROGAM combination of wide FoV ($>$2.5 sr), energy coverage and high sensitivity (Fig. \ref{fig:sensitivity}) will outperform the gamma-ray observatories in operation.
 The expected  detection rate of GRB prompt emission by e-ASTROGAM in coincidence with a GW detection is up to 1.5 events per year \cite{Patricelli2016}; it will double after the incorporation of KAGRA and LIGO-India into the GW network, which should happen several years before 2029. e-ASTROGAM will also play a key role in the multiwavelength study of GW events: in fact, its large FoV will maximize the detection probability and   provide accurate sky localization ($<$ 1 sq. deg at 1 MeV), thus allowing the follow-up of the GW events by other telescopes. This capability will be crucial for the identification and the multiwavelength characterization of the GW progenitor and of its host galaxy. 
\ea could associate binary systems to short GRBs,  improving the localization of sources and measuring spectral energy distributions.

 e-ASTROGAM may coincide with the third generation of ground-based interferometer projects, such as the Einstein Telescope \cite{Punturo2010} and Cosmic Explorer \cite{McClelland2015}, with an order of magnitude increase in sensitivity. Furthermore, the space detector eLISA \cite{eLISA} will open GW observations to massive (10$^4$-10$^6$) $M_\odot$ BHs, which could have magnetized circumbinary disks powering EM emission.  Simultaneous GW/EM emission will transform our understanding of the formation, evolution, properties and environment of different mass compact objects through  cosmic history.

Another important topic in the new astronomy will be neutrino astrophysics. Although astrophysical neutrinos have been detected by IceCube  \cite{astronu}, no significant cluster (in space or in time) has been found yet.
%A dominant contribution from extragalactic sources is expected from
%the diffuse nature of the observed events, with a galactic
%fraction at most of few 10\%.
%Results from more recent analyses, with different sensitivities to the two
%sky hemispheres and somewhat different energies compared to the HESE
%samples, indicate a tension in the best fit spectral index found under
%the assumption of uniform flux for the whole sky and one spectral
%index per each neutrino flavor [M. Aartsen et al., arXiv:1607.08006]. This tension may indicate either a
%break in the spectrum of spatially different fluxes. 
Coincidences in time and space of astrophysical neutrino  events with exceptional flares
of blazars have been proposed by several authors (e.g. \cite{kadler2016}). 
 Among the speculated possible origins of the IceCube events are GRBs with jets shocked in surrounding matter. Models explaining hypernovae and low
luminosity GRBs (e.g. \cite{senno2016}) 
predict neutrino and gamma-ray emission, with the highest energy $\gamma$-rays being  likely absorbed, and thus a possible cutoff in the MeV range. 

Observations by \ea could therefore open a new avenue within multimessenger astrophysics, also by allowing multimessenger coincidences for KM3NeT in the Mediterranean sea, and thus making it possible to remove the background for neutrinos in the TeV range.

%\newpage

%\subsection{Resolving the high-energy inner Galaxy and feedback on star formation (1.5 pages)}

\subsection{The origin and impact of high-energy particles on Galaxy evolution, from cosmic rays to   antimatter}

Relativistic particles permeate the interstellar medium (ISM) of galaxies and drive their evolution by providing heat, pressure and ionization to the clouds and to galactic winds and outflows. Sub-GeV particles have a particularly important role and understanding their origin and transport has profound implications. High-energy particles also signal the presence of antimatter and potential sources of dark matter. Observations with e-ASTROGAM can advance our knowledge on all fronts by observing the radiation borne from particle interactions with interstellar gas: electrons emitting bremsstrahlung $\gamma$ rays (often dominant below 50-100 MeV); nuclei producing $\pi^0$s decaying into $\gamma$ rays (with the characteristic ``pion bump'' in energy density below one GeV); nuclear excitation lines, and the 511-keV line from positron annihilation. In addition to the the gas-related emission, e-ASTROGAM will also observe the large-scale emission due to IC scattering of CR electrons on the interstellar radiation field and cosmic microwave background. This is a significant component above $\sim$100 MeV, and it is believed to be the dominant interstellar diffuse component below few tens of MeV. %The  energy  coverage  of  e-ASTROGAM  is  well  suited  to probe the distribution of CR electrons in the Galaxy from the bremsstrahlung and the IC emission.

We have fair measurements of the local spectrum of Galactic cosmic-rays (CRs), from GeV to PeV energies, but not at lower energies. There is only a very coarse description of their flux both radially and vertically across the Milky Way. Their diffusion processes in and out of the spiral arms and star-forming regions are poorly understood, as is their penetration through dense clouds as a function of energy. Convincing, albeit not definitive, observational evidence is available for CR  acceleration by supernova shockwaves (via diffusive shock acceleration), but little understanding of the total energy imparted to CRs, of their escape into the ambient medium, of their diffusion through stellar-wind-driven turbulence in starburst regions, and of their role in the self-regulation of the Galactic ecosystem.

\subsubsection{What are the CR energy distributions produced inside SNRs and injected into the surrounding ISM?}

\begin{figure}[h]
\centering
\includegraphics[width=\linewidth]{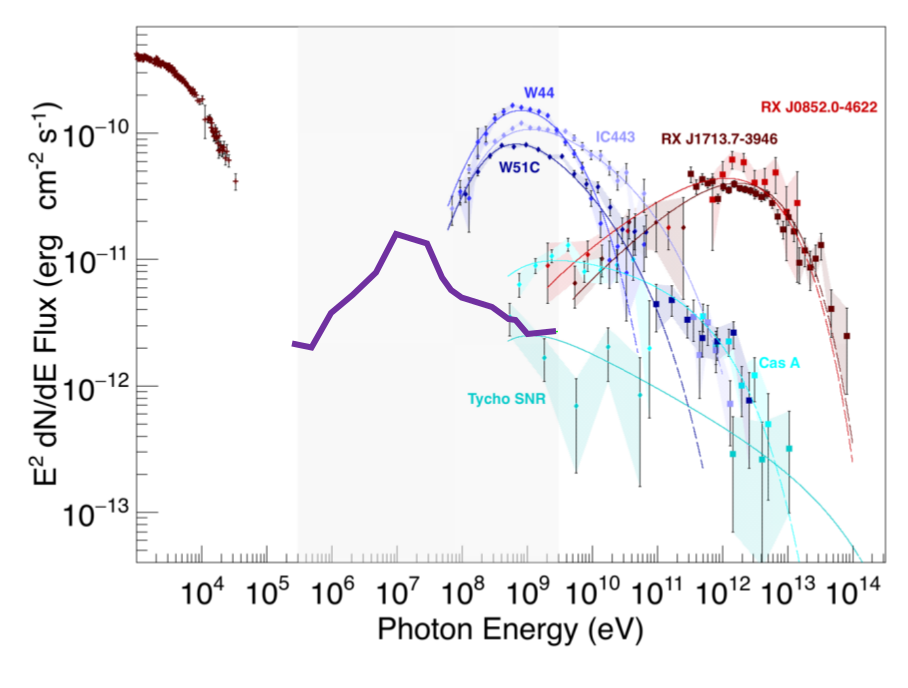}
\caption{e-ASTROGAM sensitivity for 1-year exposure (thick purple line) compared to typical $\gamma$-ray energy spectra for several SNRs; young SNRs ($<$1000 years) are shown in green. {High-energy data ($E> 100$~MeV)} are taken from \cite{funk2015}; {low-energy data} (related to RX  J1713.7-3946) from \cite{masaaki}.} \label{fig:cas-A} 
\end{figure}

 \textit{Fermi}-LAT has barely resolved a handful of SNRs, indicating that younger ones tend to emit harder gamma--rays \cite{FermiSNRcat}, and detected curved spectra compatible with pion bumps in only three sources (IC 443, W44, W51C, \cite{FermiW44IC443,FermiW51C}). The performance of e-ASTROGAM will open the way for spectral imaging of a score of SNRs, spanning ages from $10^3$ to $10^5$ years. The bremsstrahlung emitting electrons seen with e-ASTROGAM have energies close to the radio synchrotron emitting ones, and lower than those seen in synchrotron X-rays, thus permitting tomographic reconstruction of the magnetic field and electron distributions inside the remnant. Fig. \ref{fig:cas-A} indicates that e-ASTROGAM has the sensitivity in one year of exposure to detect CR electrons even for a strong mean magnetic field. The sub-GeV part of the gamma radiation is essential to separate the emission from relativistic electrons
and nuclei above 100 MeV, 
%Above 100 MeV hadronic gamma-ray emission is often dominant and has been successfully detected with  \textit{Fermi}-LAT. Separating the emission components in the GeV band is difficult though, and MeV-band studies with e-ASTROGAM would permit clean studies of electron spectra in SNRs. Figure~\ref{fig:cas-A} includes the spectral sensitivity reached in 1-yr effective time of observations of an SNR and demonstrates that eA can detect electron bremsstrahlung even for a strong mean magnetic field. 
so the new data can constrain how electrons and protons are differentially injected into the shock, how large and sometimes highly intermittent magnetic fields build up near the shock, and how the acceleration efficiency and the total CR content of a remnant evolves as the shockwave slows down. 

\begin{figure*}
\centering
\includegraphics[width=0.7\textwidth]{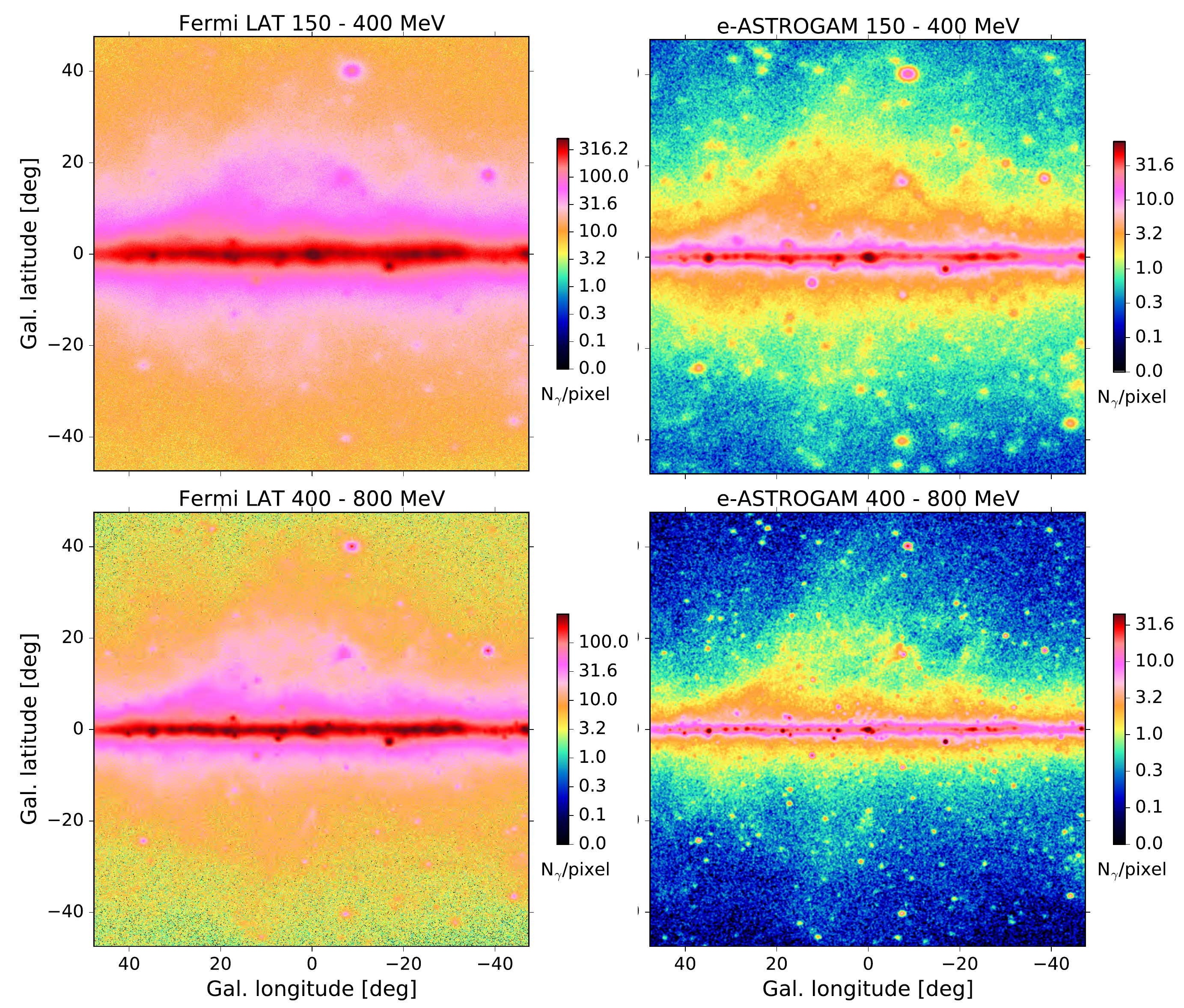}
\caption{Imaging the inner Galaxy region: simulated performance of \ea in one year 
of effective exposure (right panels) compared  to 8 years of Fermi (left panels) for
the energy ranges 150 MeV - 400 MeV (top) and 400 MeV - 800 MeV (bottom).} 
\label{fig:gcimaging}
\vspace{-10pt}
\end{figure*}

Older remnants often interact with molecular clouds that provide target gas for CRs escaping the remnant \cite{Uchiyama2012}. Resolving the diffuse pion emission produced in those clouds against the bright Galactic background is essential to probe the CR spectra that are actually injected into the ISM. Imaging the remnant and shocked clouds both require an angular resolution better than 0.2$^{\circ}$ around one GeV and a sensitivity below $10^{-11}$ erg cm$^{-2}$ s$^{-1}$ above 50 MeV that e-ASTROGAM will achieve. The instrument may also detect line emission from nuclear excitation caused by low-energy CR nuclei, opening a new and unique way to remotely measure the flux and elemental composition of low-energy particles. The comparison with direct composition measurements near the Earth would provide valuable clues as to the type of supernovae that dominate CR production in the Galaxy.

{A conservative estimate on the number of SNRs studied in detail by e-ASTROGAM can be obtained starting from the 3FGL catalog of sources in the 100 MeV--300 GeV energy range based on the first four years of science data from the $Fermi$ mission \cite{3FGL}. This catalog contains 12 SNRs, 9 PWNe, the Cygnus superbubble produced by multiple supernovae \cite{FermiCyg}, as well as 62 additional sources potentially associated to SNRs. Given that e-ASTROGAM will have a comparable or better sensitivity, FoV and observation strategy than $Fermi$-LAT in the 100 MeV--1 GeV energy range, and accounting for the measured spectra of SNRs, we expect the detection of at least the same amount of objects of this class in this energy range.

However, e-ASTROGAM will have a much improved sensitivity with respect to $Fermi$-LAT below 100 MeV. This will allow bright SNRs like IC 443, W44, W51C, and W28 to be studied for the first time with good statistics spectra in the very important energy band below the pion-decay bump, which is crucial to make firm conclusions about the hadronic and leptonic contributions to the observed emission. This is very important to constrain the CR acceleration physics. 

Another definitive advantage of e-ASTROGAM is its very good sensitivity in the MeV energy domain, where the $^{44}$Ti 1157 keV line measurements should uncover about 10 new, very young SNRs (of ages up to 500~yr) presently hidden in highly obscured clouds (see also Sect.~\ref{sec:ccn}). These objects are thought to accelerate CR up to PeV energies and will be prime targets for the Cherenkov Telescope Array (CTA) as well. }

\subsubsection{How do CR fluxes vary with Galactic environments, from passive interstellar clouds to active starburst regions and near the Galactic Center?} 
The gamma-rays produced by CR nuclei along their interstellar journey can remotely probe the CR flux and spectrum across Galactic spiral arms, inside young stellar clusters and superbubbles, in the central molecular zone, and down to parsec scales inside nearby clouds. The current picture provided by the  \textit{Fermi}-LAT analyses 
does not provide sufficient resolution to probe theoretical expectations based on spiral distributions of CR sources and environmental changes in CR transport due to their streaming in the varying Galactic magnetic field \cite{Recchia2016} or to different levels of interstellar MHD turbulence powered by massive stars and supernovae \cite{PICARD,Grenier15}. With their much improved angular resolution and lower energy band, e-ASTROGAM observations can complement the  \textit{Fermi}-LAT archival data to probe the heterogeneity of the CR population in a variety of Galactic environments, across four decades in momentum around the maximum energy density. 
%in specific Galactic environments relative to the spiral arms, relative to star-forming regions, inside young stellar clusters and superbubbles, and around the Galactic Centre. 
The \ea sub-degree resolution is essential to map the structured CR emission and avoid confusion with Galactic point sources (Fig. \ref{fig:gcimaging}). It is also key to resolve starburst regions hosting cocoons of  young energetic CRs \cite{FermiCyg} to reveal 
%The recent discovery with Fermi of a cocoon of young and energetic CRs in the Cygnus X superbubble  has disclosed an important, but unknown aspect of the interstellar CR journey (refxxx): what is 
the impact on the CR properties (energy and diffusion) of the large level of supersonic turbulence driven by the massive stars. How is the emerging CR spectrum modified by confinement, re-acceleration and enhanced losses in the turbulent medium? Since most CR sources occur in star-forming regions, these questions challenge our global understanding of the early steps of CR propagation and our use of observational diagnostics in the Galaxy at large and in starburst galaxies in particular. The sensitivity and PSF of e-ASTROGAM permit searches for Galactic CR cocoons other than Cygnus X and an extensive characterization of the particle emissions across four decades in energy (with Fermi archives and TeV data from CTA and HAWC).

With  a  sensitivity of $10^{-12}$ erg cm$^{-2}$ s$^{-1}$  around 100  MeV,  e-ASTROGAM can also detect the  pion bump from CRs in the Large Magellanic Cloud (LMC), in particular in the star-forming regions of 30 Doradus and N11 \cite{FermiLMC}, and in nearby starburst galaxies \cite{FermiStarburst}. Such observations will provide insight on the CR  in external galaxies at energies relevant for the physics of their ISM. The detection of the pion bump is crucial to disentangle the CR origin of the emission from other sources such as pulsars.

\subsubsection{Where are the low-energy CRs and how do they penetrate dense clouds?}

Locally, the CR energy density is dominated by sub-GeV and GeV protons. It is comparable to the energy densities of the interstellar gas, magnetic field, and stellar radiation. Low-energy cosmic rays (LECRs) influence galactic evolution by changing the thermodynamical state (pressure, heat) and chemical evolution (via the ionization rate) of the dense clouds that lead to star formation. They also provide critical pressure support in starburst regions to launch galactic winds into Galaxy halos \cite{Recchia16wind}. Yet, our knowledge of the production pathways and transport properties of LECRs is very rudimentary in our Galaxy, and even more so in the conditions of merger/starburst galaxies. In the Local Bubble, the pronounced break in CR momentum implied near one GeV by the Voyager and Fermi LAT data suggests that LECRs are advected off the plane  by  a  local  Galactic  wind  \cite{Schlick2014,Grenier15}. Beyond Voyager, indirect measures of the LECR flux at the low, ionizing, energies are uncertain by several orders of magnitude, even in the local ISM. Whereas multi-GeV CRs appear to penetrate deeply and rather uniformly into molecular clouds,
%and to have diffusion lengths far exceeding the cloud dimensions that ensure a rather uniform flux inside the cloud. 
molecular line observations suggest strong spatial variations in the ionization  induced by LECRs \cite{Indriolo2012}. 
%The e-ASTROGAM data can shed important light on these questions by observing the radiation borne from LECR interactions with interstellar gas. 

The  energy  band  and  performance  of  e-ASTROGAM  are  well  suited  to  probe the distribution of LECRs in different Galactic environments, both from the bremsstrahlung radiation of low-energy electrons and the pion bump from low-energy nuclei. Another long-awaited goal is to test the concentration and exclusion processes that govern the penetration of CRs into dense clouds,  which are predicted to leave an energy-dependent signature below 1 GeV \cite{Skilling1976}.% accessible to e-ASTROGAM .

%\begin{wrapfigure}{h}{0.5\textwidth}

Inelastic collisions of LECRs with interstellar gas should produce a rich spectrum of gamma-ray lines between 0.3 and 10 MeV \cite{ben13}. Spectroscopic observations of these lines with e-ASTROGAM are the only direct way to detect these elusive particles, to measure their energy density in and out of dense clouds, and to measure the production rate of light elements (Li, Be, and B) resulting from their interactions with gas. Fig. \ref{fig-low-energy-CR} shows that e-ASTROGAM should allow a firm detection of $\gamma$-ray line complexes from the inner Galaxy with a total flux of $\sim 2\times10^{-4}$ cm$^{-2}$ s$^{-1}$ sr$^{-1}$ in the (0.3-10) MeV band, and possibly also from superbubbles, hypernovae, and active star forming regions. 
%This will allow us for the first time to obtain the long-sought details of injection of low-energy particles into CR acceleration process and at the same time put new constraints on the yields of supernova nucleosynthesis and on the global dynamics of active regions in the Galaxy.

\subsubsection{The origin and energy content of  Galactic wind and Fermi bubbles}  
There is increasing evidence, observationally (e.g. \cite{Crocker11,kre13}) and theoretically, for the emergence from the inner 200 pc of the Galaxy of a Galactic wind flowing to large height ($\sim 10$ kpc) into the halo and partly accelerated by the pressure gradient supplied by CRs \cite{Breit91,Everett08}. The relation with the Fermi Bubbles seen in gamma rays and possibly in microwaves and polarized radio waves is unclear despite their biconical structure and the finding of gas at large velocities ($>$ 900 km/s) in their direction \cite{Fox15}. The Bubbles may be the few-Myr-old relics of past  accretion-driven outflow(s)  from Sgr A*  or the CR-driven Galactic wind powered by the starburst activity in the central nucleus. 
They  are  filled  with relativistic  particles  of  unknown  nature  (electrons  or  nuclei)  and  origin. The bubble shapes near the Galactic disc and their spectrum below 0.5 GeV are quite uncertain because of the large confusion with Galactic foregrounds \cite{FermiBubbles}. The improved angular resolution of e-ASTROGAM will reveal the geometry and sub-GeV spectrum of the Bubbles down to their base to help identify the dominant particles, study particle ageing in the outflow, distinguish between impulsive and wind models, and estimate the total power expelled from the modest nucleus of our Galaxy.

\begin{figure}
\centering
\includegraphics[width=0.8\linewidth]{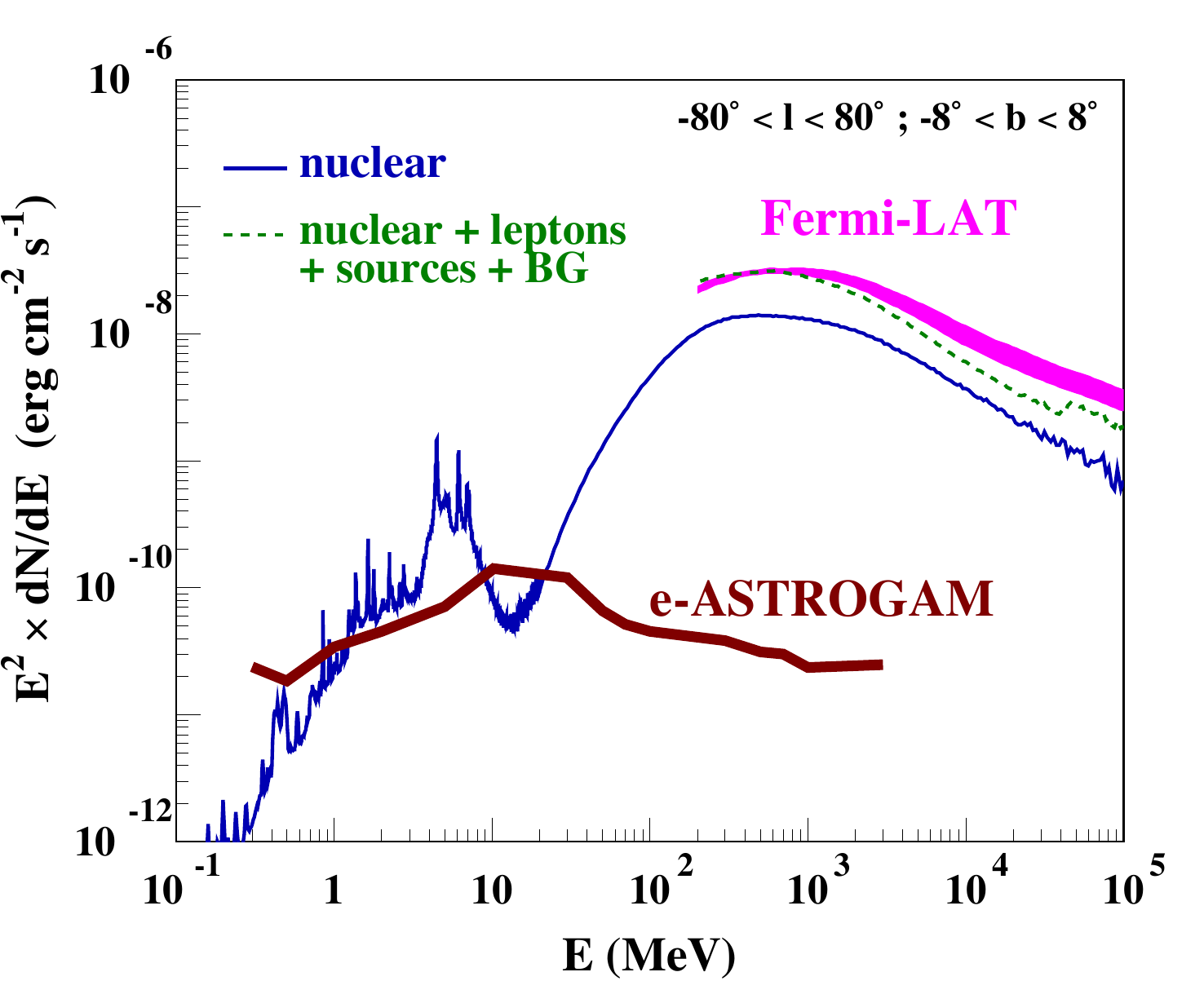}
\caption{Predicted gamma-ray emission due to nuclear interactions of CRs in the inner Galaxy. The gamma-ray line emission below 10 MeV is due to LECRs, {whose properties in the ISM have been adjusted such that the mean CR ionization rate deduced from H$_3^+$ observations and the  \textit{Fermi}-LAT data (magenta band) at 1 GeV are simultaneously reproduced (adapted from \cite{ben13})}. The 1-year sensitivity of \ea (for Galactic background) is superimposed.} \label{fig-low-energy-CR} 
\end{figure}

\subsubsection{Antimatter and WIMP Dark Matter}

The detection of a bright gamma-ray 
$e^+e^-$ annihilation line at 511 keV places strong
constraints in many emission models for high-energy
astrophysical sources. The detection of
$e^+e^-$ annihilation radiation can thus serve as an important and unambiguous calorimeter, e.g.,  to track supernova nucleosynthetic activity or constrain the presence of dark matter at the center of our Galaxy.
e-ASTROGAM will have significantly better 
sensitivity than INTEGRAL/SPI in the 511 keV line (Table~\ref{table:sensitivity_line}).
It would represent a major step forward,
for example,  in answering the many questions posed by INTEGRAL's detection
of spatially extended $e^+e^-$ annihilation in our Galaxy. Note
also that e-ASTROGAM's energy coverage is  ideally suited (compared
to Fermi) to capture the radiation from decaying pions and anti-proton
annihilation, especially if the relevant sources are cosmological and
redshifted. By either resolving the extragalactic gamma-ray
background at $\sim$1-100 MeV or constraining its angular anisotropy,
e-ASTROGAM could thus directly detect or, at least,  significantly
improve on COMPTEL constraints for anti-matter domains in the Universe
left over from the Big Bang and the process of baryogenesis.

% In the presently favored dark matter (DM) scenario, DM is a weakly interacting matter particle (WIMP) of mass O(100 GeV/$c^2$) subject to self-annihilation or decay. WIMPs are present in many  extensions of the Standard Model of particle physics (e.g. the neutralinos in supersymmetry, or the  Kaluza-Klein states
% in theories with extra dimensions, see e.g. \cite{book}). 
% Annihilation of pairs of WIMPs or WIMP decays would produce via hadronization a flux of $\gamma$-rays, and an excess of antimatter (positrons in particular). e-ASTROGAM may  be sensitive to such fluxes. In particular,   for  $m_{\rm WIMP} < 100$ GeV/$c^2$  the bulk of photons is expected below 1 GeV \cite{photonsfromdm}. 

 The theoretical paradigm about dark matter (DM) most popular today is motivated by the so-called "WIMP miracle" (i.e., the fact that we can explain the missing mass with just one more particle, and without the need for new interactions). Such a Weakly Interacting Massive Particle, or WIMP, is present in many  extensions of the Standard Model (SM) of particle physics (e.g. the neutralinos in supersymmetry, or the  Kaluza-Klein states in theories with extra dimensions, see e.g. \cite{book}). To account for the missing matter, a WIMP that comes from an equilibrium state in the early Universe should have mass O(100 GeV/$c^2$), and a self-annihilation cross section $\sigma$ and a typical velocity $v$ such that $<\sigma v> \sim 3 \times 10^{-26}$ cm$^3$/s (this value is a kind of ``thermal benchmark''). The annihilation of pairs of WIMPs in high-density DM regions could result in an excess of photons and antimatter coming mainly from the hadronic cascades; for  $m_{\rm WIMP} < 100$ GeV/$c^2$  the bulk of photons is expected below 1 GeV \cite{photonsfromdm}. Monochromatic lines from 2-body final states (either of which being a photon, or both) are not excluded.

In final states characterized by a photon continuum or by $\gamma$-ray lines, the e-ASTROGAM sensitivity for the detection of DM is complementary to Fermi and CTA because it covers with larger sensitivity the low-mass interval. Moreover, the possibility of Fermi (and in the future of CTA) to detect DM requires that these experiments demonstrate that there is an excess of photons. Modeling the photon background from astrophysical sources is where e-ASTROGAM is fundamental, since the MeV/GeV range is probably the most constraining for the SED of astrophysical sources. In addition, e-ASTROGAM can constrain the signal from nearby pulsars.

The region of the Galactic Center (GC) is expected to host the highest density of DM in the vicinity of the Earth, but the many astrophysical processes at work in the crowded inner Galaxy make it extremely difficult to disentangle the possible DM signal from conventional emissions~\cite{Prada:2004}. e-ASTROGAM will  improve our understanding of the origin of particles in the inner Galaxy, therefore reducing the uncertainties 
associated to DM searches. 

Other targets for the search of DM are dwarf spheroidal galaxies (dSphs, \cite{Walker:2009zp}), whose otherwise mysterious dynamics can be simply explained if they are highly DM-dominated ($M/L\sim10^3 M_\odot/L_\odot$).
dSph allows an almost background free observation, since they are not expected to be $\gamma$-ray emitters unless a sizable WIMP pair annihilation takes place. There are currently about 30 known dSphs, with new objects in this class being discovered. For a low WIMP mass, \ea will have a discovery potential comparable to  \textit{Fermi}-LAT --  stronger if the DM mass is on the few GeV scale.

In some extensions of the SM, DM is on the MeV scale. Theories predict a stable relic particle, in thermal equilibrium during the early Universe, with mass between 1 and 100 MeV~\cite{Boehm:2002yz}. Such models attracted some interest about a decade ago because they would naturally explain the 511-keV emission line toward the galactic bulge~\cite{Boehm:2003bt,diehl2016}. In this case e-ASTROGAM could make a direct detection or provide constraining limits~\cite{Boddy:2015fsa}. e-ASTROGAM outperforms Fermi and has a sensitivity much better than the "thermal benchmark'' for masses below 5 GeV for the search of a continuum signal in dSphs  and below 200 MeV for line searches \cite{FermiDM}.

Besides the popular paradigms outlined above, other scenarios are discussed in the literature for DM, in which e-ASTROGAM has the best sensitivity, in particular:
\begin{itemize}
\item Monochromatic photons at MeV energies may result from DM annihilation to quarkonium \cite{Rudaz:1986db}, as well as step-like features from the decay $b\to s+\gamma$ or $b'\to b+\gamma$, where $b'$ is a hypothetical 4th generation quark  \cite{Bergstrom:1988jt}. Another possibility is the decay of DM candidates like the gravitino, which has motivated line searches with the Fermi Large Area Telescope down to energies of 100 MeV \cite{Albert:2014hwa}. It was also pointed out that for DM lighter than around 100 MeV, the only kinematically accessible non-leptonic states  are photons and neutral pions, leading to clear gamma-ray signatures to look for \cite{Boddy:2015fsa}. Due to the gap in sensitivity at the MeV, very weak limits on DM signals exist in this range \cite{Essig:2013goa}.
\item A new class of potential smoking-gun signatures for DM signals in the range 10 MeV - 100 MeV is pointed out in  \cite{mevdm} and involve transitions between meson states and, in their simplest realization, do not require any new physics (beyond, obviously, the DM particle itself)  but inevitably arise in certain kinematical situations for GeV-scale DM annihilating or decaying to heavy quarks. Unlike direct detection or collider experiments, these signatures are thus very sensitive to DM coupling with third or second generation quarks.
\item Axion-Like-Particle DM (of extremely low mass); this is discussed in the "observatory science" (Sect. \ref{sec:obse}).
\end{itemize}

e-ASTROGAM can also shed new light on dark matter by  the study of antimatter -- an excess of antimatter is   expected from DM annihilation. The case of signals from excesses of antimatter is particularly intriguing: the presently measured flux of mildly relativistic cosmic rays (anti-electrons in particular) cannot be explained on the basis of present knowledge, and the data show an excess with respect to known astrophysical sources (PAMELA, AMS02; see \cite{pdg2015} for a review). Is this excess due to presently unknown sources, e.g. as yet unknown  pulsars or past activity of the GC \cite{Petrovic14,CarlsonProfumo}, or are we detecting evidence of new physics at the fundamental scale? This question can be answered by observations of nearby pulsars with e-ASTROGAM.

Finally  an  improved  angular  resolution  with respect  to  AGILE   and  {\textit{Fermi}}-LAT  in the  inner Galaxy region  and in regions closer  to Earth  in the  5  MeV--100 MeV energy range can disentangle the possible contributions  from the diffuse background, from point sources, and  other possible emitters. 
Overall, a large class of  spectral
features  in the MeV-GeV  range can result			
in indications for WIMP DM particles,
or significantly reduce
the astrophysical background uncertainties to identify genuine DM
signatures in VHE photon spectra \cite{mevdm}.

\subsection{Nucleosynthesis and the chemical evolution  of our Galaxy}

The origins of the cosmic atomic nuclei and their variety is one of the main themes of astrophysical research, and has been studied through nuclear gamma--ray emission for several decades. The Compton Gamma Ray Observatory (CGRO) provided the first sky survey of nuclear emission from cosmic sources \cite{scho96,die98}, INTEGRAL added high-resolution spectroscopy \cite{die13}. From these missions, however, only the brightest sources of their class have been seen, and a deeper survey as proposed with e-ASTROGAM will address fundamental issues in nuclear astrophysics, exploring the variability of nuclear emission from supernovae (SNe), localized stellar groups, and other transients related to compact stars and nuclear processes therein. Key science goals of e-ASTROGAM focus on the astrophysics of SNe, for which a validated and self-consistent model has not yet been established, neither for thermonuclear supernovae (SN Ia) nor for core collapse supernovae (SN Ib/c and II).  

\subsubsection{What are the progenitor system(s) and explosion mechanism(s) of thermonuclear SNe? Can we use SN~Ia for precision cosmology?}

SN Ia are the outcome of a thermonuclear burning front that sweeps a carbon/oxygen white dwarf in a close binary system. But exactly how the ignition conditions are obtained, and on which white dwarfs, and more so how the thermonuclear runaway proceeds through the white dwarf and turns it into a variety of isotopes that are ejected, are all questions that are subject to considerable debate (e.g. \cite{hil00,hi13} and references therein). It seems that several candidate evolutionary channels may all contribute, from the \emph{double degenerate} variant of merging white dwarf binaries disrupting one of the dwarfs through tidal forces or a hard collision, to a variety of \emph{single degenerate} models where accretion of material from a companion star may lead to either the white dwarf reaching the critical Chandrasekhar mass stability limit, or be ignited earlier through a surface explosion from a helium flash.  

Such uncertainties are troublesome for cosmology since the use of SN Ia as standard candles depends on an empirical relationship between the shape and the maximum of the light curve \cite{phi93}. Although useful up to now, in view of the development of \textit{precision cosmology}, a better, astrophysically supported understanding of thermonuclear SNe, as well as their evolutionary effects at large distances and low metallicities, are mandatory. The brightness-decline relation \cite{phi93} is closely related to the mass of synthesized $^{56}$Ni, and factors like the progenitor evolution, ignition density, flame propagation, mixing during the burning, completeness of burning in outer, expanding regions, all lead to different amounts of $^{56}$Ni, which is measured directly through gamma--ray lines. On the other hand, radiation transport from radioactivity to optical light and their spectra depend on complex atomic line transitions in the expanding supernova as well as total mass burned, the amount and distribution of radioactive nickel and intermediate mass elements, all of which must combine in quite a tight way to reproduce the observations \cite{woo07,ker14}. Some of these factors depend on the evolution of the white dwarf prior to the explosion and cast some doubt on the use of SN Ia as high precision distance indicators. It is thus of critical importance to disentangle the role of these factors to understand the limits of the Phillips \cite{phi93} relation. 

\begin{figure}
\centering
\includegraphics[width=0.9\linewidth]{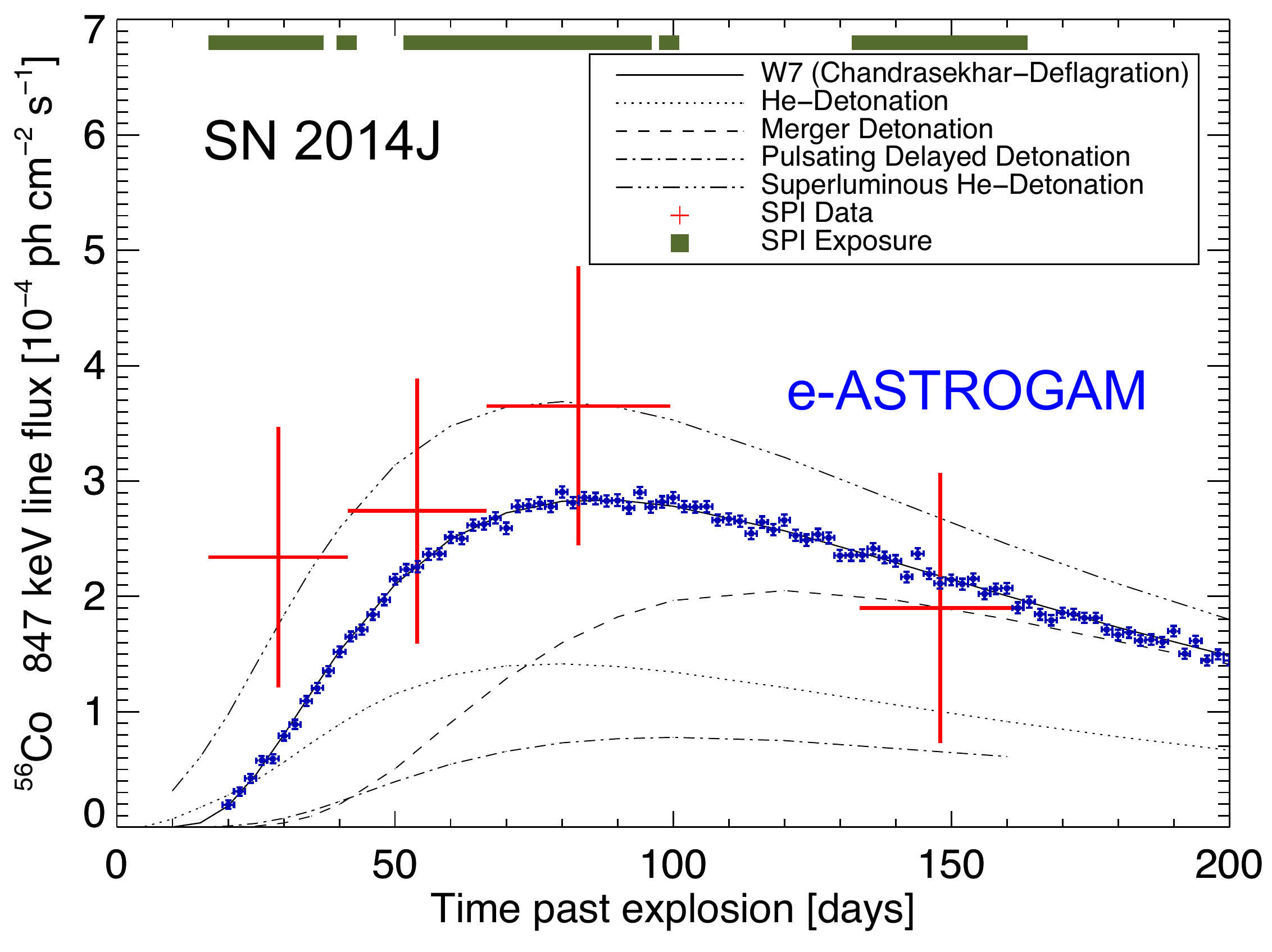}
\caption{The evolution of the 847 keV line brightness from $^{56}$Co decay reflects how radioactive energy is deposited within a supernova to make it shine. Here, INTEGRAL data from the Type Ia supernova SN~2014J (adapted from Fig.~4 in Ref.~\cite{die15}, red data points) are compared to a family of candidate models \cite{the14}. A simulation of the  response to a time evolution such as in the W7 model \cite{nom84} shows that the sensitivity improvement by \ea   (blue points) will lead to breakthrough science.}
\label{fig:sn2014j}
%\vspace{-10pt}
\end{figure}

With SN2014J, for the first time a SN Ia occurred close enough for current generation $\gamma$ray telescopes, at 3.5 Mpc in the starburst galaxy M82. INTEGRAL  could detect the long awaited gamma--ray signatures of the thermonuclear runaway, through the early emission from the decay of $^{56}$Ni (mean lifetime $\tau\simeq$8.8 days) about 20 days after the explosion, and the main gamma-ray lines at 847, 1238, and 511 keV from the decay of $^{56}$Co ($\tau\simeq$111 days). These data suggested  either a surface explosion or some unusual morphology of the runaway \cite{die14,chu14,die15,chu15,ise16}, either case in stark contrast to the conventional Chandrasekhar model. Clearly, the glimpse offered by SN2014J observations underlines the importance of gamma--ray line diagnostics in these systems and emphasize that more and better observations hold the key to a deeper  understanding of how the thermonuclear explosion of a white dwarf star unfolds. 

e-ASTROGAM will achieve a major gain in sensitivity compared to INTEGRAL for the main gamma--ray lines arising from $^{56}$Ni and $^{56}$Co decays. Thus with the expected sensitivity of $3.5 \times 10^{-6}$ ph~cm$^{-2}$~s$^{-1}$ for the 847~keV line in 1~Ms of integration time (see Table~\ref{table:sensitivity_line}), e-ASTROGAM will detect ($10 \pm 3$) SNe~Ia in 3~years of nominal mission lifetime, up to a distance of $\sim$35~Mpc (for the brightest SNe). As illustrated in Fig.~\ref{fig:sn2014j}, e-ASTROGAM will provide much better data than we have now with INTEGRAL for SN~2014J from similarly nearby events. These data will allow us to probe the explosion mechanism in detail, and compare with astrophysical models for each event to better understand the progenitor system(s) and the thermonuclear explosion process.

Gamma rays escaping from supernovae contribute to the diffuse background in the MeV range. The contribution from Type Ia supernovae dominates, potentially providing a significant fraction of the observed flux in the MeV band. (see, e.g., \cite{ruizl}).

\subsubsection{How do core-collapse supernovae (CCSNe) explode? What is the recent history of CCSNe in the Milky Way?\label{sec:ccn}}

Similar to SN Ia, core collapse physics is also not well understood in terms of an astrophysical model. But these events are more common, being the end states of the evolution of massive stars, and are key to understanding the diversity of elements in the universe. Also here, deviations from spherical symmetry are the rule. The goal is to explain a tremendous variety of core collapse events, e.g. electron capture supernovae such as the Crab, clumpy explosions such as Cas A, collapsars that appear as GRB sources and produce stellar mass black holes, superluminous supernovae that may be powered entirely differently by magnetar rotational energy, or pair instability supernovae that create huge amounts of radioactive $^{56}$Ni. 

% The formation of stellar mass black holes is one of the outcomes, depending on how the collapse occurs, also the properties of neutron stars and the role of magnetars when magnetic fields are amplified through the collapse. 

Stellar rotation is known to exist but is complicated to track in its effects on stellar evolution, yet important for many of the above outcomes: nucleosynthesis, pre-supernova structure, core collapse. Measuring nucleosynthesis products such as $^{56}$Ni, $^{56}$Co and $^{44}$Ti is one of the more direct ways to extract information on the inner processes triggering the explosion near the newly forming compact stellar remnant (e.g., \cite{gre14}) -- other observables are indirect, and mostly reflect interactions within the envelope, or with circumstellar, pre-explosively ejected, or ambient gas. 
e-ASTROGAM will detect the signatures of $^{56}$Ni and $^{56}$Co decay from several CCSNe in nearby galaxies. Comparing $\gamma$-ray characteristics of different classes of CCSNe, possibly including the pair instability SNe with their order of magnitude higher $^{56}$Ni production \cite{gal09}, will probe potentially large variations in their progenitors and offer a direct view of their central engines.  

With a gain in sensitivity for the $^{44}$Ti line at 1157~keV by a factor of 27 compared to INTEGRAL/SPI (see Table~\ref{table:sensitivity_line}), e-ASTROGAM should also detect the radioactive emission from $^{44}$Ti ($\tau\simeq$87 years) from most of the young (age $\lesssim$ 500 yr) SNRs in the Milky Way, thus uncovering about 10 new SNRs in the Galaxy, as well as the youngest SNR(s) in the LMC. {Among the youngest Galactic SNRs, only Cassiopeia A has been firmly detected in $^{44}$Ti surveys carried out up to now \cite{tsy16}, which is surprising in view of an otherwise inferred rate of one CCSN every 50 years in our Galaxy. e-ASTROGAM will measure in particular the amounts of $^{44}$Ti in the SNRs G1.9+0.3, Tycho and SN~1987A (LMC), which are currently disputed in the literature (see, e.g., \cite{tsy16}).} Besides giving new insights on the dynamics of core collapse at the mass cut, e-ASTROGAM observations of young SNR will be very important for cosmic-ray physics. Indeed, only young SNRs can accelerate CR up to PeV energies and above, and the identification of new Pevatrons would be invaluable when the Cherenkov Telescope Array (CTA) will be fully operational.

%The pre-supernova star and its structure is important for if and how the star may explode at the end of its evolution. Binaries may play a role here, in particular in peeling off the envelope of the star prior to explosion. But strong mass loss may also occur otherwise. Currently, the different supernova light curve shapes with long lasting emission reflect interactions of the supernova with such pre-supernova mass loss materials, but none of the mass loss models seems to fit. Moreover, it seems necessary that the neutrino powered explosion may need help from secondary energy supplies such as stellar rotation and density gradients in the outer regions of the star. Here, radioactivity gamma rays from $^{26}$Al and $^{60}$Fe can help: massive stars produce both those radioactive ejecta, but only $^{60}$Fe is contributed from He and C shell burning stages. Ejected in the core collapse event, the ratio of those radioactive isotopes can be measured in gamma rays, and reflects the nucleosynthesis conditions in those critical outer shells of a massive star during the last hunderd thousand years before the core collapse event. 

\subsubsection{Nova explosions\label{sec:novae}}

e-ASTROGAM will also contribute to the study of novae, which are responsible for the enrichment of the Galaxy in some species and for the peculiar isotopic signatures found in some pre-solar meteoritic grains \cite{jos07}. Emission is expected \cite{gom98,her04} from positron-electron annihilation (e$^+$ being emitted by the $\beta^+$-unstable short-lived isotopes $^{13}$N and $^{18}$F), and from the decay of the medium-lived isotopes $^{22}$Na, which is produced in ONe novae, and $^{7}$Be, produced in CO novae. The first type of emission consists of a 511~keV line plus a continuum between about 20 and 511 keV, whereas the second, long-lasting emission consists of two $\gamma$-ray lines at 1275~keV ($^{22}$Na) and 478~keV ($^{7}$Be). Both   provide a direct insight on the amount of radioactive nuclei in the expanding nova envelope, an information only obtainable through the observation of $\gamma$-rays. The large FoV of e-ASTROGAM is crucial for these observations, since the 511 keV line emission happens before the nova is discovered optically, preventing pointed observations. The detectability distance with e-ASTROGAM is around 3~kpc for both the 478 and 1275~keV line, and 4--6 novae at distance $D < 3$~kpc are expected in the 3 years of nominal mission lifetime. 

e-ASTROGAM will also help to disentangle the current puzzle posed by the detection of novae at GeV energies 
with {\it Fermi}-LAT \cite{Ferminovae,cheung16}. Some novae have been identified theoretically as sites of particle acceleration, in the shocks within the ejecta and/or between the ejecta and the circumstellar matter, making them responsible for a fraction of the Galactic cosmic rays \cite{tat07}. These ``miniature supernovae'' are key systems to study the time dependence of diffusive shock acceleration of CR. An important consequence of the production of high-energy particles is that photons with energies higher than about 100 MeV are emitted, both via neutral pion decay and IC processes. 
%This theoretical prediction has been confirmed by the {\it Fermi}-LAT detection of several novae \cite{Ferminovae}. 
With the e-ASTROGAM sensitivity, it will be possible to disentangle the origin of this high-energy emission, hadronic and/or leptonic, and thus understand better the properties of the nova ejecta and the shocks.
 
\subsubsection{How are cosmic isotopes created in stars and distributed in the interstellar medium?}

The cycle of matter proceeds from the formation of stars through nuclear fusion reactions within stars during their evolution, towards the ejection of stellar debris into interstellar space in winds and SN explosions. Interstellar gas, enriched with some newly produced nuclei, eventually cools down to form new stars, closing and starting the cycle again. The cooling down of hot nucleosynthesis ejecta and their trajectories towards new star formation are particularly hard to constrain through observations. The recycling time scale in the interstellar medium is of the order of tens of Myr, but SNRs can be seen over time scales of few 10$^5$~yr at most. Long-lived radioactive gamma--ray emitters $^{26}$Al ($\tau\simeq$1.0$\times$10$^6$~yr) and $^{60}$Fe ($\tau\simeq$3.8$\times$10$^6$~yr) can trace mixing processes of ejecta into the next generation star forming regions over much longer time, testing, among other things, molecular-cloud lifetime and models for stimulated/triggered star formation. 

INTEGRAL/SPI data for the 1809~keV $^{26}$Al line suggest that on the global, Galactic scale, superbubbles are key structures in the transport of fresh ejecta towards new star forming regions \cite{kre13,kra15}. With its huge increase in sensitivity, e-ASTROGAM will provide a detailed view of the morphology of this emission, with high precision measurements of the line flux from many regions of the Galaxy. Thus, e-ASTROGAM will observe the $^{26}$Al radioactivity from dozens of nearby ($\sim$kpc) stellar objects and associations. In particular, it will measure precisely the amount of $^{26}$Al ejected by the Wolf-Rayet star WR11 in the $\gamma^2$-Velorum binary system (expected line flux of $\sim$10$^{-5}$~ph~cm$^{-2}$~s$^{-1}$), thus providing a unique calibration of the $^{26}$Al production during the Wolf-Rayet phase of a massive star. e-ASTROGAM has also the capability of detecting $^{26}$Al emission from  the LMC (expected line flux of $\sim$10$^{-6}$~ph~cm$^{-2}$~s$^{-1}$), thus providing new insight into stellar nucleosynthesis outside the Milky Way. 

For the first time, e-ASTROGAM will provide the sensitivity needed to  establish the Galactic $^{60}$Fe emission and build an accurate map of the $^{60}$Fe flux in the Milky Way, enabling its comparison with the $^{26}$Al map to gain insight into the stellar progenitors of both radioisotopes. In particular, measuring $\gamma$-ray line ratios for specific massive-star groups will constrain $^{60}$Fe production in massive stars beyond $\sim$40~$M_\odot$, which directly relates to stellar rotation and uncertain convective-layer evolution in massive star interiors  \cite{lim06}.

\subsection {Observatory science in the MeV - GeV domain\label{sec:obse}}

During the first phase  \ea will collect data especially for the core science topics, as described above. However, given the very large sky coverage  and the accumulated exposure, a very large number of sources can be detected and monitored. \ea has the capability of studying thousands of sources both Galactic and extragalactic of which many are expected to be new detections. Therefore, a very large community of astronomical users will benefit from \ea data available for multifrequency studies through a Guest Investigator programme managed by ESA.

\ea will detect with highly improved sensitivity in the MeV-GeV domain phenomena characterized by: 
(1) rapid and very rapid variability timescales (sub-second, second, minutes, hours), and (2) steady sources. The \ea sensitivity to pointlike sources varying on timescales of seconds (for GRBs) and minutes-hours-days (compact objects, novae, magnetars, blazars) will provide unique information about outstanding physical processes including jet processes, shock accelerations and magnetic field reconnection.  The study of steady sources (diffuse emission, pulsars, PWNe, SNRs, extragalactic background) will provide a detailed diagnostic of fundamental processes that operate in quasi-stable regimes. The \ea Observatory science program will emphasize multifrequency response to both variable and steady sources in a decade that will benefit from the operations of many other observatories planned to be operative in the 2030s, that include LIGO-Virgo-GEO600-KAGRA, SKA, ALMA, E-ELT, LSST, ATHENA, CTA and possibly e-LISA. \ea will provide unique data for multifrequency science, triggering other instruments and reacting rapidly to transient detections. 

We summarize here the most relevant classes of phenomena or sources in addition to the \ea ``core science"  topics. 
\begin{itemize}
\item {\bf Diffuse Galactic gamma-ray background}, for which \ea is in a position to determine the underlying cosmic ray population and spatial and spectral variations across the Galaxy. 
%\item {\bf Diffuse extragalactic gamma-ray background}, 
\item {\bf Pulsars and millisecond pulsars both isolated and in binaries}, whose (pulsed or unpulsed) emission will be observable in a spectral range rich in information to discriminate between  particle acceleration models.
\item {\bf Pulsar wind nebulae}, a product of the interaction between shocked relativistic pulsar winds and the ISM, for which \ea will obtain crucial data on particle acceleration and propagation. 
\item {\bf Magnetars}, enigmatic and strongly variable compact stars characterized by very strong magnetic fields that exhibit special phenomena exclusively in the MeV energy range.
\item {\bf Galactic compact binaries}, including white dwarfs, neutron stars and stellar mass black holes whose spectral transitions and outbursts in the MeV range will be systematically monitored by \eap.
\item {\bf Novae}, that in addition to line emission in the MeV range can also be studied for their surprising and poorly understood $\gamma$-ray emission up to hundreds of MeV, a product of shock interaction within the nova ejecta and/or of the nova ejecta with the circumstellar matter (see Sect. \ref{sec:novae}).
\item {\bf Massive binary stars with colliding winds of the Eta-Carinae type}, whose MHD shocks are predicted to produce particle acceleration and $\gamma$-ray emission, a topic of great interest and yet unsettled. 
\item {\bf Interstellar shocks}, such as the Cygnus cocoon showing the existence of particle acceleration over large distances in the ISM, for which the spectral and angular resolution of \ea will be unique.
%\item {\bf Seyfert galaxies}, for which MeV data are absolutely crucial for modeling.
\item {\bf Blazar population studies in the MeV range}, to be obtained by the detection capability of thousands of sources by \eap. 
\item {\bf Studies of the propagation of $\gamma$-rays over cosmological distances}, for which the attenuation is predicted to be negligible in standard QED - effects of absorption might indicate new physics at work, possibly the existence of axion-like-particles (ALPs) coupling to gamma--rays \cite{dmr2008}.  ALPs $a$ are spin-0, neutral and very light particles predicted by many extensions of the SM~\cite{alp1}, coupling to two photons through an amplitude $g_{a \gamma} \, {\bf E} \cdot {\bf B}$ (presently bound by $g_{a \gamma} < 0.66 \cdot 10^{- 10} \, {\rm GeV}^{- 1}$~\cite{cast,pdg2016} for a wide mass range.) In a  photon beam emitted by a far-away source (blazar, GRB),  $\gamma \to a$ and $a \to \gamma$ conversions can take place, resulting in photon-ALP oscillations. These may show up  in the e-ASTROGAM energy spectrum about $E_L = 2.56 \cdot 10^{21} \left(m/{\rm eV} \right)^2 \, \xi^{- 1} \, {\rm GeV}$ for an ALP mass   $3.42 \cdot 10^{- 13} \, \xi^{1/2} \, {\rm eV} < m < 3.42 \cdot 10^{- 11} \, \xi^{1/2} \, {\rm eV}$, where $\xi \equiv \left(g_{a \gamma} \, 10^{11} \, {\rm GeV} \right) \left(B/{\rm nG} \right)$~\cite{dmr2008}. For a $100 \%$ polarized beam the amplitude of the fluctuations can be twice then for an unpolarized beam~\cite{wb2012,gr2013}.
CCSNe can also be 
a source of ALPs,  produced in the few seconds after
the explosion inside the core by  Primakoff effect, and reconverted to photons of the same energy (peaking at $\sim$50 MeV) in the Milky Way
magnetic field. The arrival time of these photons would be the same as for neutrinos, providing a clear signature; e-ASTROGAM would have a sensitivity better than \textit{Fermi}-LAT  \cite{meyer} and access  to much smaller mass/coupling values
than dedicated laboratory experiments.
\item {\bf  Solar flares and contribution to ``Space Weather"}, that will be studied with unprecedented line emission and continuum capability for theoretical modeling as well as fast reaction for alerts.

The Sun is an efficient particle accelerator during flares. A possible mechanism requires that ions and
electrons are accelerated by the release of energy during magnetic reconnection in the upper corona; they can then propagate into
the chromosphere where they interact to produce a strong X/$\gamma$-ray bremsstrahlung continuum, 
nuclear lines (through nuclear interactions with the solar atmosphere:
prompt deexcitation lines from e.g. $^{12}$C* at 4.44 MeV and $^{16}$O* at 6.2 MeV but also a neutron capture line at
 2.2 MeV from $n$ H $\rightarrow$ D + 2.2 MeV photons), and pion-decay components.

After a solar flare, the Sun is predicted to produce $\gamma$-ray emission through radioactive decays (e.g. lines at
847 and 1434 keV from the decay of $^{56}$Co and $^{52}$Mn, respectively). The detection of these lines would provide
additional information on energy spectra and composition of flare-accelerated particles,  and on mixing
processes in the solar atmosphere. e-ASTROGAM can study for the first time solar-flare 
radiation from 300 keV to $>$100 MeV.%; the expected detection rate is $\sim ??$ flares/ year.

To summarize, the  continuum and line sensitivity of e-ASTROGAM in the (0.3-100) MeV energy range will provide a
 diagnostic for acceleration mechanisms  with simultaneous
information at GeV energies. MeV polarization will add important information on the
in-situ magnetic properties.

The  solar emission produced by interactions of CRs with the surface and the heliosphere \cite{or3,or1,or2} will be also observed. This will allow  precise studies on the CRs and their propagation very close to the Sun.

\item {\bf Terrestrial Gamma-Ray Flashes}, an atmospheric phenomenon with possible environmental impact for which \ea can provide continuous monitoring (including the 511-keV line detection). 

Atmospheric lightning and thunderstorms can produce particle acceleration in  TGFs up to the GeV; the overall expected TGF detection rate by e-ASTROGAM is of about 60 TGFs per day, improving by
a factor of 30 $Fermi-$LAT. Observations   in the Compton and pair production regime
will provide TGF imaging in the MeV - 100 MeV range. The optimal sensitivity of e-ASTROGAM in the MeV-GeV range will clarify the physics and the many implications of this puzzling atmospheric phenomenon.
\end{itemize}

%\newpage

\section{Scientific Requirements}\label{sec:scirec}

e-ASTROGAM's requirements to achieve its core science objectives, such as the angular and energy resolution, the field of view, the continuum and line sensitivity, the polarization sensitivity, {and the timing accuracy}, are summarized in Table~\ref{table:requirements}. 

\begin{table*}
\centering
\caption{e-ASTROGAM scientific requirements.}
\vspace{-0.1cm}
\includegraphics[width=0.9\textwidth]{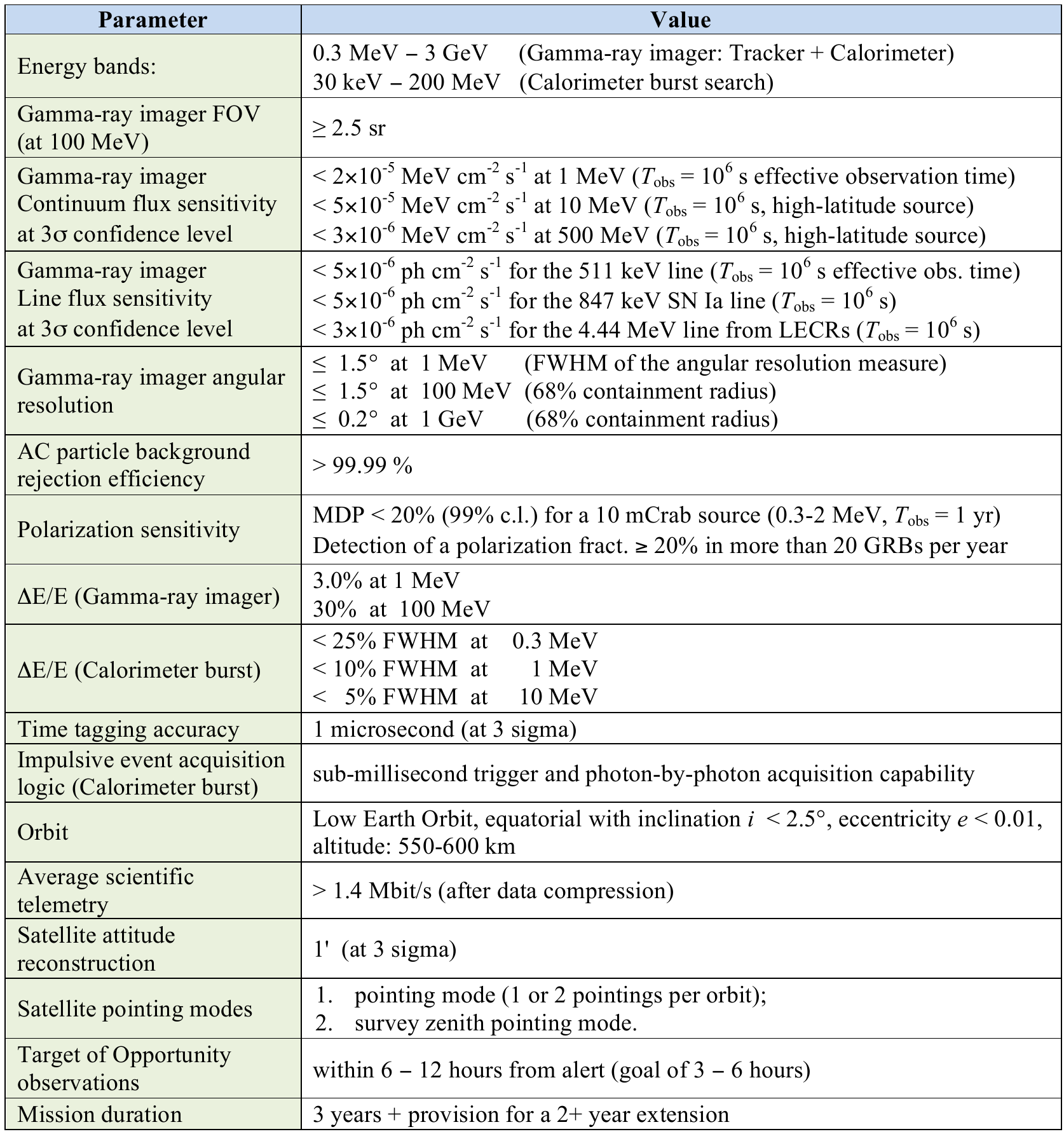}
\label{table:requirements}
\end{table*}

{ 
\begin{itemize}
\item The very large spectral band covered by the telescope in the standard gamma-ray acquisition mode will give a complete view of the main nonthermal processes at work in a given astrophysical object, for the first time with a single instrument. The e-ASTROGAM energy band includes the 511~keV line from $e^+e^-$ annihilation, the nuclear de-excitation lines, the characteristic spectral bump from pion decay, the typical domains of nonthermal electron bremsstrahlung and IC emission, as well as the high-energy range of synchrotron radiation in sources with high magnetic field ($B \gsim 1$~G). The designed wide energy band is particularly important for the study of blazars, GRBs, Galactic compact binaries, pulsars, as well as the physics of CRs in SNRs and in the ISM. 
\item The large energy band covered by the Calorimeter in the burst search mode of data acquisition (Sect.~\ref{sec:trigger}) is primarily designed for the triggering and study of GRBs. It is also well adapted to the broadband emissions of TGFs and solar flares. 
\item The wide field of view of the telescope is especially important to enable the measurement of source flux variability over a wide range of timescales both for a-priori chosen sources and in serendipitous observations. Coupled with the scanning mode of operation, this capability enables continuous monitoring of source fluxes that will greatly increase the chances of detecting correlated flux variability with other wavelengths. The designed wide field of view is particularly important for the study of blazars, GRBs, Galactic compact objects, supernovae, novae, and extended emissions in the Milky Way (CRs, radioactivity). It will also enable, for example, searches of periodicity and orbital modulation in binary systems. 
\item One of the main requirements of \ea is to improve dramatically the detection sensitivity in a region of the electromagnetic spectrum, the so-called MeV domain, which is still largely unknown. The sensitivity requirement is relevant to all science drivers discussed above. Thus, the goal of detecting a significant number ($N > 5$) of SN~Ia in gamma rays after 3 years requires a sensitivity in the 847~keV line $<5 \times 10^{-6}$ ph~cm$^{-2}$~s$^{-1}$  in 1~Ms of integration time (Table~\ref{table:requirements}). 
\item Another major requirement for a future gamma-ray observatory is to improve significantly the angular resolution over past and current  missions, which have been severely affected by a spatial confusion issue. Thus, the e-ASTROGAM angular resolution will be excellent in the MeV range and above a few hundreds of MeV, improving $CGRO$/COMPTEL and $Fermi$-LAT by almost a factor of 4 at 1 MeV and 1 GeV, respectively. The targeted angular resolution given in Table~\ref{table:requirements} is close to the physical limits: for Compton scattering, the limit is given by the Doppler broadening induced by the velocity of the atomic electrons, while for low-energy pair production, the limit is provided by the nuclear recoil. e-ASTROGAM angular resolution will allow a number of currently unidentified gamma-ray sources (e.g. 992 sources in the 3FGL catalog \cite{3FGL}) to be associated with objects identified at other wavelengths. The GC region is the most challenging case, for which the e-ASTROGAM capability will be fully employed. 
\item The polarization sensitivity of e-ASTROGAM is designed to enable measurements of the gamma-ray polarization fraction in more than 20 GRBs per year (GRBs being promising candidates for highly gamma-ray polarized sources, see, e.g., \cite{mcc16}). Such measurements will provide important information on the magnetization and content (leptons, hadrons, Poynting flux) of the relativistic outflows, and, in the case of GRBs at cosmological distance, will address fundamental questions of physics related to vacuum birefringence and Lorentz invariance violation (e.g., \cite{got14}). With the designed polarization sensitivity, e-ASTROGAM will also be able to study the polarimetric properties of more than 50 pulsars, magnetars, and black hole systems in the Galaxy.
\item The spectral resolution of e-ASTROGAM is well adapted to the main science drivers of the mission. Thus, the main gamma-ray lines produced in SN explosions or by LECR interactions in the ISM are significantly broadened by the Doppler effect, and a FWHM resolution of 3\% at 1 MeV is adequate. In the pair production domain, an energy resolution of 30\% will be more than enough to measure accurately putative spectral breaks and cutoffs in various sources and  distinguish the characteristic pion-decay bump from leptonic emissions. 
\item The timing performance of e-ASTROGAM is mainly driven by the physics of magnetars and rotation-powered pulsars, as well as by the properties of TGFs. The targeted microsecond timing accuracy is already achieved in, e.g., the AGILE mission \cite{tav09}. 
\end{itemize}
}

The e-ASTROGAM requirements reflect the dual capacity of the instrument to detect both Compton scattering events in the 0.3 (and below) -- 10 MeV range and pair-producing events in the 10 MeV -- 3 GeV energy range; a small overlap around 10 MeV allows (although in a limited energy band)   cross-calibration, thus reducing systematic uncertainties. The main instrument features of e-ASTROGAM necessary to meet the scientific requirements in Table~\ref{table:requirements}, are described in Sect. \ref{sec:perfass}.

The sensitivity performance is consistent with the requirement of an equatorial low-Earth orbit (LEO)  of altitude in the range 550~--~600 km.  Such an orbit  is preferred for a variety of reasons. It has been demonstrated to be only marginally affected by the South Atlantic Anomaly and is therefore a low-particle background orbit, ideal for high-energy observations. The orbit is practically unaffected by precipitating particles originating from solar flares, a virtue for background rejection. Finally, both ESA and ASI have satellite communication bases near the equator (Kourou and Malindi) that can be efficiently used as mission ground stations.

Table~\ref{table:requirements} also includes the most important system requirements such as the satellite attitude reconstruction, telemetry budget, and pointing capability. e-ASTROGAM is a multi-purpose astrophysics mission with the capability of a very flexible observation strategy. Two main scientific observation modes are to be managed by the Mission Operation Center (MOC):
\begin{compactitem}
\item[$\bullet$] pointing mode;
\item[$\bullet$] survey mode. 
\end{compactitem}

The pointing mode can be implemented either in a fixed inertial pointing or in the more efficient double-pointing per orbit mode. In the latter case, the e-ASTROGAM satellite is required to be able to perform two sky pointings  per orbit, lasting approximately 40 minutes each. The survey mode consists in a continuous pointing to the zenith to perform a scan of the sky at each orbit. This mode can be activated at any time in principle, and depending on the scientific prioritization and on the mission schedule foreseen by the Science Management Plan, can lead to an optimized  all-sky survey.

Requirements for the Ground Segment are standard for an observatory-class mission. Target of Opportunity observations (ToOs) are required to follow particularly important transient events that need a satellite repointing. The e-ASTROGAM mission requirement for ToO execution is within 6--12 hours, with the goal of reaching 3--6 hours. The speed of repointing depends on the torque of the reaction wheels. We expect a repointing velocity similar to Fermi ($\sim 30$ degrees/min, which grants to have a visible object in FoV within less than 5'). 

e-ASTROGAM  does not use any consumable and could in principle be operated for a duration up to 10-20 years (well within the foreseen operation duration of 3 years
with a possible extension of two), limited mainly by orbital instabilities and by the risk of accidents. Radiation damage in LEO, with almost equatorial inclination, is negligible. As an example, the degradation of Fermi, whose inclination implies significant crossing of the South Atlantic Anomaly, is negligible for what concerns electronics, negligible for what concerns Tracker aging, and around  1\%/year in terms of loss in light yield of the Calorimeter crystals.

%\end{document}

%\newpage

Table \ref{tab:nevents} summarizes our conservative estimates of the number of sources detectable by e-ASTROGAM in 3 years, based on current knowledge and $\log N - \log S$ determinations of Galactic and extragalactic sources, {including} GRBs. It takes information from the {the \textit{Swift}-BAT 70-Month Hard X-ray survey catalog \cite{b70h}, the 4th \textit{INTEGRAL}-IBIS catalog \cite{ibisc}, and the 3rd \textit{Fermi}-LAT catalog \cite{3FGL}. Noteworthy, the latter catalog contains more than 1000 unidentified sources in the 100 MeV -- 300 GeV range with no counterparts at other wavelength, and most of them will be detected by e-ASTROGAM, in addition to  a relevant number of new unidentified sources. The discovery space of e-ASTROGAM for new sources and source classes is very large.}

\vskip 2mm
\begin{table}
\begin{center}
\begin{tabular}{| l | l | l |}\hline
Type & 3 yr & New sources\\ \hline
%All sky (above 100 MeV) & $> 3000$ & $\sim$1800 (including GRBs)  \\
{Total} & {3000 -- 4000} & $\sim$1800 (including GRBs)  \\
Galactic & $\sim1000$ & $\sim$400 \\
%Galactic sources $(>$ 30 MeV) & &\\
MeV blazars  & $\sim350$ & $\sim350$ \\
GeV blazars  & {1000 -- 1500} & $\sim350$ \\
Other AGN  ($< $10 MeV)& {70 -- 100} & {35 -- 50}\\
Supernovae  & {10 -- 15} & {10 -- 15}\\
Novae & 4 -- 6 &   4 -- 6 \\
GRBs  & $\sim$600 &  $\sim$600\\ \hline
\end{tabular}
\end{center}
\caption{Estimated number of sources {of various classes} detectable by e-ASTROGAM in 3 years. {The last column gives the number of sources not known before in any  wavelength.} \label{tab:nevents}}
\end{table}

\section{The Scientific Instrument}
\subsection{Measurement principle and payload overview}

The e-ASTROGAM payload is shown in Figure~\ref{fig:payload}. It consists of three main detectors: 
\begin{itemize}
\item A {\bf silicon Tracker} in which the cosmic gamma-rays undergo a first Compton scattering or a pair conversion; it is based on the technology of double sided Si strip detectors to measure the energy and the 3D position of each interaction with an excellent energy and spatial resolution;
\item  A 3D-imaging {\bf Calorimeter} to absorb and measure the energy of the secondary particles; it is made of an array of small scintillation crystals (33,856 CsI (Tl) bars of 5$\times$5$\times$80 mm$^3$) read out by silicon drift photodetectors to achieve the required energy resolution (4.5\% at 662 keV);
\item  An {\bf Anticoincidence system} (AC), composed of a standard AC shielding { surrounding the top and four lateral sides of the instrument, and a Time-of-Flight unit located below the instrument, to veto the particle background arising from the platform}; it is made of plastic scintillator tiles with a detection efficiency exceeding 99.99\%.
\end{itemize}

\begin{figure}
\centering
\includegraphics[width=0.9\linewidth]{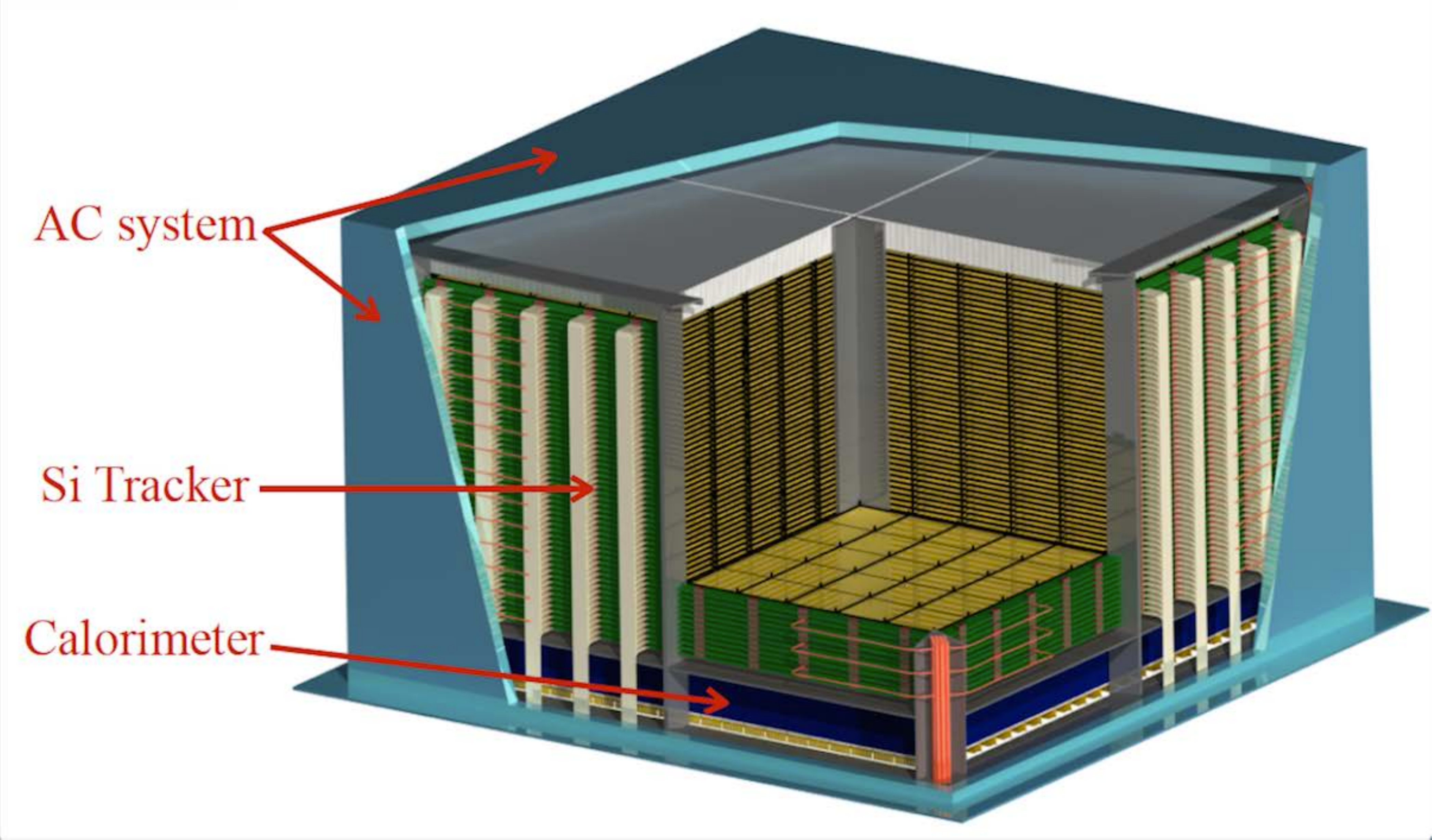}
\caption{Overview of the e-ASTROGAM payload showing the silicon Tracker, the Calorimeter and the Anticoincidence system.}
\label{fig:payload}       
\end{figure}

The payload is completed by a Payload Data Handling Unit (PDHU) and a Power Supply Unit (PSU) located below the Calorimeter inside the platform together with the back-end electronics (BEE). The PDHU is in charge of the payload internal control, the scientific data processing, the operative mode management, the on-board time management, and the telemetry and telecommand management. The total payload mass and power budget (including maturity margins) are 999~kg and 1340~W, respectively. 

Interactions of photons with matter in the e-ASTROGAM energy range is dominated by Compton scattering from (below) 0.2~MeV up to about 15 MeV in silicon, and by $e^+e^-$ pair production in the field of a target nucleus at higher energies. e-ASTROGAM maximizes its efficiency for imaging and spectroscopy of energetic gamma-rays by using both processes. Figure~\ref{fig:evt-top} shows a schematic representation of topologies for Compton and pair events.

For pair-production events, e-ASTROGAM is similar in design to AGILE and $Fermi$-LAT, but optimized for lower energy. This goal is achieved by eliminating the passive  converters used in both these instruments. This approach reduces gamma-ray conversion efficiency, but it improves the instrument point-spread function (PSF) by reducing absorption and multiple Coulomb scattering of the electron and positron.
The broad PSF is a primary limiting factor in the science that can be done at energies below 100 MeV by AGILE and $Fermi$-LAT. Pair events produce two main tracks from the created electron and positron. Tracking of the initial opening angle and of the plane spanned by the electron and positron tracks enables direct back-projection of the source position. Multiple scattering of the pair in the tracker material  leads to broadening of the tracks and limits the angular resolution. The nuclear recoil taking up an unmeasured momentum results in an additional small uncertainty. The energy of the gamma-ray is measured using the Calorimeter and information on the electron and positron multiple scattering in the Tracker. Polarization information in the pair domain is given by the azimuthal orientation of the electron-positron plane; in addition to improving the PSF, the use of low-mass tracker planes also enables photon polarization measurements.

%For Compton events, point interactions of the gamma-ray in the Tracker and Calorimeter produce spatially resolved energy deposits, which have to be reconstructed in sequence using the redundant kinematic information from multiple interactions. Once the sequence is established, two sets of information are used for imaging: the total energy and the energy deposit in the first interaction measure the first Compton scatter angle. The combination with the direction of the scattered photon from the vertices of the first and second interactions generates a ring on the sky containing the source direction. Multiple photons from the same source enable a full deconvolution of the image, using probabilistic techniques. For energetic Compton scatters (above $\sim$1 MeV), measurement of the track of the scattered electron becomes possible, resulting in a reduction of the event ring to an arc, hence further improving event reconstruction. Compton scattering angles depend on polarization of the incoming photon, hence careful statistical analysis of the photons for a strong (e.g., transient) source yields a measurement of the degree of polarization of its high-energy emission (e.g. \cite{for08}).

\begin{figure}
\centering
\includegraphics[width=\columnwidth]{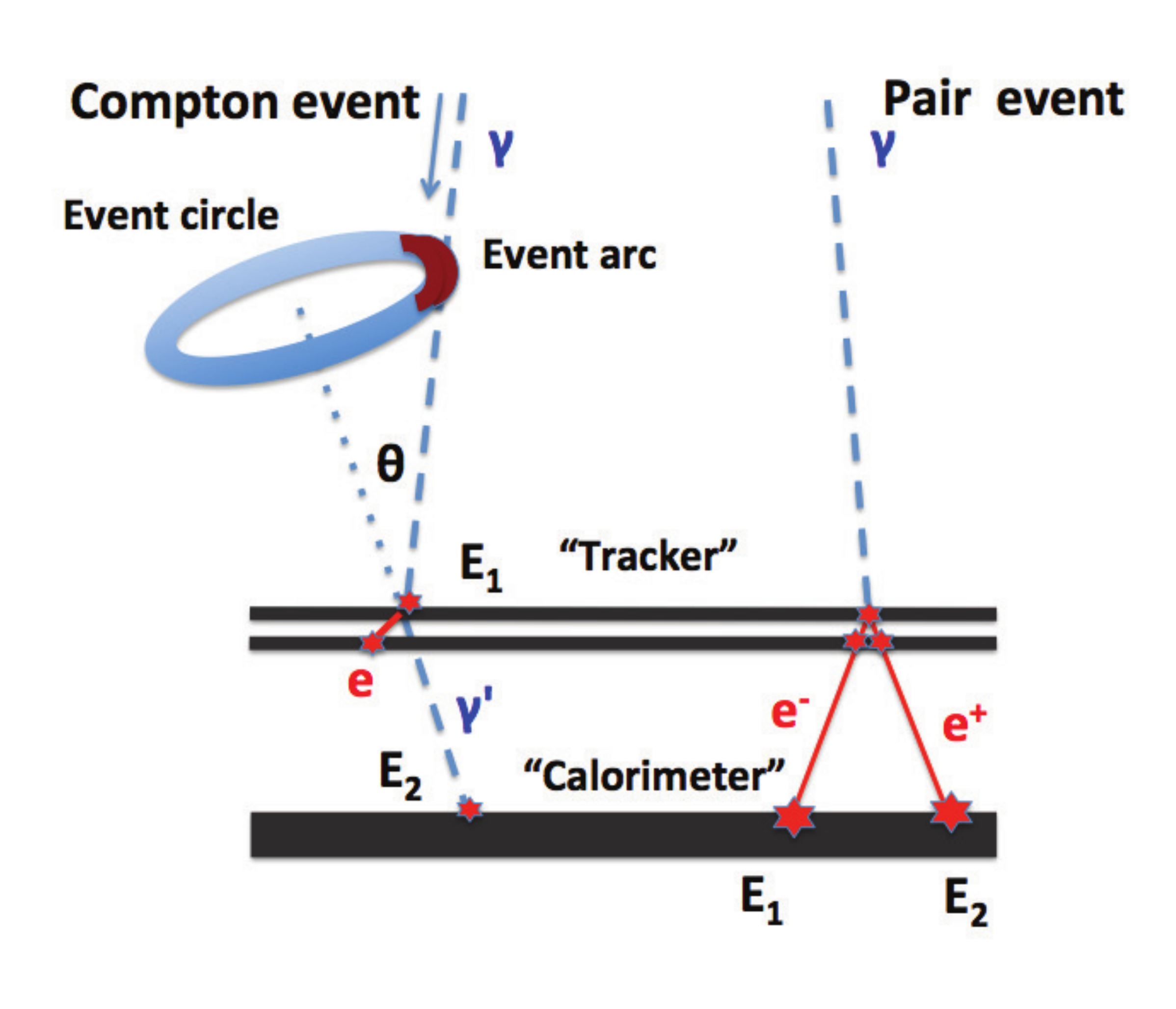}
\caption{Representative  topologies for a Compton event  (left) and  for a pair event (right). Photon tracks are shown in pale blue, dashed, and electron and/or positron tracks in red, solid. From \cite{alexnew}.}
\label{fig:evt-top}       
\end{figure}

Detecting gamma-rays by Compton scattering  is more complicated than for pair production, because the scattered photon carries a significant amount of the information about the incident photon and thus it needs to be detected too. In practice, a Compton telescope requires two separate photon interactions in order to have a clear detection. The first Compton scattering of the incident photon occurs in one of the Tracker planes, creating an electron and a scattered photon. The Tracker measures the interaction location, the electron energy, and in some cases the electron direction. The scattered photon can be absorbed in the Calorimeter or (with smaller probability) scattered a second time in the Tracker before being absorbed in the Calorimeter where its energy and absorption position are measured.

The basic principle of the Compton mode of operation is illustrated in Figure \ref{fig:evt-top}, left. An incident gamma-ray Compton scatters by an angle $\Theta$ in one layer of the Tracker, transferring energy $E_{1}$ to an electron. The scattered photon is then absorbed in the Calorimeter, depositing an energy $E_{2}$, and 
the scattering angle is given by $\cos\Theta = {m_{e}c^{2}}/{E_2} + {m_{e}c^{2}}/{(E_1+E_2)} $, where $m_e$ is the electron mass. 
With this information, one can derive an ``event circle" from which the original photon arrived. We will call ``untracked" this sort of Compton events. The uncertainty in the event circle reconstruction is reflected in its width and is due to the uncertainties in direction reconstruction of 
the scattered photon and the energy measurements of the scattered electron $(E_1)$ and the scattered photon $(E_2)$. 
Multiple photons from the same source enable a full deconvolution of the image, using probabilistic techniques. 

For energetic Compton scatters (above $\sim$1 MeV), measurement of the track of the scattered electron becomes possible, resulting in a reduction of the event ring to an arc, hence further improving event reconstruction. 
If the scattered electron direction is measured, the event circle reduces to an event arc with length due to the uncertainty in the electron direction reconstruction, allowing improved source localization. This event is called  ``tracked", and its direction reconstruction is somewhat similar to that for pair event -- the primary photon direction is reconstructed from the direction and energy of two secondary particles: scattered electron and photon.  Redundant kinematic information from multiple interactions could also help. Compton scattering angles depend on polarization of the incoming photon, hence careful statistical analysis of the photons for a strong (e.g., transient) source yields a measurement of the degree of polarization of its high-energy emission (e.g. \cite{for08}).

Especially for the Compton mode at low energies, but also more broadly over the entire energy range covered by e-ASTROGAM, it is important to keep the amount of passive materials on the top and at the sides of the detector to a minimum, to reduce background  in the field of view and to optimize angular and energy resolutions. In addition, the passive materials between the Tracker layers, and between the Tracker and the Calorimeter, must be minimized for best performance. 

\subsubsection{Silicon Tracker\label{sec:tracker}}

The Si Tracker is the heart of the e-ASTROGAM payload. It is based on the silicon strip detector technology widely employed in medical imaging and particle physics experiments (e.g. ATLAS and CMS at LHC), and already applied to the detection of gamma-rays in space with the AGILE and Fermi missions. The e-ASTROGAM Tracker needs double sided strip detectors (DSSDs) to work also as a Compton telescope. 

The essential characteristics of the e-ASTROGAM Tracker are: 
\begin{itemize}
\item  its light mechanical structure minimizing the amount of passive material within the detection volume to enable the tracking of low-energy Compton electrons and $e^+e^-$ pairs, and improve the point spread function in both the Compton and pair domains by reducing the effect of multiple Coulomb scattering;
\item its fine spatial resolution of less than 40 $\mu$m ($<1/6$ of the microstrip pitch) obtained by analog readout of the signals (as in the AGILE Tracker);
\item  its charge readout with a very good spectral resolution of $\sim$6 keV FWHM (noise level in the baseline configuration; the statistical contribution to the energy resolution is negligible in the relevant range of deposited energy ($E<500$ keV), so the energy resolution is practically independent of energy and corresponds to the noise level) obtained with an ultra low-noise FEE, in order to accurately measure low-energy deposits produced by Compton events; the energy threshold is 15~keV. 
\end{itemize}

The Si Tracker comprises 5600 DSSDs arranged in 56 layers (100 DSSDs per layer). It is divided in four towers of 5$\times$5 DSSDs. The spacing of the Si layers is of 10~mm. The total detection area amounts to 9025 cm$^2$ and the total Si thickness to 2.8~cm, which corresponds to 0.3 radiation length on axis, and a probability of a Compton interaction at 1~MeV of 40\%. Such a stacking of relatively thin detectors enables an efficient tracking of the electrons and positrons produced by pair conversion, and of the recoil electrons produced by Compton scattering. The DSSD signals are read out by 860~160 independent, low-power, electronics channels with self-triggering capability.

\paragraph{Silicon detectors}
The active element is a Si DSSD of 500~$\mu$m thickness and $9.5\times9.5$~cm$^2$ area, with electrodes of 100~$\mu$m width, and 240 $\mu$m pitch (corresponding to 384 microstrips per side), a guard ring of 1.5 mm, and polysilicon resistors for the bias. It can be manufactured from high resistivity ($R \ge 5$~k$\Omega$~cm) 6'' substrate by, e.g., the Silicon Radiation Sensors$^\copyright$ group of the Fondazione Bruno Kessler FBK (SRS-FBK) or Hamamatsu Photonics$^\copyright$.  

Each layer of a tower contains $5\times5$ DSSDs, which are chained together with wire bonding strip to strip. Ladders of five Si tiles are first assembled and then bonded to five other ladders in the orthogonal direction. Si strip bonding is now a standard technology previously used in, e.g., the Fermi/LAT and AGILE Tracker and the PAMELA and AMS-02 cosmic-ray experiments. 

\paragraph{Mechanical structure}

The mechanical structure holding a tray of $5\times5$ DSSDs with the associated FEE is composed of two frames sandwiching the Si detectors, the rods direction of the upper frame being orthogonal to that of the lower frame, to form a grid. On each side, the support rods are parallel to the DSSD strips to enable the wire bonding. The DSSDs are glued onto the frames with a structural adhesive, and a Kapton$^\copyright$ foil is added in the middle of the glue thickness to ensure electrical insulation. The frames, which are 2 mm thick, are made of a polymeric resin reinforced by high modulus carbon fibers woven into fabrics. Carbon fiber spacers determine a total spacing of the Si layers of 10 mm; placed at the crossing between frames, they limit the vertical displacements under loads achieving a uniform distribution of the displacement among them and realize unilateral connections between the trays. The towers of the tracker are assembled together by a structural mainframe composed of vertical fixing aluminum columns. In order to increase the stiffness of each tower, the baseline configuration of the Tracker comprises two honeycomb panels (e.g. Hexel 3/16-5052 + 1 mm thick carbon fiber foils), one at the top (4 cm thickness) and one at the bottom (1 cm thickness).

The Tracker mechanical design is the result of detailed structural calculation based on a simplified Finite Element Model (FEM) and including both static and modal analyses. The maximum vertical displacement obtained by FEM is of 280 microns (Figure~\ref{fig:trackermechanics}), far below the rupture limit of Si detectors.

\begin{figure*}
\centering
\includegraphics[width=.7\textwidth]{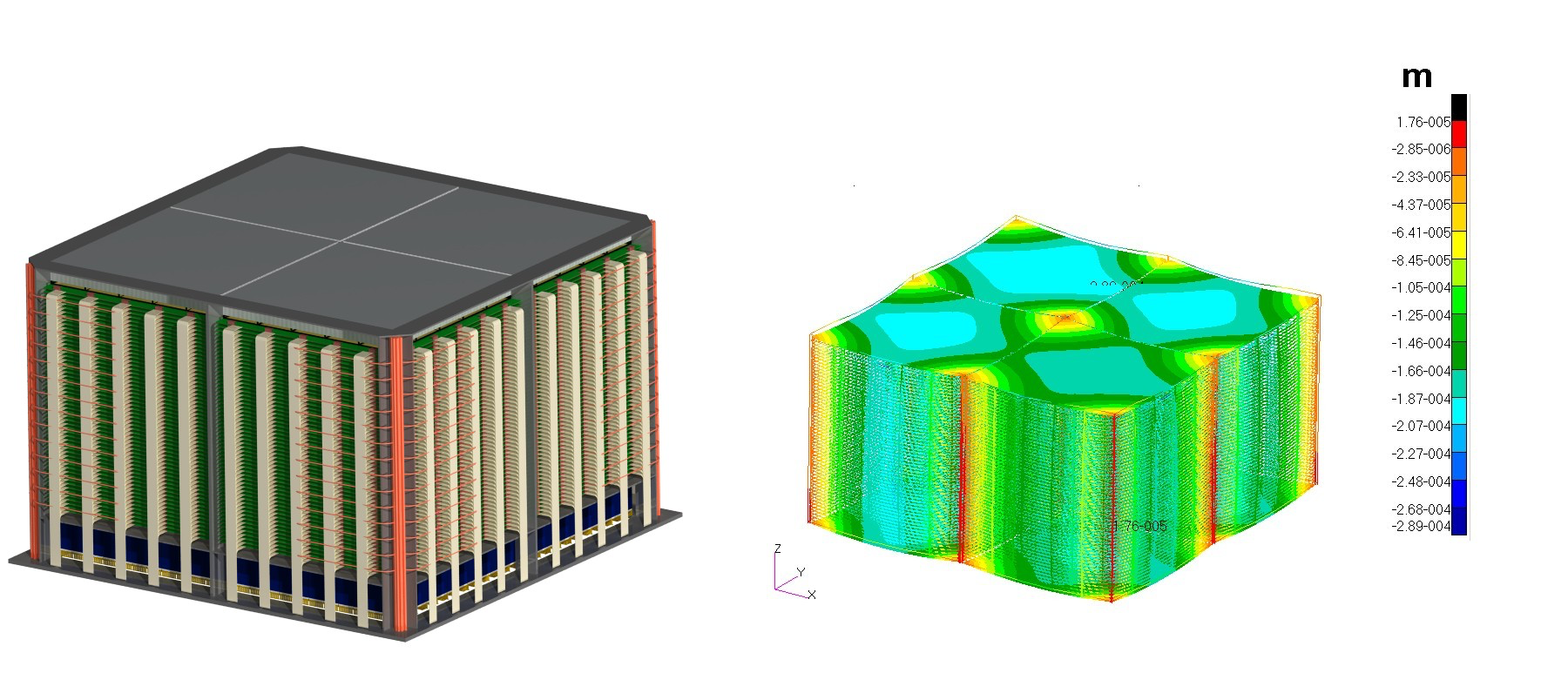}
\caption{{\it Left panel} -- Overview of the Tracker mechanical design. {\it Right panel} -- Illustration of the Tracker structural analysis showing normalized deformation of elements.}
\label{fig:trackermechanics}       
\end{figure*}

\paragraph{Front-End Electronics}

The FEE is distributed over the four sides of the Tracker (there are no electronics between the four towers), the detector microstrips being connected to the FEE ASICs through pitch adapters. The FEE ASICs are the 32-channels IDeF-X HD circuits developed at CEA/Saclay. The 860,160 DSSD signals are read by a total of 26,880 ASICs (4$\times$120 ASICs per layer). 

IDeF-X HD is a $5.8\times2.5$~mm$^2$ chip operated at 3.3~V and consuming 800 $\mu$W/channel (static dissipated power). It was built based on the AMS CMOS 0.35~$\mu$m technology using the full custom Rad-Hard libraries, and combine the most recent developments for space applications, including fully space-qualified devices \cite{lim05,gev12}. In particular, IDeF-X HD was extensively studied with respect to radiation tolerance (Single Event Latchup (SEL) -- free; Single Event Effect (SEE) $> 9$~MeV~cm$^2$~mg$^{-1}$; Total Ionizing Dose (TID) $>300$~krad w/o effect on noise response), as it was selected for the FEE of the STIX instrument of the Solar Orbiter mission. 

Each individual channel of IDeF-X HD is made of: a charge sensitive preamplifier (CSA) optimized for low current ($<1$~nA) and capacitance around 10 pF, a variable gain (inverting or non inverting) stage, a pole-zero cancellation stage, an adjustable shaper in the range from 1 to 11~$\mu$s (peaking time), a baseline holder, providing a stable offset whatever the leakage current into a channel, a peak detector and hold, and a discriminator (each channel, 6 bit DAC) in the range from 0 to 13 keV for Si (the settings is not linear to allow fine tuning in the low range and coarse in the high range). Together with the 32 inputs (DC or AC coupled), the other interfaces of IDeF-X HD are two differential analog output and two independent slow controls. The CSA includes a continuous reset feedback circuitry, which includes a "non stationary noise suppressor" to optimize the noise response into the whole dynamic range. The dynamic range is 10 fC (but is programmable up to 40 fC).

The analog output signals of IDeF-X will be converted to digital signals with the OWB-1 ADC integrated system. OWB-1 is a low noise (0.6 LSB), low power (1~mW per active channel) chip of 11.55 mm$^2$ area, including parallel Wilkinson ADC for 16 differential channels with a real 13 bits resolution at a conversion rate of 2.8~$\mu$s (and 11 bits resolution at 0.9~$\mu$s conversion rate). The chip was also built from the AMS CMOS 0.35~$\mu$m technology and it is radiation hard by design (SEL hardened + Single Event Upsets flag). Qualification procedure of this ASIC is on going. 

\subsubsection{Calorimeter}

The e-ASTROGAM Calorimeter is a pixelated detector made of a high-$Z$ scintillation material -- Thallium activated Cesium Iodide -- for an efficient absorption of Compton scattered gamma-rays and electron-positron pairs. It consists of an array of 33,856 parallelepiped bars of CsI(Tl) of 8~cm length and 5$\times$5~mm$^2$ cross section, read out by silicon drift detectors (SDDs) at both ends \cite{gat84}, arranged in an array of 529 ($=23 \times 23$) elementary modules comprising each 64 crystals (see Figure~\ref{fig:calorimeteroverview}). The Calorimeter thickness -- 8 cm of CsI(Tl) -- makes it a 4.3 radiation-length detector having an absorption probability of a 1-MeV photon on axis of 88\%.

The Calorimeter detection principle and architecture are based on the heritage of the space instruments {\it INTEGRAL}/PICsIT, {\it AGILE}/MCAL and {\it Fermi}/LAT, as well as on the particle physics experiment LHC/ALICE at CERN. However, the e-ASTROGAM calorimeter features two major improvements with respect to the previous instruments:
\begin{compactitem}
\item[$\bullet$] the energy resolution is optimized to a FWHM of 4.5\% at 662 keV (scaling with the inverse of the square root of the energy) by the use of low-noise SDDs for the readout of the scintillation signals, combined with an appropriate ultra low-noise FEE;
\item[$\bullet$] the spatial resolution is improved by measuring the depth of interaction in the detector from a suitable weighting function of the recorded scintillation signals at both ends; the position resolution along the CsI(Tl) bars is $\sim$5~mm FWHM, i.e. comparable to the resolution in the X--Y plane given by the crystal cross section ($5\times5$~mm$^2$). Accurately measuring the 3D position and deposited energy of each interaction is essential for a proper reconstruction of the Compton events.
\end{compactitem}

\begin{figure}
\centering
\includegraphics[width=\columnwidth]{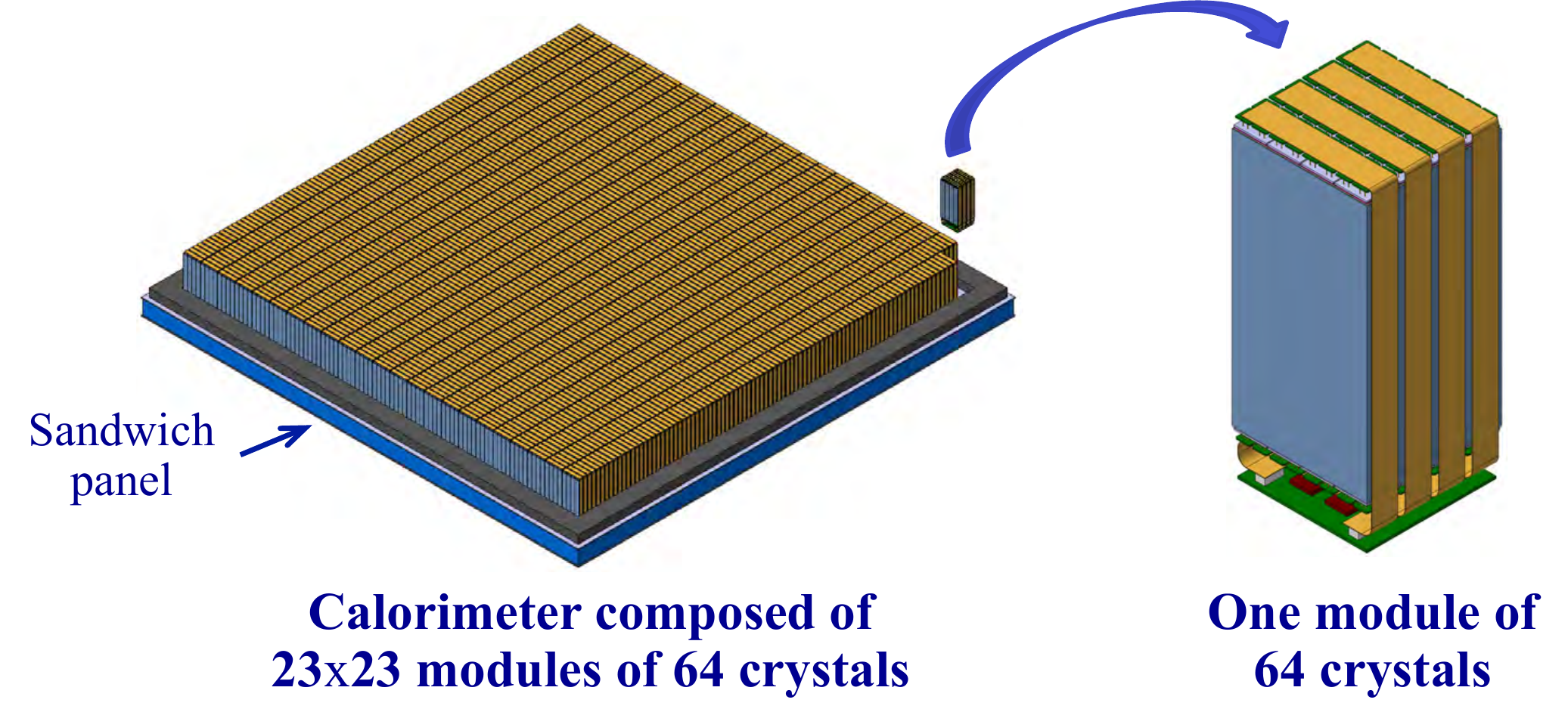}
\caption{Overview of the Calorimeter and of one of its 529 ($=23 \times 23$) basic modules comprising 64 CsI(Tl) crystals.}
\label{fig:calorimeteroverview}       
\end{figure}

The simultaneous data set provided by the Silicon Tracker, the Calorimeter and the Anticoincidence system constitutes the basis for the gamma-ray detection. However, thanks to the detector excellent granularity, Calorimeter-only events can be used on board to provide a burst notice and a first approximate localization via fast onboard reconstruction even in the absence of a signal from the Tracker. This data acquisition mode is dedicated to the search for fast transient events such as GRBs and TGFs. The corresponding trigger condition is explained in Sect. \ref{sec:trigger}.

\paragraph{Scintillation crystals and housing}

The choice of CsI(Tl) as the scintillation material in the baseline design is motivated by the facts that (i) it is one of the brightest scintillator (light yield of 54 photons per keV), (ii) it matches well to the response of Si photodiodes (broad emission spectrum with a maximum at ~540 nm), (iii) it is only slightly hygroscopic (much less than, e.g., LaBr$_3$ and CeBr$_3$), (iv) it has good radiation hardness properties, and (v) it has already flown on, e.g., the INTEGRAL, AGILE and Fermi satellites. The required 33,856 CsI(Tl) bars could be produced by, e.g., Saint Gobain Crystals$^\copyright$ or Detec Europe$^\copyright$.

The possibility of using a new scintillator technology that could offer an improved energy resolution and a reduced scintillation decay time will be studied during the assessment phase. Several potential candidates exist, including RGB (RbGd$_2$Br$_7$:Ce) and lutetium-based scintillators. New developments in nano-crystalline glass ceramic scintillators also offer sufficient promise as to be worthy of further investigation. There is a very active development ongoing within the e-ASTROGAM consortium on this topic, with ESA support, and the results are encouraging.

Each crystal will be wrapped with a reflective material in order to optimize the light collection and reduce the optical cross talk. The selected reflective material is the Radiant Mirror Film ESR from 3M, with a thickness of 63~$\mu$m. This wrapping material, which is a non-metallic multilayer polymer, was previously employed for the CsI crystals of the Fermi/LAT calorimeter, as well as, e.g., for the calorimeters of the JLAB/DVCS and GSI/PANDA particle physics experiments.

\paragraph{Photosensors: silicon drift detectors}

Silicon Drift Detectors (SDDs) are solid-state devices suitable for both direct X-ray and scintillation light detection \cite{gat84}. Their peculiar characteristic is the fact that the signal charge is driven toward a small collecting anode by means of a suitably tailored electric field within the depletion region. Under these conditions, a very small output capacitance can be obtained, almost independent from the active area, and hence an improvement of more than an order of magnitude in noise performance with respect to PIN photodiode of equivalent active area. When SDDs are used as readout devices for CsI(Tl) scintillation crystals, the improved noise performance results in a very good energy resolution (4.5\% FWHM at 662 keV) and an energy threshold lower than 30 keV \cite{mar05}. For the e-ASTROGAM calorimeter we foresee to use SDDs designed and developed at INFN Trieste, and fabricated at the FBK (SRS-FBK). A 6" wafer processing line was successfully set up at SRS-FBK, with mass production capabilities. 

In the e-ASTROGAM calorimeter, the basic detector element is formed by the coupling of four CsI(Tl) bars to two square arrays of 2$\times$2 SDDs of 5~mm side. The CsI(Tl) scintillator bars will be optically coupled to two SDD arrays, one on top and the other at the bottom of the crystals, to enable the reconstruction of the energy and position of interaction along the bars by means of a suitable weighting function of the recorded signals at both ends. This hodoscopic architecture was already employed for the AGILE and Fermi LAT calorimeters, but using much larger crystals and PIN photodiodes as readout devices. The improved noise performance of the SDDs will allow us to obtain a depth-of-interaction resolution of about ~5 mm FWHM when coupled to 80 mm long scintillator bars, as already demonstrated at laboratory level \cite{lab08}. 

\paragraph{Mechanical structure}

The e-ASTROGAM Calorimeter is made of 529 ($=23 \times 23$) elementary modules, each of which comprising 16 basic elements of four CsI(Tl) bars coupled to two square arrays of 2$\times$2 SDDs. This gives a total of 33,856 identical crystals and 67,712 SDDs. The surface covered by the calorimeter is about 100$\times$100 cm$^2$. The spacing between the Calorimeter and the Tracker (i.e. the distance between the Calorimeter and the bottom honeycomb panel of the Tracker structure) is 2~mm.

\begin{figure*}
\centering
\includegraphics[width=0.75\textwidth]{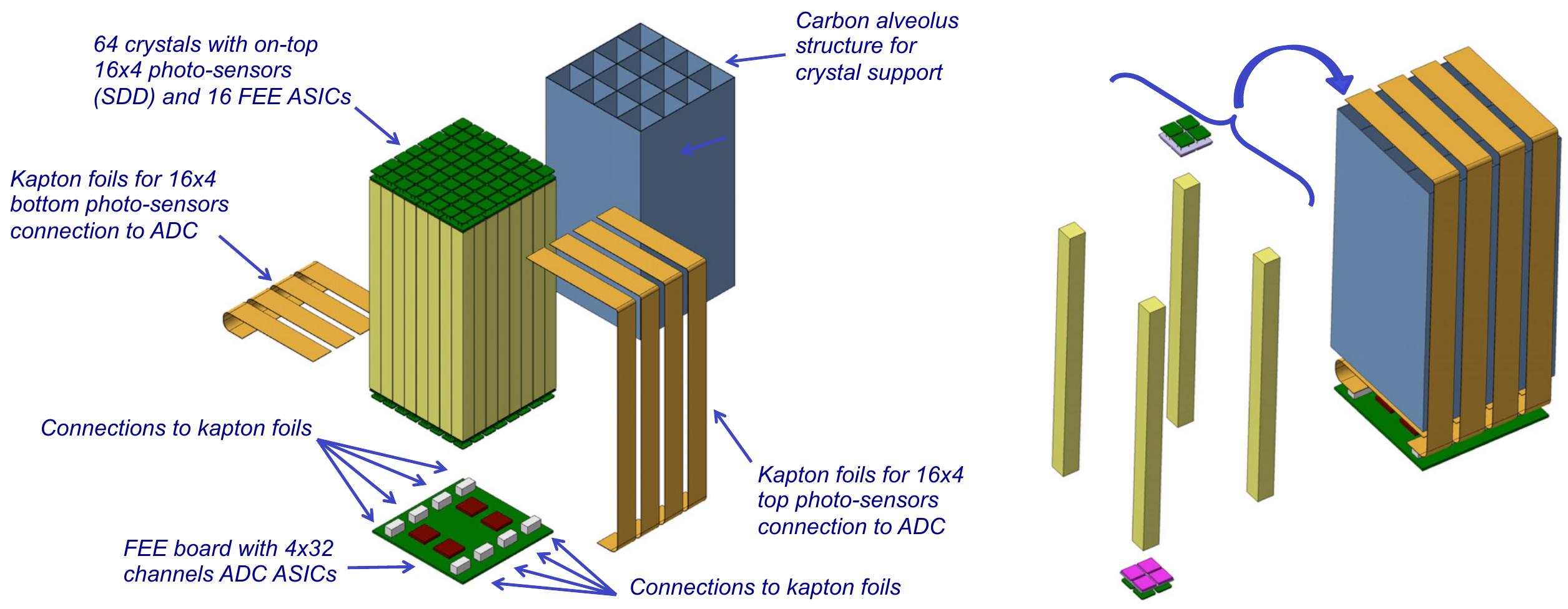}
\caption{Exploded view drawing of a elementary module of the Calorimeter with its alveolus structure.}
\label{fig:calorimetermodule}       
\end{figure*}

An exploded view of a basic module of the Calorimeter is shown in Figure~\ref{fig:calorimetermodule}. The crystals are supported by an alveolar structure made of a polymer resin reinforced by high modulus carbon fibers, with four crystals (i.e. one detector element) contained in each cell. The thickness of the structure walls is 200~$\mu$m. Each alveolus is epoxy-glued to an insert in carbon fibers, which makes the interface with the support elements at the bottom of the module. Each module is fixed to a bottom sandwich panel composed of a 3~cm honeycomb core covered by carbon fiber skins. The sandwich panel is stiff enough to hold the weight of the 33,856 crystals of the calorimeter (305~kg). At the top of the alveolar structure, a carbon plate is added to further limit the displacement of the crystals.

The FEE boards with the OWB-1 ADC ASICs (see below) are positioned under the crystals on dedicated supporting circuits. The SDD boards with the VEGA analog ASICs in contact of the collecting anodes are connected to the ADC ASIC boards via Kapton foils (Figure~\ref{fig:calorimetermodule}). The low heat generated by the SDD boards ($2\times27$~mW per module) is also evacuated by the Kapton circuits. 

\paragraph{ Front-End Electronics }

We foresee using for the FEE ASICs a modified version of the VEGA circuits developed by Politecnico di Milano and Universit\`a di Pavia. With the development of a new chip of 4 channels, a total of 16,928 ASICs will be needed to read the 67,712 SSD signals.

The VEGA ASIC \cite{aha15}, which was specifically developed for SDD readout, is realized in CMOS AMS 0.35 $\mu$m technology. It is a low power circuit, consuming 420 $\mu$W/channel, which contains a preamplifier, a programmable shaper-amplifier, a discriminator and a peak stretcher. It provides an Equivalent Noise Charge (ENC) of 16 electrons rms at 3 $\mu$s shaping time when coupled to a 350~fF input capacitance SDD. With this system, an energy resolution of 183 eV FWHM at 5.9 keV has been obtained. The ASIC includes a linear array of 32 independent channels with dimensions of 200$\times$500 $\mu$m per channel, each with a dynamic range of 60~keV for X-ray direct detection in Si (equivalent to about 16,400 electrons).

A minimal set of optimization steps will be carried out to tailor the characteristics of the VEGA ASIC to the e-ASTROGAM calorimeter requirements. In particular, considering a typical photoelectron yield of about 30 electron per keV obtainable for CsI(Tl) scintillators coupled to silicon devices, the shaping-amplifier dynamic range should be increased by a factor of $\sim$20 (equivalent to $\sim$10 MeV detection in CsI(Tl)).  

The analog output signals of the new VEGA ASICs will be converted to digital signals with the same OWB-1 ADC integrated system as used for the Tracker FEE. The 2$\times$64 signals of a module will be processed by four 32-channels OBW-1 ASICs placed on a dedicated board below the crystals (Figure~\ref{fig:calorimetermodule}).

\subsubsection{Anticoincidence System} \label{sec:anticoincidence}\label{acs}

The third main detector of the e-ASTROGAM payload consists of an Anticoincidence system composed of two main parts: (1) a standard Anticoincidence, named Upper-AC, made of segmented panels of plastic scintillators covering the top and four lateral sides of the instrument, requiring a total active area of about 5.2~m$^2$, and (2) a Time of Flight (ToF), aimed at rejecting the particle background produced by the platform. The Upper-AC detector is segmented in 33 plastic tiles (6 tiles per lateral side and 9 tiles for the top). All scintillator tiles are coupled to silicon photomultipliers (SiPM) by optical fibers. The architecture of the Upper-AC detector is fully derived from the successful design of the {\it AGILE} \cite{per06} and {\it Fermi}/LAT \cite{moi07} AC systems. In particular, their segmentation has proven successful at limiting the ``backsplash'' self-veto, therefore the dead time of the instrument. The Upper-AC particle background rejection is designed to achieve a relativistic charged particle detection inefficiency lower than 10$^{-4}$, a standard value already realized in current space experiments. In addition to the panel segmentation, providing coarse information on the part of the detector that has been hit, we are also considering the possibility of even finer position resolution based on the analysis of the relative light output of multiple fibers.

In the baseline design, the Upper-AC system covers the entire instrument from five sides, leaving open the bottom for design considerations of cabling to the S/C bus, the layout of heat pipes, etc.  
The bottom side of the instrument is protected by the ToF to discriminate the particles coming out from the instruments {from those entering the instrument from below}. The ToF is composed by two scintillator layers separated by 50 cm. The required timing resolution is of 300 ps. The readout will be performed by SiPM connected with Time Digital Converter (TDC). The ToF will be based on technologies well proven in space (AMS and PAMELA satellites).

The plastic scintillator type that was selected for both Upper-AC and ToF is the BC400 or BC408 from Saint Gobain Crystals$^\copyright$ or the equivalent EJ212 or EJ200 from Eljen Technology$^\copyright$. For all these materials, the scintillator peak emission is around 425 nm. The scintillator thickness has been set to a minimum of 5~mm for the Upper-AC lateral panels and 6~mm for the Upper-AC top panel and ToF in order to get enough light to detect more than 99.99\% of the passing through relativistic charged particles. The energy threshold of the AC detectors is set to 100~keV. The light emitted by the plastic is transferred to the SiPM through wavelength shifting optical fibers. The combination of optical fibers and SiPM provides the best solution to collect the scintillator optical light, as it has the higher gain and gives the best efficiency of charged particle rejection. SiPM has also the advantage over traditional photomultiplier tubes to work at low bias (few tens of volts), so without the need of high voltage. The selected SiPM could be from the B-Series (blue sensitive) of the SensL$^\copyright$ company. The peak sensitivity of these photosensors is at 420 nm (within a sensitivity range of 300 - 800 nm), which copes well with the plastic scintillator emission. SensL$^\copyright$ SiPMs were recently studied with respect to radiation tolerance in an ESA-led program and are now at technology readiness level (TRL) 5. For the Upper-AC readout we foresee the selection of the VATA64 ASIC from the Ideas Company, this chip being optimized for SiPM readout and already space qualified (see Ref.~\cite{bag11}). Alternative options will be considered for the ToF, such as the MUSIC ASIC, which has also been developed for SiPM readout, and has a Single Photon Resolution Time of about 100 ps rms \cite{gom16}. 

\subsubsection{Data Handling and Power Supply}

The e-ASTROGAM payload is completed by a Payload Data Handling Unit (PDHU) and a Power Supply Unit (PSU). The PDHU is in charge of carrying out the following principal tasks: (i) payload internal control; (ii) scientific data processing; (iii) operative modes management; (iv) on board time management; (v) Telemetry and Telecommand management. The main functions related to the scientific data processing are: (i) BEE interfacing through dedicated links to acquire the scientific data; (ii) the real-time software processing of the collected silicon Tracker, Anticoincidence and Calorimeter scientific data aimed at rejecting background events to meet the telemetry requirements; (iii) scientific data compression; (iv) formatting of the compressed data into telemetry packets. 

The heart of the PDHU architecture is based on a powerful Digital Signal Processor (DSP) running the payload on-board software. Considering the large amount of events to be processed, a possible DSP could be the HiRel component C6727B-250 produced by Texas Instruments$^\copyright$. This floating point CPU running at 250 MHz is capable of 500 MMACS (2000 MIPS / 1500 MFLOPS) by executing up to 8 instructions in parallel (6 of which floating points).

The PSU is in charge of generating the required payload voltages with high DC/DC conversion efficiency and distributing them to the other sub-systems.

\subsubsection{Trigger logic and data flow architecture\label{sec:trigger}}

The e-ASTROGAM on-board scientific data processing is composed of two main trigger pipelines, the gamma-ray acquisition mode and the Calorimeter burst search. Both are based on the experience of the AGILE and Fermi missions. The simultaneous data sets provided by the silicon Tracker, the Calorimeter and the AC constitute the basis for the gamma-ray detection and processing. The gamma-rays trigger logic is structured on two main levels: Level-1 (fast: 5-10 $\mu$s logic, hardware); and  Level-2 (asynchronous, 50 $\mu$s processing, software). Figure~\ref{fig:triggerrate} shows the expected data rates at the input of the Level-1 and at the output of the Level-1 and Level-2 trigger stages. 

Level-1 is a hardware trigger logic with fast response implemented in the silicon Tracker BEE providing a preliminary discrimination between Compton and pair-producing photon events and a first cut of background events. Discrimination criteria based on the hit multiplicity in the Tracker and in the Calorimeter can provide optimal algorithms to identify Compton events. The Level-1 trigger configuration is defined to save the largest possible number of potential Compton events.

\begin{figure}
\centering
\includegraphics[width=\linewidth]{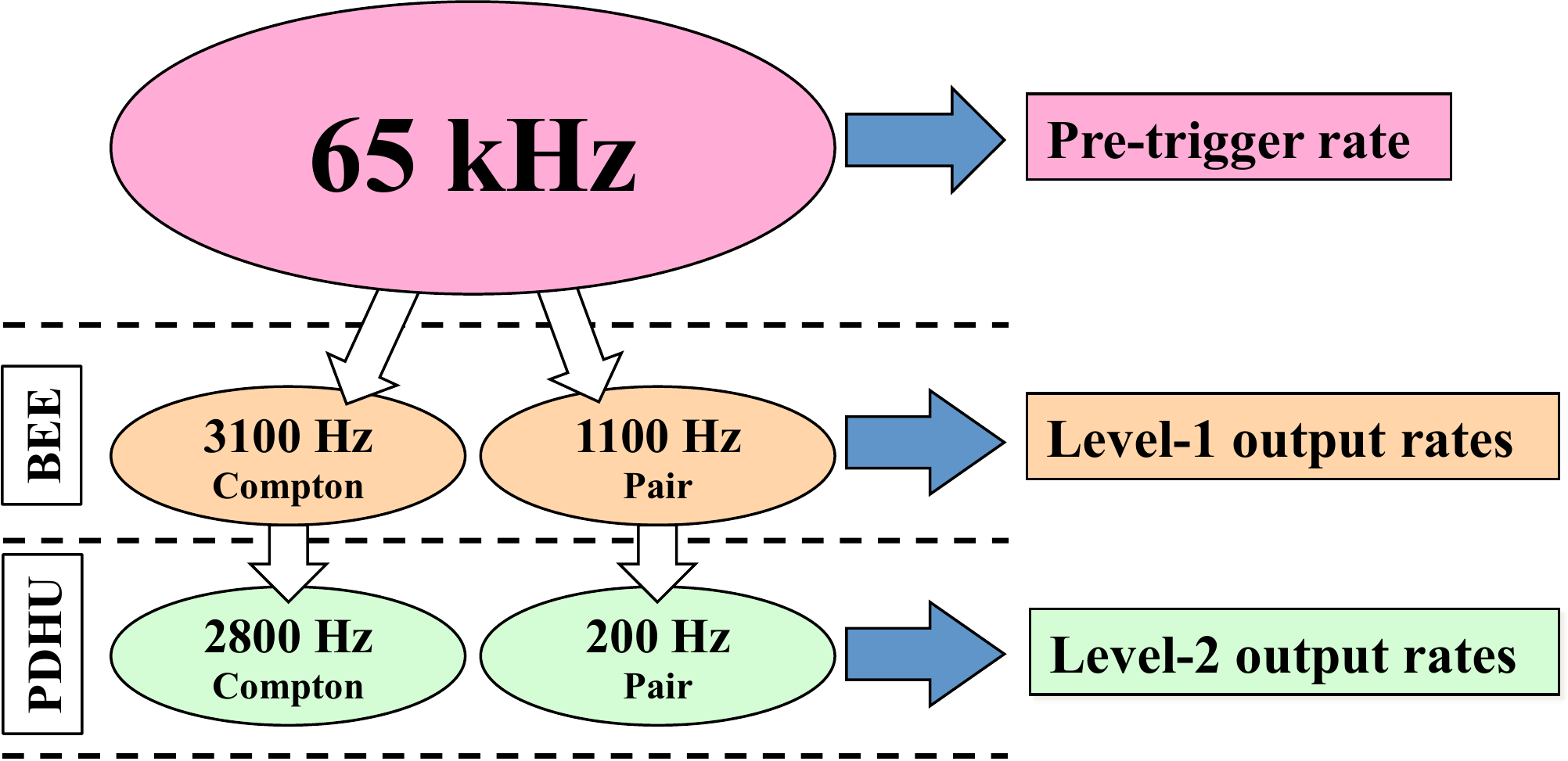}
\caption{Expected data flow of the on-board e-ASTROGAM gamma-ray data acquisition system.}
\label{fig:triggerrate}       
\end{figure}

Level-2 is a software trigger stage carried out by the PDHU and aimed, at further reducing the residual particle and photon background of the pair data set and at finalizing the selection of the Compton events. The heart of the Level-2 trigger stage consists of the track reconstruction of the candidate pair events implemented with Kalman Filter techniques. The Level-2 trigger is a full asynchronous processing stage and does not increase the dead time of the instrument. At the end, the Compton events and the pair events surviving the Level-2 trigger are collected in dedicated telemetry packets and sent to the ground.

Level-1 is has a rate tuned to limit the dead time and to fit the accepted event rate into the acceptance capability of the onboard reconstruction stage. Level-2, which is fed by events passing Level-1, is tuned so that the rate of accepted events fits into telemetry bandwidth. Only events passing Level-2 are downlinked.

The Calorimeter burst search is a software algorithm implemented by the PDHU. 
%The burst search is based on the integration and processing of a proper set of rate meters measuring the trend of the background and foreground counting rates. Since the expected impulsive signals (gamma-ray bursts and terrestrial gamma-ray flashes) are strongly energy and timescale dependent, the rate meters are integrated on different timescales (in the range 100 $\mu$s -- 10~s) and energy ranges (in the overall range 30~keV -- 200~MeV). 
The burst search is based on the integration and processing of a proper set of rate meters measuring the trend of the background and foreground counting rates. Since the expected impulsive signals (GRBs and TGFs) are strongly energy- and timescale-dependent, the rate meters are integrated on a wide range of timescales (0.1 ms -- 10 s) and energy ranges (in the overall range 30 keV -- 200 MeV). 
Assuming a typical GRB spectrum, the energy channels are defined in order to obtain a homogeneous threshold over the whole energy range.  

The burst search logic routinely compares the background and foreground rate meters. Elementary triggers are generated when foreground rate meters show an over-threshold counting excess respect to the background estimation. Finally, the triggers have to be validated by satisfying coincidence conditions on the different energy channels aimed at spurious trigger rejection. These conditions are implemented using look-up tables fully programmable from ground. 

A cyclic buffer is required to routinely save the events; the size of this buffer is defined in order to store 100 s of background. Following a burst trigger, the logic will be able to identify the beginning and the end of the segment of data acquisition and then transfer it to ground by telemetry.

\subsection{Performance assessment}\label{sec:perfass}

The scientific performance of the e-ASTROGAM instrument was evaluated by detailed numerical simulations with the software tools MEGAlib and BoGEMMS. The MEGAlib package \cite{zog06} was originally developed for analysis of simulation and calibration data related to the Compton scattering and pair-creation telescope MEGA \cite{kan05}. It has then been successfully applied to a wide variety of hard X-ray and gamma-ray telescopes on ground and in space, such as COMPTEL, NCT, and {\it NuSTAR}. BoGEMMS (Bologna Geant4 Multi-Mission Simulator) is a software for simulation of payload of X- and gamma-ray missions, which has been developed at the INAF/IASF Bologna \cite{bul12}. It has already been applied to several hard X-ray/gamma-ray instruments and mission projects, including Simbol-X, NHXM, Gamma-Light, {\it AGILE}, and GAMMA-400. Both software packages exploit the Geant4 toolkit to model the geometrical and physical parameters of the detectors and simulate the interactions of photons and particles in the instrument.

The numerical mass model of e-ASTROGAM used to simulate the performance of the instrument is shown in Figure~\ref{fig:massmodel}. An accurate mass model that includes passive material in the detector and its surroundings, true energy thresholds and energy and position measurement accuracy, as well as a roughly accurate S/C bus mass and position are crucial to the modeling. In particular, care was taken to include all passive materials close to the Si and CsI(Tl) detectors.

\subsubsection{Background model}

\begin{figure}
%\vspace{-15pt}
\centering
\includegraphics[width=0.8\linewidth]{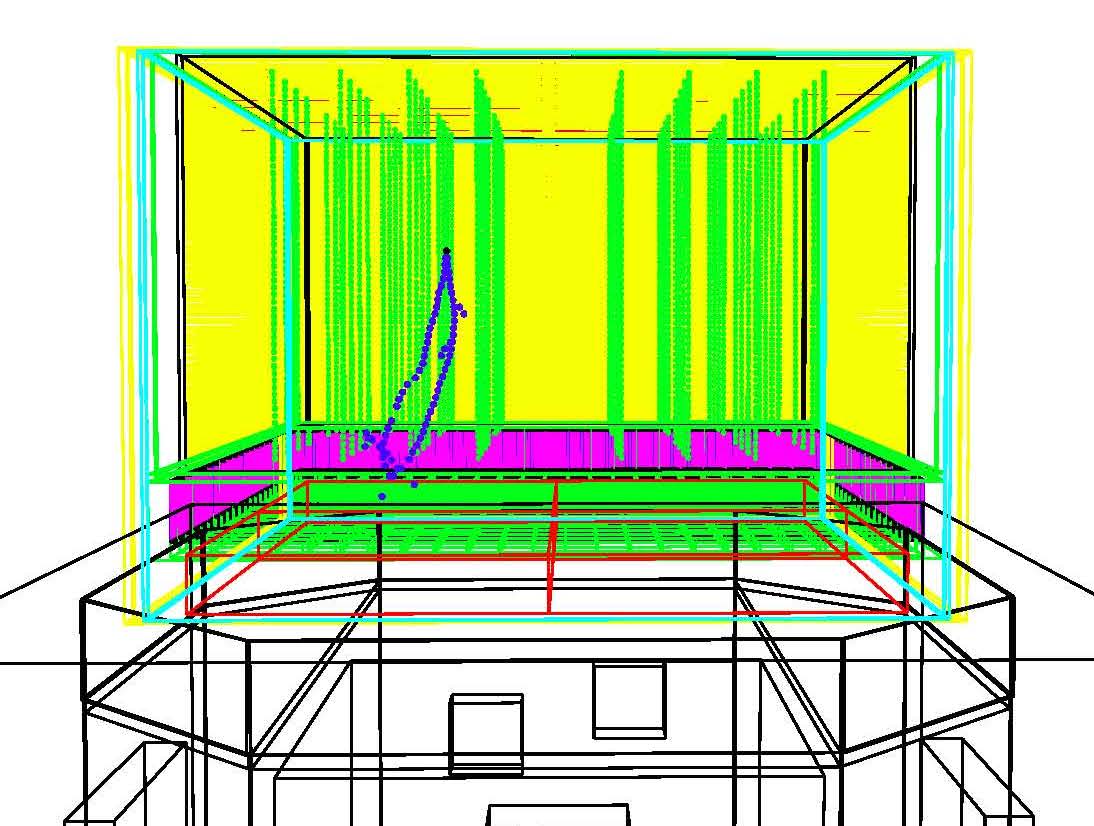}
\caption{Geant4/MEGAlib mass model of the e-ASTROGAM telescope, with a simulated pair event (in blue) produced by a 30-MeV photon. The vertical green lines represent the carbon fiber spacers placed between the DSSD layers to increase the stiffness of the Tracker (see Sect.~\ref{sec:tracker}).}
\label{fig:massmodel}    
\end{figure}

For best environmental conditions, e-ASTROGAM should be launched into a quasi-equatorial (inclination $i < 2.5^\circ$)  LEO at a typical altitude of 550~km. The background environment in such an orbit is now well-known (Figure~\ref{fig:background}), thanks to the Beppo-SAX mission, which measured the radiation environment on a low-inclination ($i \sim 4^\circ$), 500 -- 600 km altitude orbit almost uninterruptedly during 1996 -- 2002 \cite{cam14} and the on-going {\it AGILE} mission, which scans the gamma-ray sky since 2007 from a quasi-equatorial orbit   at an average altitude of 535~km \cite{tav09}. The dominant sources of background for the e-ASTROGAM telescope in the MeV domain are the cosmic diffuse gamma-ray background, the atmospheric gamma-ray emission, the reactions induced by albedo neutrons, and the background produced by the radioactivity of the satellite materials activated by fast protons and alpha particles. All these components were  modeled in detail using the MEGAlib environment tools. In the pair domain above 10 MeV, the background is mainly induced by fast particles (mainly leptons) impinging the spacecraft, as well as by the cosmic diffuse radiation and the atmospheric gamma-ray emission.

\begin{figure}
\centering
\includegraphics[width=0.8\linewidth]{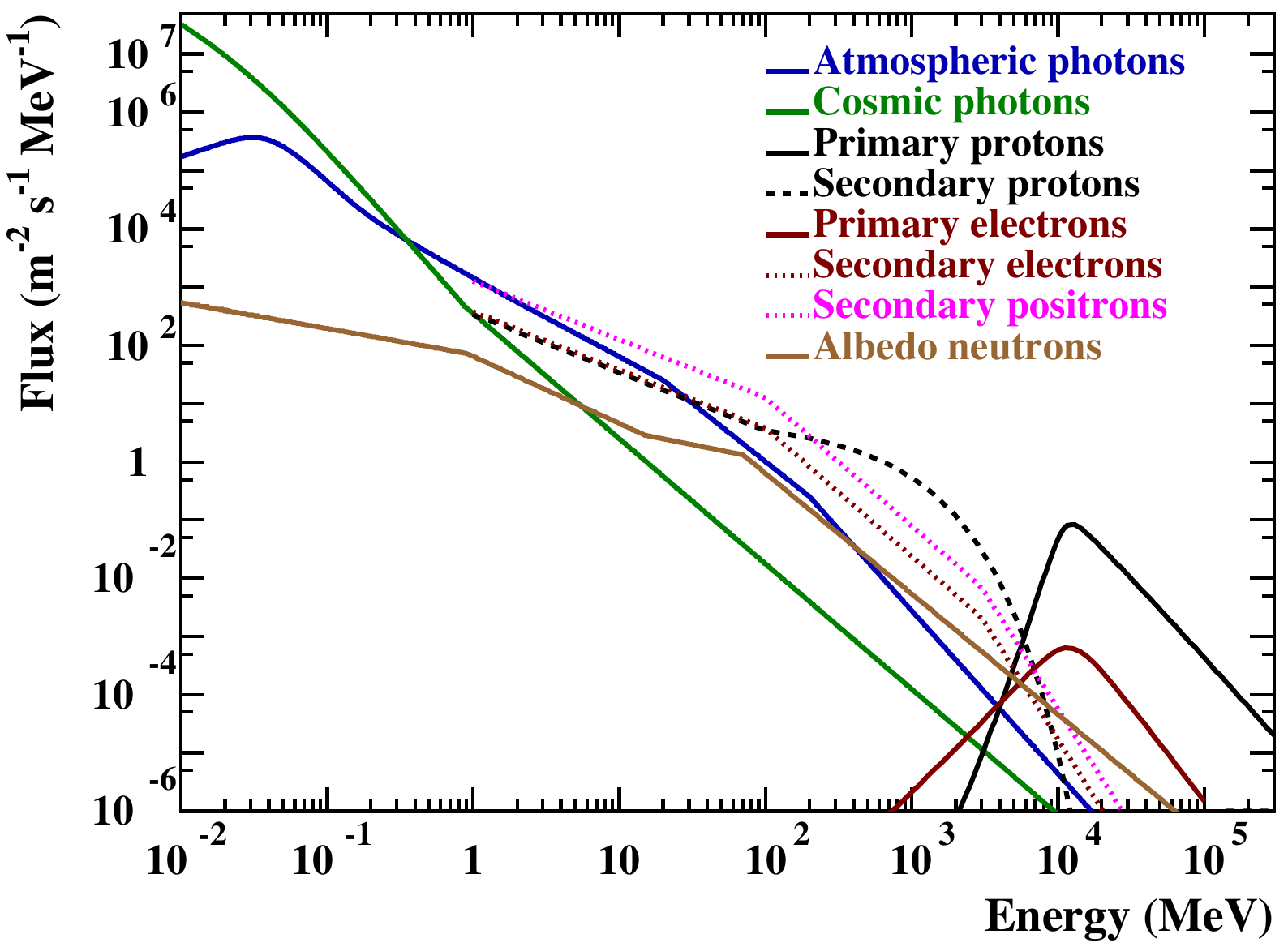}
%\vspace{-0.4cm}
\caption{Background environment of e-ASTROGAM on its orbit. The satellite will be exposed to Galactic CR (mainly protons and electrons) modulated by the geomagnetic field, semi-trapped secondary protons and leptons, as well as to albedo neutrons and atmospheric gamma rays. The cosmic diffuse X- and gamma-ray radiation (in green) is the dominant background component below a few hundred keV, but it is also a fundamental science topic for e-ASTROGAM above a few MeV.
%(see Sect. \ref{sec:scirec}).
}
\label{fig:background}
\end{figure}

\subsubsection{Angular and spectral resolution}

e-ASTROGAM will image the Universe with substantially improved angular resolution both in the MeV domain and above a few hundreds of MeV, i.e. improving the angular resolution of the {\it CGRO}/COMPTEL telescope and that of the {\it Fermi}/LAT instrument by a factor of $\sim$4 at 1 MeV and 1 GeV, respectively.

In the pair production domain, the PSF improvement over {\it Fermi}/LAT is due to (i) the absence of heavy converters in the Tracker, (ii) the light mechanical structure of this detector minimizing the amount of passive material within the detection volume and thus enabling a better tracking of the secondary electrons and positrons, and (iii) the analog readout of the DSSD signals allowing a fine spatial resolution of about 40~$\mu$m ($\sim$1/6 of the microstrip pitch). In the Compton domain, thanks to the fine spatial and spectral resolutions of both the Tracker and the Calorimeter, the e-ASTROGAM angular resolution will be close to the physical limit induced by the Doppler broadening due to the velocity of the target atomic electrons.

\begin{figure}%[t!]
%\hspace{1.5cm}
\begin{minipage}{0.49\linewidth}
\includegraphics[scale=0.35]{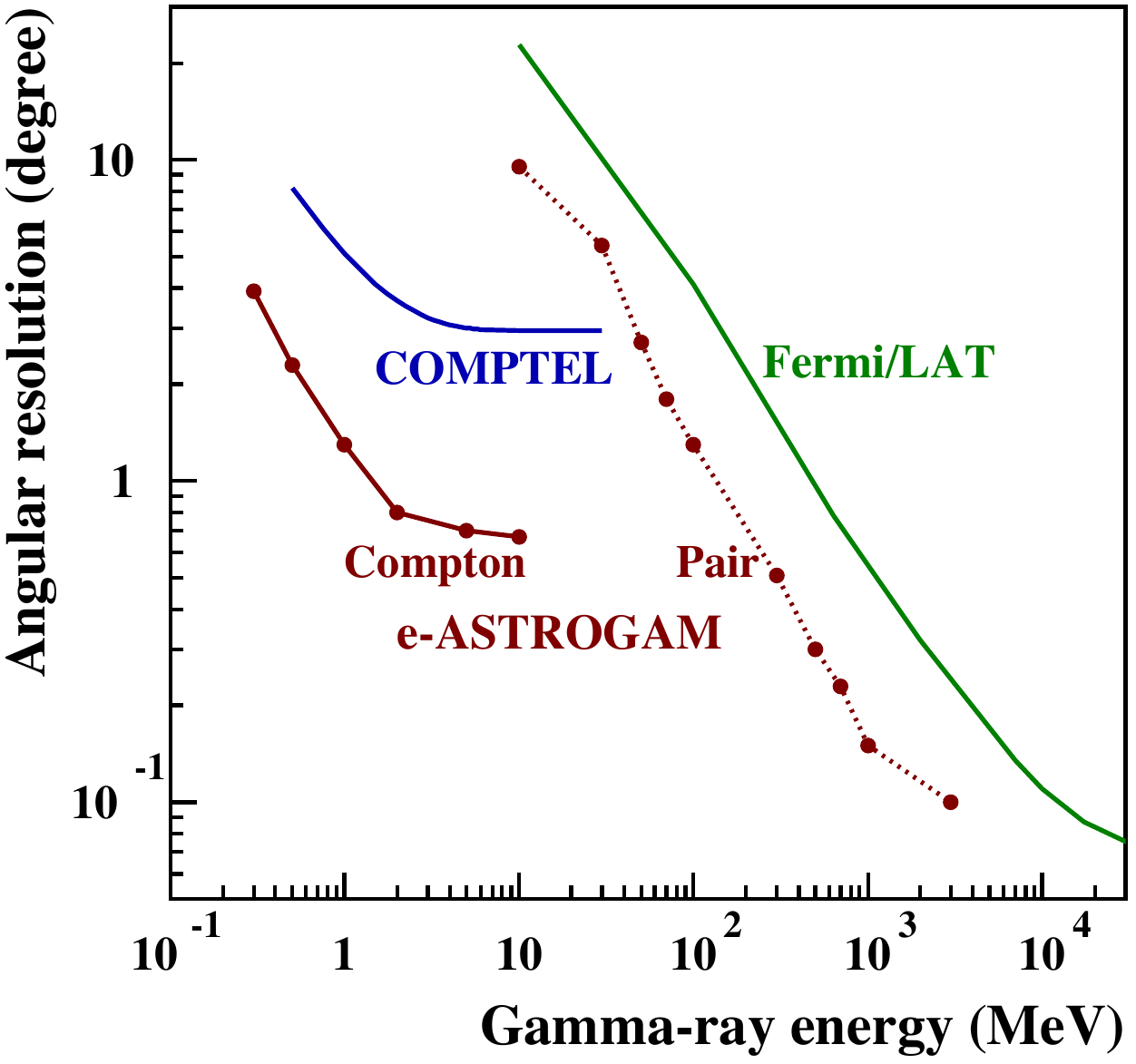}
\end{minipage}
\begin{minipage}{0.49\linewidth}
\includegraphics[scale=0.35]{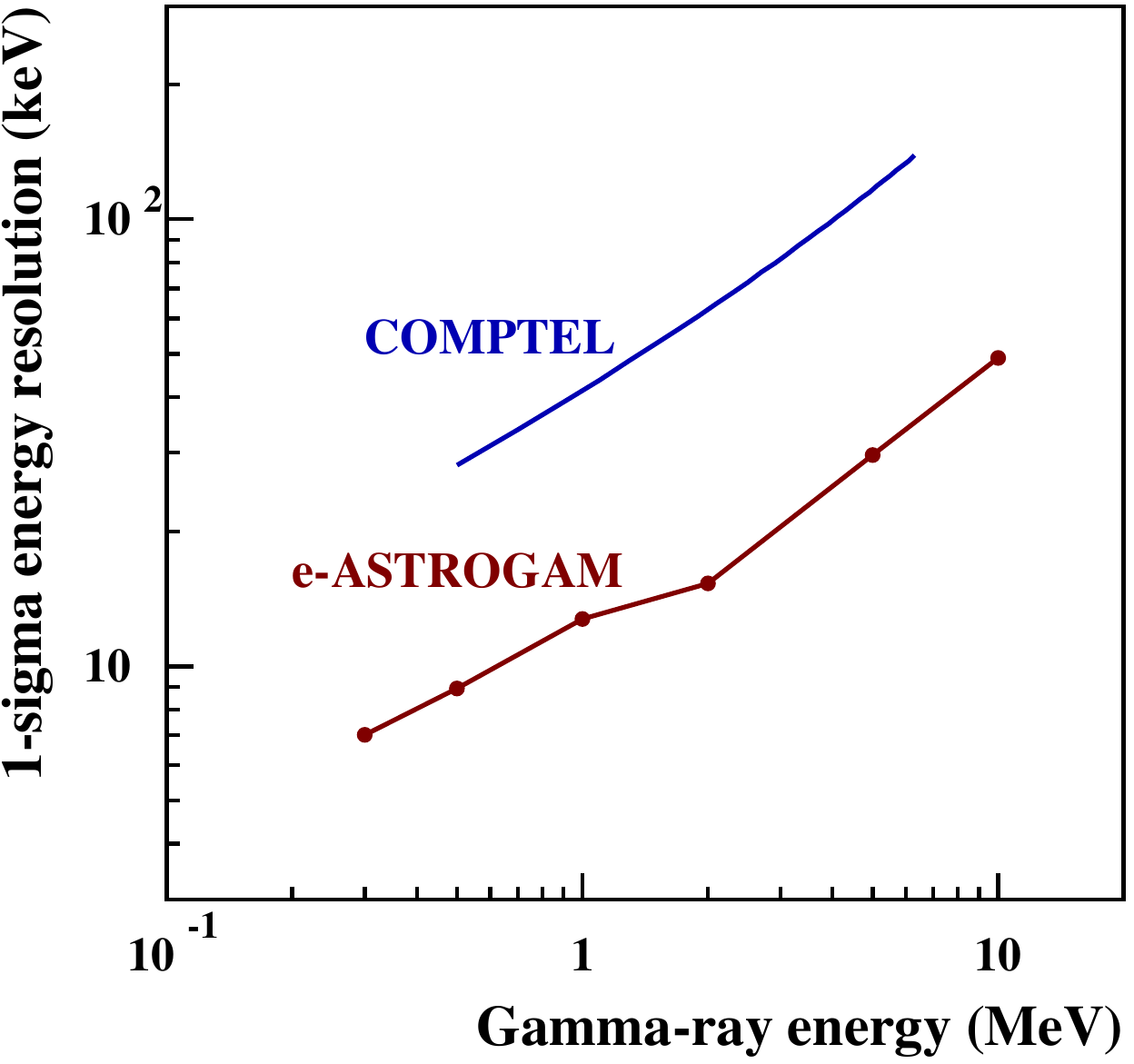}
\end{minipage}
\caption{{\it Left panel} -- e-ASTROGAM on-axis angular resolution compared to that of COMPTEL and {\it Fermi}/LAT. In the Compton domain, the presented performance of e-ASTROGAM and COMPTEL is the FWHM of the angular resolution measure (ARM). In the pair domain, the point spread function (PSF) is the 68\% containment radius for a 30$^\circ$ point source. The Fermi/LAT PSF is from the Pass 8 analysis (release 2 version 6) and corresponds to the FRONT and PSF event type. {\it Right panel} -- 1$\sigma$ energy resolution of COMPTEL and e-ASTROGAM in the Compton domain after event reconstruction and selection on the ARM.}
\label{fig:perf}
\end{figure}

Figure~\ref{fig:Jurgen}  shows an example of the e-ASTROGAM imaging capability in the MeV domain compared to COMPTEL. The e-ASTROGAM synthetic map of the Cygnus region was produced from the third {\it Fermi} LAT (3FGL) catalog of sources detected at photon energies $E_\gamma > 100$~MeV \cite{3FGL}, assuming a simple extrapolation of the measured power-law spectra to lower energies. It is clear from this example that e-ASTROGAM will substantially overcome (or eliminate in some cases) the confusion issue that severely affected the previous and current generations of gamma-ray telescopes. The e-ASTROGAM imaging potential will be particularly relevant to study the various high-energy phenomena occurring in the GC region.

e-ASTROGAM will also significantly improve the energy resolution with respect to COMPTEL, e.g. by a factor of $\sim$3.2 at 1 MeV, where it will reach a 1$\sigma$ resolution of $\Delta E/E=1.3$\% (Figure~\ref{fig:perf}). In the pair production domain above 30~MeV, the simulated spectral resolution is within 20--30\%.

\subsubsection{Field of View}

The e-ASTROGAM field of view was evaluated from detailed simulations of the angular dependence of the sensitivity. Specifically, the width of the field of view was calculated as the half width at half maximum (HWHM) of the inverse of the sensitivity distribution as a function of the polar, off-axis angle, for a constant azimuthal angle $\phi=22.5^\circ$. In the Compton domain, the sensitivity remains high within $40^\circ$ to $50^\circ$ off-axis angle and then degrades for larger incident angles. For example, the field of view at 1~MeV amounts to 46$^\circ$ HWHM, with a fraction-of-sky coverage in zenith pointing mode of 23\%, corresponding to $\Omega = 2.9$~sr.

In the pair-production domain, the field-of-view assessment is also based on in-flight data from the {\it AGILE} and {\it Fermi}-LAT gamma-ray imager detectors. With the e-ASTROGAM characteristics (size, Si plane spacing, overall geometry), the field of view is found to be $> 2.5$~sr above 10~MeV.  

\subsubsection{Effective area and continuum sensitivity}

Improving the sensitivity in the medium-energy gamma-ray domain (1--100~MeV) by one to two orders of magnitude compared to previous missions is the main requirement for the proposed e-ASTROGAM mission. Such a performance will open an entirely new window for discoveries in the high-energy Universe. Tables~\ref{table:sensitivity_Compton} and \ref{table:sensitivity_pair} present the simulated effective area and continuum sensitivity in the Compton and pair-production domains. The sensitivity below 10 MeV is largely independent of the source location (inner galaxy vs. high latitude), because the diffuse gamma-ray background is not a major background component in the Compton domain.

\begin{table*}
\centering
\caption{e-ASTROGAM performance in the Compton domain simulated with MEGAlib v2.26.01. The 3$\sigma$ continuum sensitivity is for the detection of a point source on axis after an observation time $T_{\rm obs}=10^6$~s.}
\vspace{-0.1cm}
\includegraphics[width=0.9\textwidth]{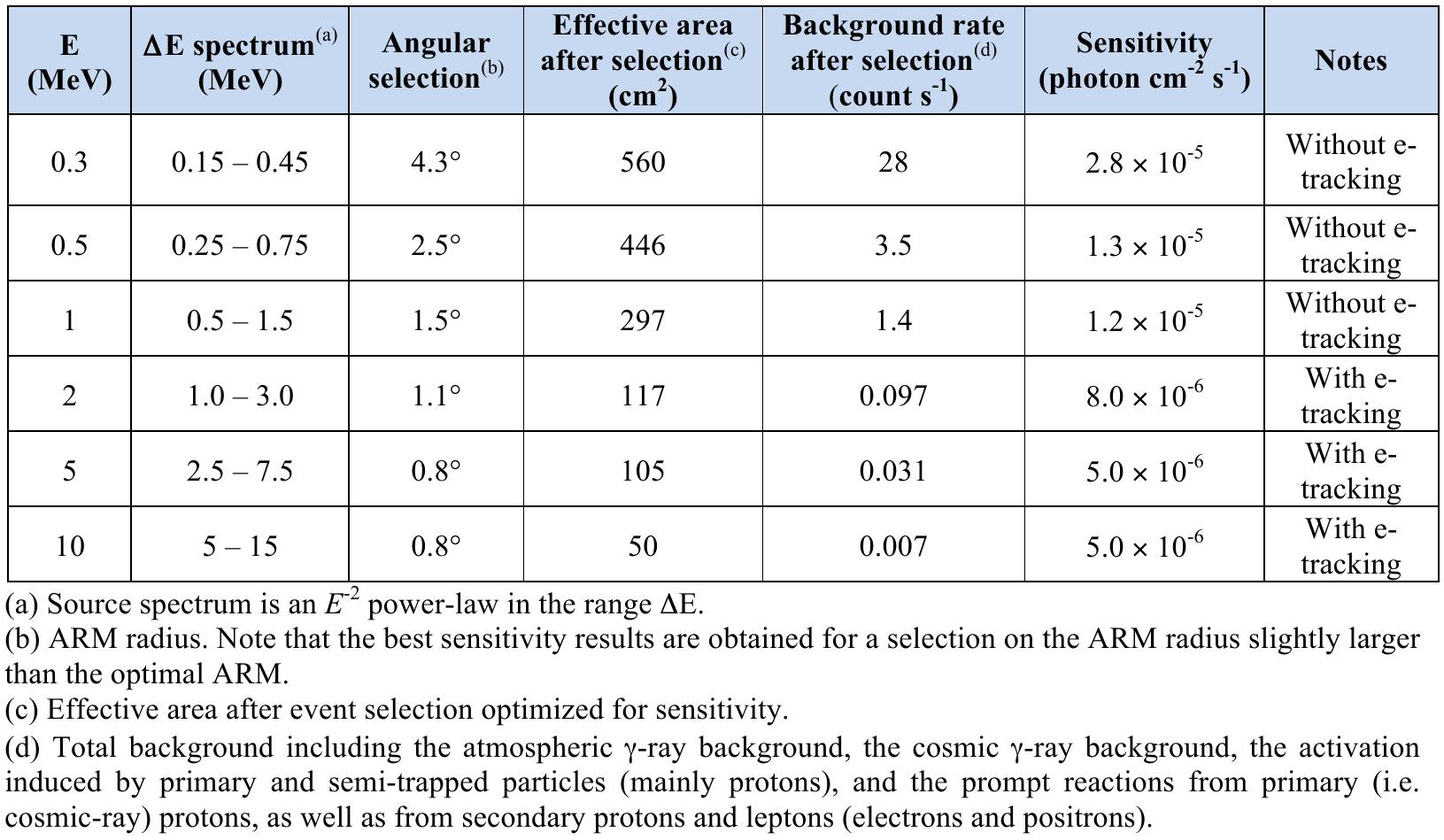}
\label{table:sensitivity_Compton}
\end{table*}

\begin{table*}
\centering
\caption{e-ASTROGAM performance in the pair-production domain simulated with BoGEMMS v2.0.1, together with Kalman v1.5.0 and Trigger v1.0.0. All results are for a 30$^\circ$ off-axis source and for $T_{\rm obs} = 10^6$~s. The King function used to fit the PSF, derived from the model of XMM data, is defined, e.g., in \cite{xmm04}.}
\vspace{-0.1cm}
\includegraphics[width=0.9\textwidth]{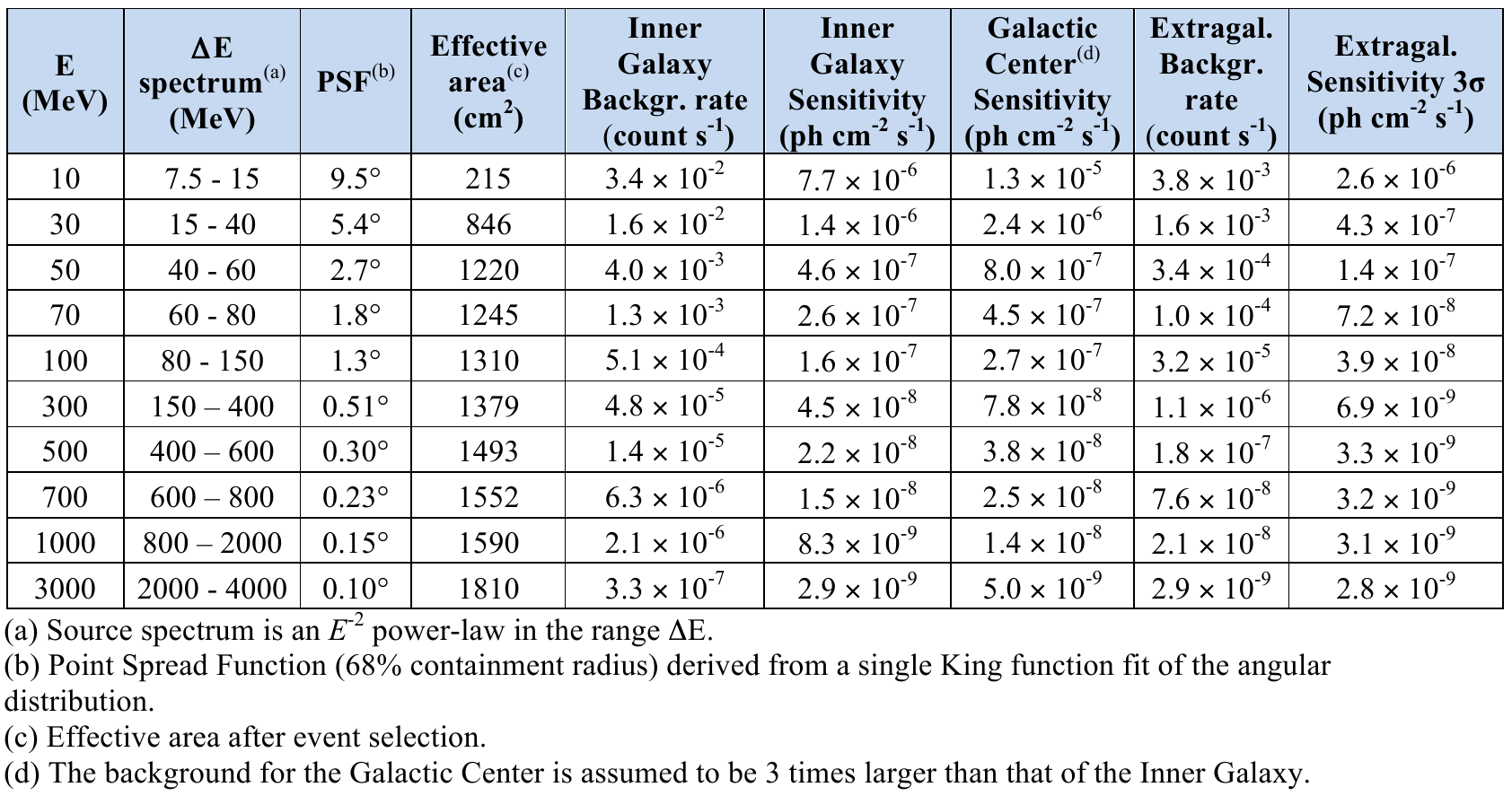}
\label{table:sensitivity_pair}
\end{table*}

Figure~\ref{fig:sensitivity}  shows the e-ASTROGAM continuum sensitivity for a 1-year effective exposure of a high Galactic latitude source. Such an effective exposure will be reached for broad regions of the sky after 3 years of operation, given the very large field of view of the instrument. We see that e-ASTROGAM would provide an important leap in sensitivity over a wide energy band, from about 200 keV to 100 MeV. At higher energies, e-ASTROGAM would also provide a new vision of the gamma-ray sky thanks to its  angular resolution, which would reduce the source confusion that plagues the current {\it Fermi}-LAT and {\it AGILE} images near the Galactic plane (see, e.g., the 3FGL catalog \cite{3FGL}).

\subsubsection{Line sensitivity}

Table~\ref{table:sensitivity_line} shows the e-ASTROGAM 3$\sigma$ sensitivity for the detection of key gamma-ray lines from pointing observations, together with the sensitivity of the {\it INTEGRAL} Spectrometer (SPI). The latter was obtained from the {\it INTEGRAL} Observation Time Estimator (OTE) assuming 5$\times$5 dithering observations. The reported line widths are from SPI observations of the 511 and 847 keV lines (SN 2014J), and from theoretical predictions for the other lines. Noteworthy, the neutron capture line from accreting neutron stars can be significantly redshifted and broadened (FWHM between 10 and 100 keV) depending on the geometry of the mass accretion \cite{bil93}.

\begin{table*}
\centering
\caption{e-ASTROGAM line sensitivity (3$\sigma$ in 10$^6$ s) compared to that of {\it INTEGRAL}/SPI\cite{roq03}.}
\vspace{-0.1cm}
\includegraphics[width=0.9\textwidth]{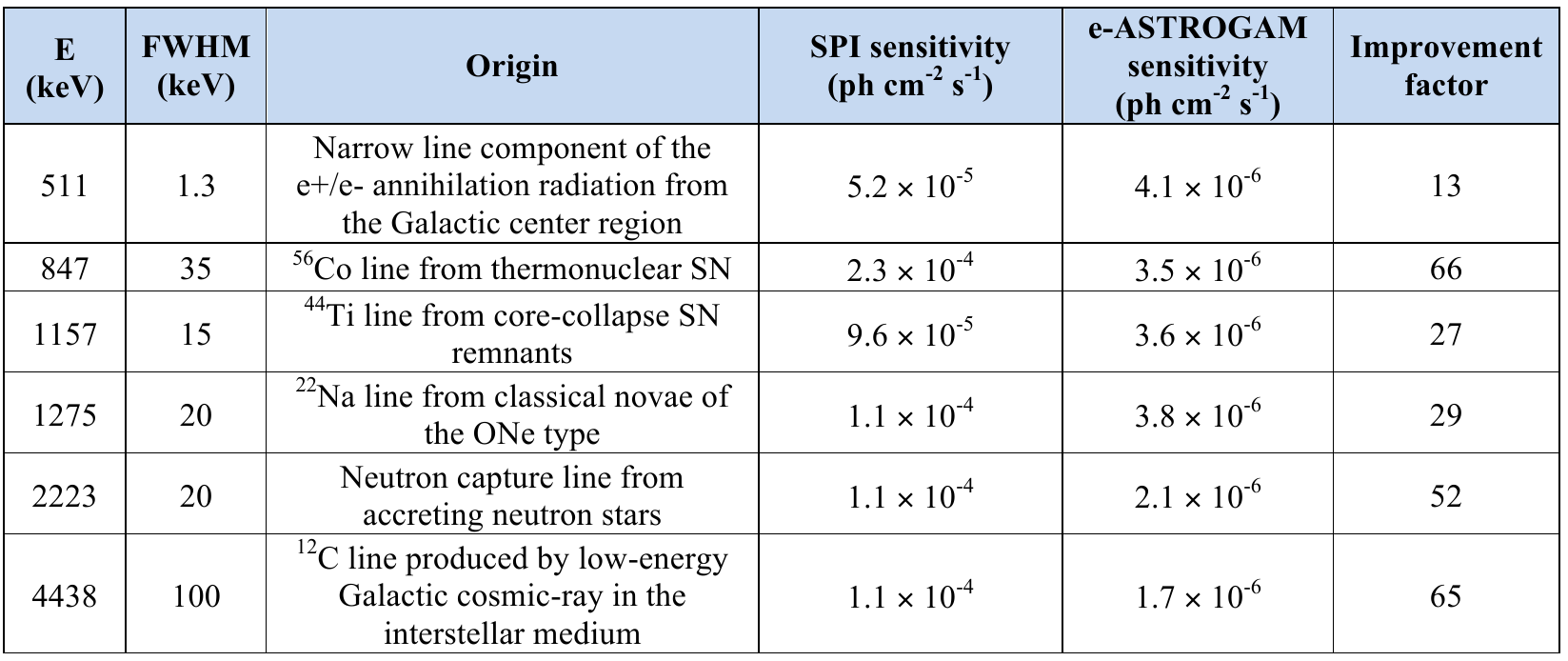}
\label{table:sensitivity_line}
\end{table*}

We see that e-ASTROGAM will achieve a major gain in sensitivity compared to SPI for all gamma-ray lines, the most significant improvement being for the 847~keV line from Type Ia SNe (see Sect. \ref{sciencecase}). With the predicted line sensitivity, e-ASTROGAM will also (i) provide a much better map of the 511~keV radiation from positron annihilation in the inner Galaxy, (ii) uncover $\sim$10 young, $^{44}$Ti-rich SN remnants in the Galaxy and thus provide new insight on the explosion mechanism of core-collapse SNe (iii) detect for the first time the expected \cite{cla74} line from $^{22}$Na decay in novae hosted by ONe white dwarfs, (iv) provide a new constraint on the nuclear equation of state of neutron stars by detecting the predicted \cite{bil93} redshifted 2.2~MeV line from Scorpius X-1, and (iv) measure the energy density of low-energy cosmic rays in the inner Galaxy to better understand the role of these particles in the Galactic ecosystem.

\subsubsection{Polarization response}\label{sec:polargrb}

\begin{figure*}%[b!]
\centering
\includegraphics[width=0.8\textwidth]{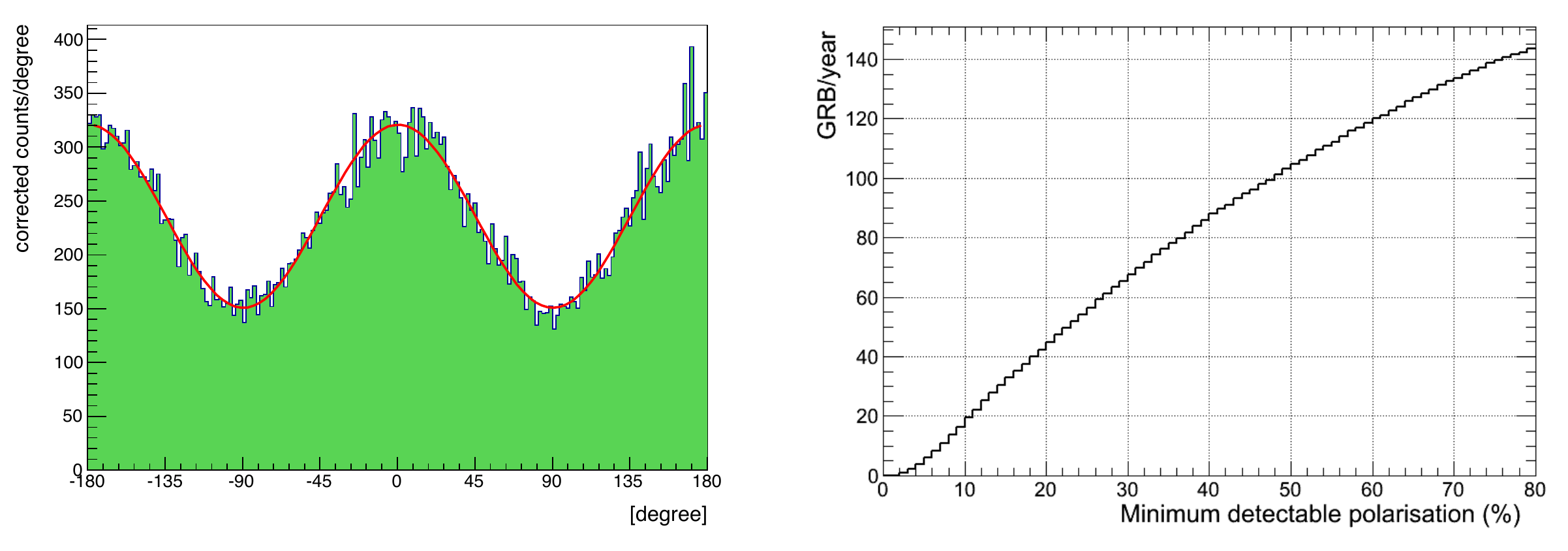}
\caption{{\it Left panel} -- e-ASTROGAM polarization response (polarigramme) in the 0.2 -- 2 MeV range for a 100\% polarized, 10 mCrab-like source observed on axis for 10$^6$ s. The corresponding modulation is $\mu_{100}$ = 0.36. {\it Right panel} -- Cumulative number of GRBs to be detected by e-ASTROGAM as a function of the minimum detectable polarization at the 99\% confidence level.}
\label{fig:polarization}
\end{figure*}

Both Compton scattering and pair creation partially preserve the linear polarization information of incident photons. In a Compton telescope, the polarization signature is reflected in the probability distribution of the azimuthal scattering angle. In the pair domain, the polarization information is given by the distribution of azimuthal orientation of the electron-positron plane (as discussed since the 1950s; see, e.g., \cite{ber50,yang50,wick51}). e-ASTROGAM will be able to perform for the first time at these energies polarization measurements thanks to the fine 3D position resolution of both the Si Tracker and the Calorimeter, as well as the light mechanical structure of the Tracker, which is devoid of any heavy absorber in the detection volume.

The left panel of Figure~\ref{fig:polarization} shows an example of a polarigramme in the 0.2 -- 2 MeV range (i.e. in the Compton domain), simulated with MEGAlib. The calculations assume a 100\% polarized emission from a 10 mCrab-like source observed on axis. The systematic effects of instrumental origin were corrected by simulating the azimuthal response of the instrument to an unpolarized source with the same spectral distribution and position in the field of view as the polarized source. From the obtained modulation ($\mu_{100} = 0.36$), we find that at low energies (0.2 -- 2~MeV), e-ASTROGAM will be able to achieve a Minimum Detectable Polarization (MDP) at the 99\% confidence level as low as 0.7\% for a Crab-like source in 1~Ms (statistical uncertainties only). After one year of effective exposure of the GC region, the achievable MDP$_{99}$ for a 10~mCrab source will be 10\%. With such a performance, e-ASTROGAM will be able to study the polarimetric properties of many pulsars, magnetars, and black hole systems in the Galaxy.

The right panel of Figure~\ref{fig:polarization} shows the number of GRBs detectable by e-ASTROGAM as a function of MDP$_{99}$ in the 150--300 keV band. The total number of GRBs detected by e-ASTROGAM will be $\sim$600 in 3 years of nominal mission lifetime. Here, the GRB emission spectrum has been approximated by a typical Band function \cite{ban93} with $\alpha=-1.1$, $\beta=-2.3$, and $E_{\rm peak}=0.3$~MeV, and the response of e-ASTROGAM to linearly polarized GRBs has been simulated at several off-axis angles in the range $[0^\circ;90^\circ]$. The number of GRBs with polarization measurable with e-ASTROGAM has then been estimated using the Fourth BATSE GRB Catalog \cite{pac99}. We see in Figure~\ref{fig:polarization} that e-ASTROGAM should be able to detect a polarization fraction of 20\% in about 42 GRBs per year, and a polarization fraction of 10\% in $\sim$16 GRBs per year. This polarization information, combined with spectroscopy over a wide energy band, will provide unambiguous answers to fundamental questions on the sources of the GRB highly relativistic jets and the mechanisms of energy dissipation and high-energy photon emission in these extreme astrophysical phenomena.

The measurement of polarization using the azimuthal orientation of the electron-positron plane is complex and a precise evaluation of the unfolding procedures and performance requires accurate simulation and testing \cite{dbpol,epol}.  Some estimates are however possible. 
The expected performance for GRBs has been already discussed in \ref{sec:grb}. 
From the formula derived in \cite{cpol} and taking into account the angular opening of pairs [82], the multiple scattering of the secondary electron and positron \cite{dpol}, one can estimate, using a simplified model for the conversions of photons in the energy range from 20 MeV to 200 MeV, that $\sim 20\%$ polarization from Vela could be detected at 3$\sigma$  in 15 months of data. Also using a simplified model for pair production and multiple scattering of electrons and positrons, multiple scattering simplified model), a MDP of $\sim$45\% at 3$\sigma$ has been estimated for the Crab Nebula in $10^6$ s in the range from 10 to 100 MeV. These should be considered as lower limit MDP figures.

\begin{figure*}%[b!]
\centering
\includegraphics[width=0.65\textwidth]{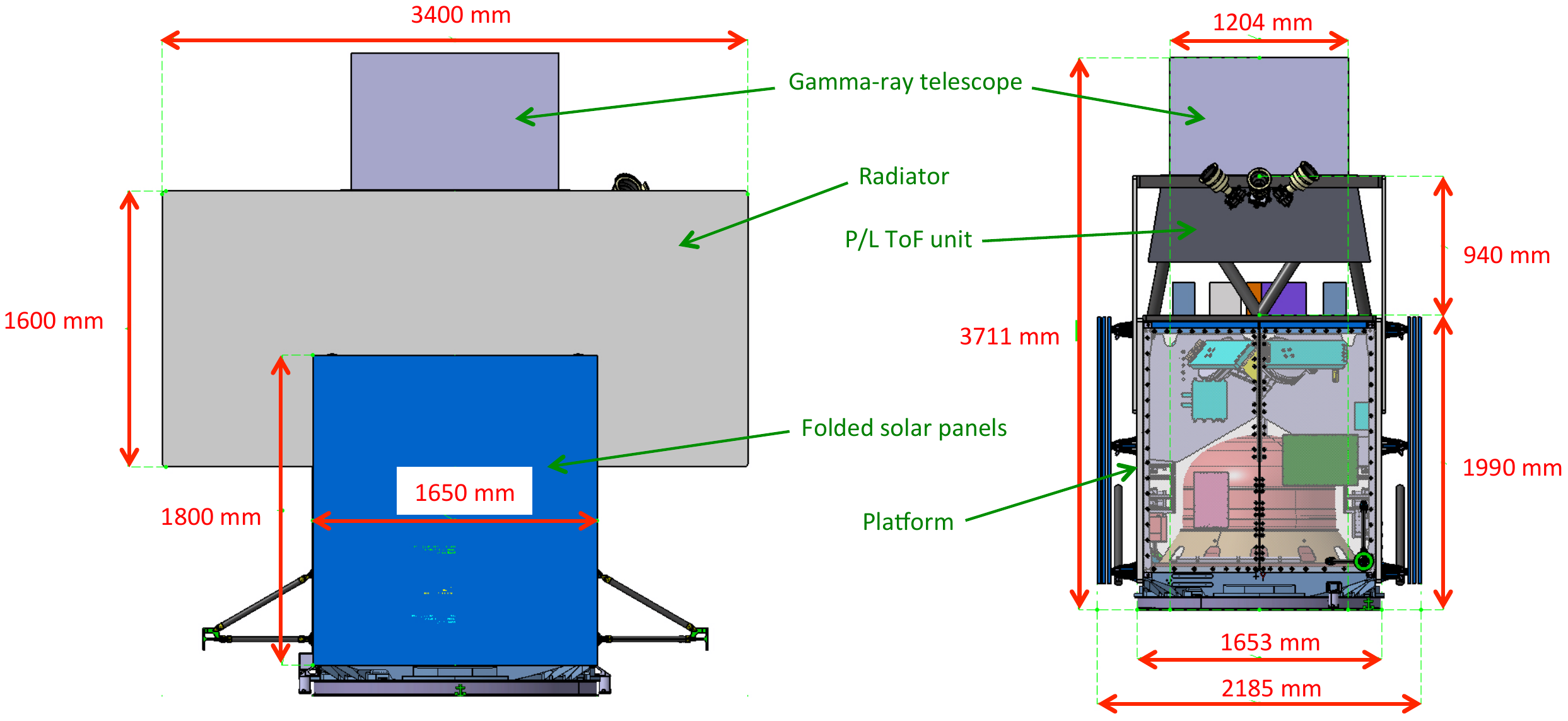}
\caption{e-ASTROGAM dimensions. The space between the payload (P/L) ToF unit and the spacecraft hosts the P/L PSU, PDHU and BEE modules.}
\label{fig:satellitedimensions}
\end{figure*}

\subsection{Technology readiness}

e-ASTROGAM is based on the heritage of AGILE and of the Fermi-LAT, with an overall simpler structure (the division into 4 tower units instead of the 16 of LAT greatly reduces complexity); it uses technology developed for previous space-based detectors. Most components are at or above TRL  6. 

The DSSDs have proven their performance in PAMELA \cite{pamelad} and AMS-02 \cite{ams02d}. Low-power ASICs suitable for the readout of the were already qualified and used on ASTRO-H/HXI \cite{astroh}. 
The low-power SiPM sensors selected for the ACD and the solid-state detectors foreseen for the CsI calorimeter have not yet flown on satellite missions but are in use in suborbital prototypes \cite{ballo} and are planned for use on future space-borne platforms. 

\section{Mission Configuration and Profile}
\subsection{Orbit and launcher}

The best orbit for e-ASTROGAM is an equatorial LEO (required to have an inclination $i < 2.5^\circ$, and eccentricity $e < 0.01$) of altitude 550~--~600 km: as already discussed, particle background properties are optimum for this orbit.
%, as already determined by the AGILE mission (which has an orbit of altitude 520~--~550 km and 2.5$^\circ$ inclination with respect to the equator). 
Such an orbit allows  making use of the ESA ground station at Kourou as well as  of the ASI Malindi station in Kenya.

The foreseen launcher for e-ASTROGAM is Ariane~6.2. The launcher fairing (5.4 m diameter) and performance for the targeted orbit ($>5$~tons) are well adapted to the spacecraft dimensions ($\rm{L} \times \rm{W} \times \rm{H} = 340 \times 218 \times 371$~cm$^3$; Figure~\ref{fig:satellitedimensions}) and separated mass (2680 kg, including maturity margins at sub-system level and an additional system margin of 20\%)

\subsection{Spacecraft and system requirements}

The e-ASTROGAM system is composed of a satellite and a ground segment that includes the ESA ground station at Kourou and possibly the ASI Malindi station in Kenya. These stations are in charge of performing the spacecraft control, monitoring, and the acquisition of scientific data. Communication with the ground is ensured by an X-band telecommand and telemetry subsystem. The average orbital contact time with the two ground stations is about 10 min for each of them. The average data generation of both payload (P/L) and platform amounts to 1382 kbps, taking into account a compression of the raw P/L data by a factor of 2.6, as obtained with e-ASTROGAM representative science data and standard compression programs. The overall data generation rate is then about 8.0~Gbit per orbit, which can be transmitted using the two ground stations at a downlink rate of $\sim$6.6~Mbps, in agreement with the bandwidth limit in X-band for e-ASTROGAM mission category (maximum downlink rate of 8.5 Mbps).

The e-ASTROGAM spacecraft is observing the sky according to a predefined pointing plan uploaded from ground. Different pointing profiles can be selected in order to observe selected sky regions or to perform a scanning that can cover a large fraction of the sky at each orbit.

The spacecraft platform is made of a structure that mechanically supports the e-ASTROGAM instrument and hosts internally the payload electronic units and all the platform subsystems.  The payload is attached to a mechanical structure at a distance of about 94~cm from the top of the platform. The  space between the payload and the platform is used to (i) host the time-of-flight (ToF) unit of the P/L Anticoincidence (AC) system, (ii) host the P/L PSU, PDHU and BEE modules (Tracker, Calorimeter and AC BEEs), and (iii) accommodate the two fixed radiators of the thermal control system. In addition, this mechanical design has the advantage of significantly reducing the instrument background due to prompt and delayed gamma-ray emissions from fast particle reactions with the platform materials. 

\begin{figure*}
\centering
\includegraphics[width=0.7\textwidth]{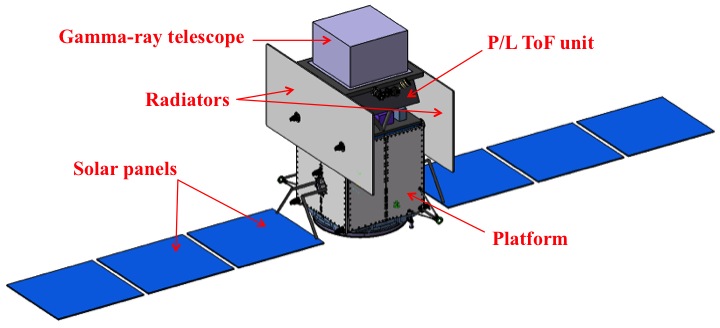}
\caption{e-ASTROGAM spacecraft in deployed configuration.}
\label{fig:deployed}
\end{figure*}

Deployable and steerable solar panels are required to support the payload operating profile and the platform pointing and communication requirements. Figure~\ref{fig:deployed} shows the spacecraft configuration in flight with deployed solar arrays. Figure~\ref{fig:ariane62} shows e-ASTROGAM under Ariane 6.2 fairing in upper position of a dual launch configuration. 

A precise timing of the payload data (1 $\mu$s at 3$\sigma$) is required to perform a proper on ground data processing able to guarantee the scientific performance of the mission. The required timing performance is obtained by a GPS delivering a pulse-per-second (PPS) signal both to the P/L PDHU and the Spacecraft Management Unit time management in order to allow a fine synchronization with the time reference. 

\subsubsection{Attitude and Orbital Control Systems}\label{sec:aocs}

\begin{figure}
\centering
\includegraphics[width=0.3\linewidth]{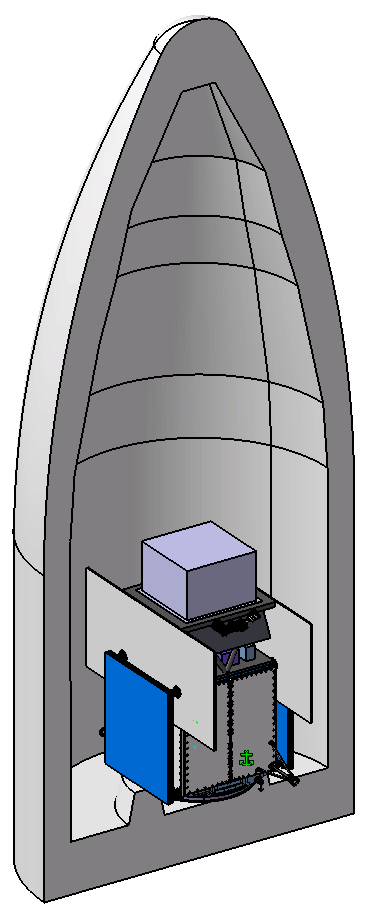}
\caption{e-ASTROGAM under Ariane 6.2 fairing in upper position.}
\label{fig:ariane62}       
\end{figure}

The Attitude and Orbital Control Systems (AOCS) performance requirements for the e-ASTROGAM mission are not critical, with an absolute Pointing Error of less than 1 degree, a relative Pointing Error of less than 0.01 deg/s and an Absolute knowledge Error of less than 30~arcseconds. In addition, to fulfill the scientific requirements, the spacecraft shall be placed in a LEO without the need to perform orbit maintenance during nominal lifetime. The propulsion subsystem is therefore provided with the only purpose to have the possibility to correct for launcher dispersion, perform debris avoidance and, as required by the international regulations and because of the construction of the instrument, to execute a direct controlled re-entry at the end of the mission. The proposed subsystem design is a monopropellant hydrazine, blow-down mode propulsion system. The hydrazine is contained in a diaphragm tank, together with the pressurant. From a preliminary estimation of the propellant budget the amount of hydrazine required to fulfill the mission needs are about 266~kg, among which more than 190~kg are allocated to the end of mission disposal.

The spacecraft is able to provide the following attitude pointings to support the payload observation requirements:
\begin{itemize}
\item  zenith pointing to perform at each orbit a scan of the sky; 
\item nearly inertial pointing (with the possibility to slowly rotate around the payload boresight) to observe continuously a selected area of the sky;
\item fast payload repointing during eclipse periods to avoid the presence of the Earth in the payload FoV (allowing 2 pointings per orbit).
\end{itemize}

\subsubsection{Thermal control system}

The required pointing accuracy ($\pm 1^\circ$), stability (0.01$^\circ$/s), and attitude knowledge of 30 arcsec (to be reached after ground processing) can be obtained using standard class sensors and actuators. The 3-axis stabilized attitude control is achieved mainly using a set of four reaction wheels used in zero momentum mode ensuring the possibility to perform fast repointing manoeuvres. Magnetic torquers are provided to perform wheels desaturation and to support a safe attitude pointing based on a basic subset of AOCS items.

The e-ASTROGAM P/L detector has an optimal performance in the temperature range -10$^\circ$~--~0$^\circ$. In order to guarantee the required environment, the P/L power dissipation is evacuated towards external space by two fixed large radiators located on the two solar array panels, below the instrument. The total radiative area is 11.6~m$^2$. The radiator is a heat transfer device based on a Loop Heat Pipe (LHP) with a condenser as a part of a radiation heat exchanger. Large radiators are necessary because of the large payload thermal dissipation, variable external environmental conditions, and limited heat transport capability within the payload. However, thanks to the large Ariane 6.2 fairing (4.5~m diameter), a deployable radiator can be avoided, as shown in Figure~\ref{fig:ariane62}.

\section{Summary}

e-ASTROGAM is a concept for a gamma-ray space observatory that can revolutionize the astronomy
of medium/high-energy gamma rays by increasing the number of known sources in this field by more than
an order of magnitude and providing polarization information for many of these sources -- thousands of
sources are expected to be detected during the first 3 years of operations. Furthermore, the proposed wide-field
gamma-ray observatory will play a major role in the development of time-domain astronomy, and provide valuable
information for the localization and identification of gravitational wave sources.

The instrument is based on
an innovative design, which minimizes any passive material in the detector volume. The instrument performance
has been assessed through detailed simulations using state-of-the-art  tools and the results fully meet
the scientific requirements of the proposed mission. 

e-ASTROGAM will operate as an observatory open to the
international community.
The gamma-ray observatory will be complementary to ground and space instruments, and multifrequency observation
programs will be very important for the success of the mission. In particular, e-ASTROGAM will be
essential  for investigations jointly done with radio (VLA, VLBI, ALMA, SKA), optical (JWST,
E-ELT and other ground telescopes), X-ray and TeV ground instrument (CTA, HAWC, LHAASO and other
ground-based detectors). Special emphasis will be given to fast reaction to transients and rapid communication
of alerts. New astronomy windows of opportunity (sources of gravitational waves, neutrinos, ultra high-energy
cosmic rays) will be fully and uniquely explored.

\begin{acknowledgements}
The contribution by P. Couzin (TAS-F), G. Cluzet (TAS-F), X. Roser (TAS-F), A. Laurens (CNES), D. Delrieu (CNES), M.-F. DelCastillo (CNES), C. Contini (CGS), P. Lattanzi (CGS), B. Morelli (CGS), A. Spalla (CGS), is acknowledged.

Comments from E. Orlando were appreciated.

The research leading to these results has received funding from the European Union's Horizon 2020 Programme under the AHEAD project (grant agreement n. 654215).
\end{acknowledgements}

% convenience commands added by IR to make the bibliography uniform
\newcommand*\etal{{\it et al.}}
\newcommand*\aap{A\&A}
\let\astap=\aap
\newcommand*\aapr{A\&A~Rev.}
\newcommand*\aaps{A\&AS}
\newcommand*\actaa{Acta Astron.}
\newcommand*\aj{AJ}
\let\applopt\ao
\newcommand*\apjl{ApJL}
\let\apjlett\apjl
\newcommand*\apjs{ApJS}
\let\apjsupp\apjs
\newcommand*\aplett{Astrophys.~Lett.}
\newcommand*\apspr{Astrophys.~Space~Phys.~Res.}
\newcommand*\apss{Ap\&SS}
\newcommand*\araa{ARA\&A}
\newcommand*\azh{AZh}
\newcommand*\baas{BAAS}
\newcommand*\bac{Bull. astr. Inst. Czechosl.}
\newcommand*\bain{Bull.~Astron.~Inst.~Netherlands}
\newcommand*\caa{Chinese Astron. Astrophys.}
\newcommand*\cjaa{Chinese J. Astron. Astrophys.}
\newcommand*\fcp{Fund.~Cosmic~Phys.}
\newcommand*\gca{Geochim.~Cosmochim.~Acta}
\newcommand*\grl{Geophys.~Res.~Lett.}
\newcommand*\iaucirc{IAU~Circ.}
\newcommand*\icarus{Icarus}
\newcommand*\jcap{J. Cosmology Astropart. Phys.}
\newcommand*\jgr{J.~Geophys.~Res.}
\newcommand*\jqsrt{J.~Quant.~Spec.~Radiat.~Transf.}
\newcommand*\jrasc{JRASC}
\newcommand*\memras{MmRAS}
\newcommand*\memsai{Mem.~Soc.~Astron.~Italiana}
\newcommand*\mnras{MNRAS}
\newcommand*\na{New A}
\newcommand*\nar{New A Rev.}
\newcommand*\nphysa{Nucl.~Phys.~A}
\newcommand*\pasa{PASA}
\newcommand*\pasj{PASJ}
\newcommand*\pasp{PASP}
\newcommand*\physrep{Phys.~Rep.}
\newcommand*\physscr{Phys.~Scr}
\newcommand*\planss{Planet.~Space~Sci.}
\newcommand*\procspie{Proc.~SPIE}
\newcommand*\qjras{QJRAS}
\newcommand*\rmxaa{Rev. Mexicana Astron. Astrofis.}
\newcommand*\skytel{S\&T}
\newcommand*\solphys{Sol.~Phys.}
\newcommand*\sovast{Soviet~Ast.}
\newcommand*\ssr{Space~Sci.~Rev.}
\newcommand*\zap{ZAp}


\begin{thebibliography}{33}
% IR trying to maintain
% - alphabetical order
% - common format:
% Author, et al. YEAR, Journal, coordinates
% (except in case of four or less authors)
%% Some unused refs. commented out by IR on 20161105

\setlength{\itemsep}{0pt}

\bibitem{astronu} Aartsen, M.G., \etal, 2013, Science 342, 1242856
\bibitem{Abbott2016a} Abbott, B.P., \etal, 2016, \prl, 116, 061102
%%\bibitem{Abbott2016b} Abbott, B.P., \etal, 2016b, \prl, 116, 241103
%%\bibitem{Abbott2016c} Abbott, B.P. \etal, 2016c, Living Rev. Relativity, 19, 1
\bibitem{Abdocygx3} Abdo A.A., \etal, 2009, Science, 326, 1512. 
\bibitem{FermiStarburst} Abdo, A.A., \etal, 2010, \apj, 709, 152
%%\bibitem{2010ApJ...710L..92A} Abdo, A.~A. \etal, 2010c, \apjl, 710, L92
\bibitem{Abdocrab} Abdo A.A., \etal, 2011, Science, 331, 739
\bibitem{FermiSNRcat} Acero, F., \etal, 2016, \apj, 224, 8
\bibitem{3FGL} Acero, F., \etal, 2015, \apjs, 218, 23 %, Ackermann, M., Ajello, M.,
\bibitem{FermiCyg} Ackermann, M., \etal, 2011, Science, 334, 1103
\bibitem{FermiW44IC443} Ackermann, M., \etal, 2013, Science, 339, 807
\bibitem{photonsfromdm} Ackermann, M., \etal, 2014, \prd, 89, 042001
\bibitem{Ferminovae} Ackermann, M., \etal, 2014, Science, 345, 554 %, Ajello, M., Albert, A., et al.\ 
\bibitem{FermiBubbles} Ackermann, M., \etal, 2014, \apj, 793, 64
\bibitem{FermiEGB} Ackermann, M., \etal, 2015, \apj, 799, 1
\bibitem{FermiDM} Ackermann, M. et al. 2015, Phys. Rev. Lett,, 115, 231301
\bibitem{Fermi3c279} Ackermann, M., \etal, 2016, \apj, 824, 2
\bibitem{FermiLMC} Ackermann, M., \etal, 2016, \aap, 586, A71
\bibitem{pamelad} Adriani, O., \etal, 2003, Nucl. Instr. Methods A, 511, 72
\bibitem{aha15} Ahangarianabhari, M. \etal, 2015, Nucl. Instr. Methods A, 770, 155 %, Macera, D., Bertuccio, G., Malcovati, P., \& Grassi, M.
\bibitem{ajello09} Ajello, M., \etal, 2009, \apj, 699, 603
\bibitem{ajello12} Ajello, M., \etal, 2012, \apj, 751, 108.
%\bibitem{Ajello2016} Ajello, M. \etal, 2016, \apj, 826, 76 
%\bibitem{Aliu2013} Aliu, E. \etal, 2013, \apj, 775, 3 
\bibitem{Albert:2014hwa} Albert, A. \etal, 2014, \jcap, 1410, 023
\bibitem{ams02d} Alcaraz, J., Alpat, B., Ambrosi, G.,  \etal,  2008, Nucl. Instr. Methods A, 593, 376
\bibitem{cast} Arik, E., \etal, 2009, \jcap, 02, 008
%\bibitem{atw12} Atwood, W.~B., 2012, American Astronomical Society Meeting Abstracts \#219, 219, 200.02 
\bibitem{bag11} Bagliesi, M.~G. \etal, 2011, Nuclear Physics B Proceedings Supplements, 215, 344 %Avanzini, C., Bigongiari, G.
\bibitem{ban93} Band, D., \etal, 1993, \apj, 413, 281 
\bibitem{b70h} Baumgartner, W.H., \etal, 2013 ApJS, 207, 19
\bibitem{ben13} Benhabiles-Mezhoud, H., \etal, 2013, \apj, 763, 98 %, Kiener, J., Tatischeff, V., \& Strong, A.~W.\
\bibitem{Bergstrom:1988jt} Bergstr{\"o}m, L., 1988, Nucl. Phys., B325, 647
\bibitem{ber50} Berlin, T.~H. \& Madansky, L., 1950, Phys.~Rev., 78, 623 
\bibitem{dbpol} Bernard, D., 2013,  Nuclear Instr. and Methods in Phys. Res. A, 729, 765
\bibitem{bil93} Bildsten, L., Salpeter, E.~E., \& Wasserman, I., 1993, \apj, 408, 615 
\bibitem{ibisc} Bird, A.J., \etal, 2010, ApJS, 186, 1
\bibitem{ballo} Bloser, P.F., \etal, 2016, Nucl. Instr. Methods A,  812, 92
\bibitem{Boddy:2015fsa} Boddy, K.~K. \& Kumar, J. 2016, AIP Conf. Proc. 1743, 020009
  %``Minding the MeV gap: The indirect detection of low mass dark matter,''
  %doi:10.1063/1.4953276
  %arXiv:1509.03333 [astro-ph.CO].
  %%CITATION = doi:10.1063/1.4953276;%%
\bibitem{Boehm:2002yz} 
  Boehm, C.T., Ensslin, A. \& Silk, J., 2004,
  %``Can Annihilating dark matter be lighter than a few GeVs?,''
  J.\ Phys.\ G, 30, 279 
  %doi:10.1088/0954-3899/30/3/004
  %[astro-ph/0208458].
  %%CITATION = doi:10.1088/0954-3899/30/3/004;%%
  %101 citations counted in INSPIRE as of 21 Sep 2016
\bibitem{Boehm:2003bt} 
  Boehm, C.T., \etal, 2004,
  %, D.~Hooper, J.~Silk, M.~Casse and J.~Paul,
  %``MeV dark matter: Has it been detected?,''
  \prl, 92, 101301 
  %doi:10.1103/PhysRevLett.92.101301
  %[astro-ph/0309686].
  %%CITATION = doi:10.1103/PhysRevLett.92.101301;%%
  %330 citations counted in INSPIRE as of 21 Sep 2016
%\bibitem{bor10} Borkowski, K.~J., \etal, 2010, \apj, 724, L161 %Reynolds, S.~P., Green, D.~A., 
\bibitem{Breit91} Breitschwerdt, D., \etal, 1991, \aap, 245, 79B
\bibitem{mevdm}  Bringmann, T., \etal, 2016, arXiv:1610.04613 
%\bibitem{2012ApJ...749...26B} Buehler, R. \etal, 2012, \apj, 749, 26 %, Scargle, J.~D., Blandford, R.~D.,
\bibitem{cpol} Buehler, R., \etal., 2010, ``Measuring polarization of gamma-rays with Fermi'', Presented at SciNeGHE Trieste, http://scineghe2010.ts.infn.it/programmaScientifico.php
\bibitem{bul12} Bulgarelli, A., \etal, 2012, Proceedings of the SPIE 8453, 845335 %, Fioretti, V., Malaguti, P., Trifoglio, M., \& Gianotti, F.
%\bibitem{bul14} Bulgarelli, A. \etal, 2014, \apj, 781, 19 %, Trifoglio, M., Gianotti, F.
\bibitem{cam14} Campana, R., \etal, 2014, Experimental Astronomy, 37, 599 %, Orlandini, M., Del Monte, E., Feroci, M., \& Frontera, F.
\bibitem{CarlsonProfumo} Carlson, E. \& Profumo, S., 2014, \prd, 90, 023015
%\bibitem{2012ApJ...752...23C} Chang, P., Broderick, A.~E., \& Pfrommer, C.\ 2012, \apj, 752, 23 
\bibitem{cheung16}Cheung, C.C., \etal, 2016, \apj, 826, 142
\bibitem{chu14} Churazov, E., \etal, 2014, Nature, 512, 406 %, Sunyaev, R., Isern, J., et al.\
\bibitem{chu15} Churazov, E., \etal, 2015, \apj, 812, 62 %, Sunyaev, R., Isern, J., et al.\
\bibitem{cla74} Clayton, D.D. \& Hoyle, F., 1974, \apj, 187, L101 
%\bibitem{collmar2014} Collmar, W. \& Zhang, S. 2014, \aap, 565, A38
\bibitem{Crocker11} Crocker, R. \& Aharonian, F., 2011, \prl, 106, 1102
\bibitem{dmr2008} De Angelis, A., Mansutti, O. \& Roncadelli, M., 2008, Phys. Lett. B, 659, 847
\bibitem{book} De Angelis, A. \& Pimenta, M.J., 2015, ``Introduction to Particle and Astroparticle Physics -- Questions to the Universe'' (Springer)
\bibitem{physcase} De Angelis A., Tatischeff, V. \& Giusti, M. $(eds.)$, 2017, ``e-ASTROGAM scientific workshop'', eBook (Lulu), http://www.lulu.com/shop/alessandro-de-angelis/e-astrogam-scientific-workshop/ebook/product-23158421.html
\bibitem{eLISA} Danzmann, K., \etal, eLISA White Paper, \url{https://www.elisascience.org/multimedia/document/white-paper-pdf}
\bibitem{die13} Diehl, R., 2013, Rep. Progr. Phys., 76, 2, id. 026301
\bibitem{die14} Diehl, R., \etal, 2014, Science, 345, 1162 %, Siegert, T., Hillebrandt, W., et al.\
\bibitem{die15} Diehl, R., \etal, 2015, \aap, 574, A72 %, Siegert, T., Hillebrandt, W., et al.\
\bibitem{die98} Diehl, R. \& Timmes, F.X., 1998, PASP, 110, 748 
\bibitem{Essig:2013goa} Essig, R. \etal, 2013, JHEP, 1, 193
\bibitem{Everett08} Everett, J., \etal, 2008, \apj, 674, 258
\bibitem{for08} Forot, M., \etal, 2008, \apj, 688, L29 %, Laurent, P., Grenier, I.~A., Gouiff{\`e}s, C., \& Lebrun, F.\
\bibitem{Fox15} Fox, A., \etal, 2015, \apj, 799, 7
\bibitem{funk2015} Funk, S., 2016, Ann. Rev. Nucl. Part. Sci. 65, 245
\bibitem{gal09} Gal-Yam, A.,  \etal, 2009, Nature, 462, 624 % Mazzali, P., Ofek, E.~O., 
\bibitem{gr2013} Galanti, G. \& Roncadelli, M., 2013, arXiv:1305.2114
\bibitem{gat84} Gatti, E. \& Rehak, P., 1984, Nucl. Instr. Methods A, 225, 608; www.pnsensor.de/Welcome/Detectors/SDD/
%\bibitem{gatti99} Gatti E., P. Rehak, Semiconductor Drift Chamber - An Application of a Novel Charge Transport Scheme, Nucl. Instr. Methods A, 225, 1984, pp. 608-614.
\bibitem{gev12} Gevin, O. \etal, 2012, Nuclear Instruments and Methods in Physics Research A, 695, 415 %, Lemaire, O., Lugiez, F.,
\bibitem{Ghisellini2010} Ghisellini, G., \etal, 2010, \mnras, 405, 387 %, Della Ceca, R., Volonteri, M.
%\bibitem{Ghisellini2012} Ghisellini, G., 2012, \mnras, 424, L26 
\bibitem{Ghisellini2013} Ghisellini, G., \etal, 2013, \mnras, 432, 2818 %, Haardt, F., Della Ceca, R., Volonteri, M., \& Sbarrato, T.\
\bibitem{gom16} G\'omez, S. \etal, 2016,  Proceedings of the SPIE, 9899, 98990G, doi: 10.1117/12.2231095 %
\bibitem{gom98} Gomez-Gomar, J., Hernanz, M., Jose, J., Isern, J., 1998, MNRAS, 296, 913
\bibitem{got14} G{\"o}tz, D., Laurent, P., Antier, S., et al.\ 2014, \mnras, 444, 2776 
\bibitem{gre14} Grefenstette, B.W., \etal, 2014, Nature, 506, 339 %, Harrison, F.~A., Boggs, S.~E.,
\bibitem{Grenier15} Grenier, I.A., Black, J.H. \& Strong, A.W.\ 2015, \araa, 53, 199
%\bibitem{1997ApJ...476..246H} Harding, A.K., Baring, M.G., \& Gonthier, P.L.\ 1997, \apj, 476, 246 
\bibitem{her04} Hernanz, M., Jose, J., 2004, New Astron. Rev., 48, 35
%\bibitem{2012ApJ...759...89H} Hewitt, J.~W. \etal, 2012, \apj, 759, 89 %, Grondin, M.-H., Lemoine-Goumard, M., et al.
\bibitem{hi13} Hillebrandt, W., Kromer, M., R\"opke, F. \& Ruiter, A., 2013, Front. Phys. 8, 116
\bibitem{hil00} Hillebrandt, W. \& Niemeyer, J.C., 2000, \araa, 38, 191
\bibitem{Indriolo2012} Indriolo, N. \& McCall, B., 2012, \apj, 745, 91
\bibitem{ise16} Isern, J., \etal, 2016, \aap, 588, A67 %, Jean, P., Bravo, E.,
\bibitem{alp1} Jaeckel, J. \& Ringwald, A., 2010, Ann. Rev. Nucl. Part. Sci., 60, 405
\bibitem{FermiW51C} Jogler, T. \& Funk, S., 2016, \apj, 816, 100
\bibitem{jos07} Jos{\'e}, J. \& Hernanz, M., 2007, Journal of Physics G Nuclear Physics, 34, R431 
\bibitem{kadler2016} Kadler, M., \etal, 2016, Nature Physics 12, 807
\bibitem{kan05} Kanbach, G., \etal, 2005, Nucl. Instr. Methods A, 541, 310 %Andritschke, R., Zoglauer, A.,
\bibitem{ker14} Kerzendorf, W. \& Sim, S., 2014, \mnras, 440, 387
\bibitem{xmm04}  Kirsch, M.G.F., \etal, 2004, ``XMM-Newton (cross)-calibration'', arXiv:astro-ph/0407257
\bibitem{PICARD} Kissmann, R., \etal, 2015, Astroparticle Physics, 70, 39
\bibitem{Koljonen} Koljonen, K.,  \etal, 2010, \mnras, 406, 307
\bibitem{kra15} Krause, M.~G.~H., \etal, 2015, \aap, 578, A113 %, Diehl, R., Bagetakos, Y.
\bibitem{kre13} Kretschmer, K., \etal, 2013, \aap, 559, A99 %, Diehl, R., Krause, M.
\bibitem{lab08} Labanti, C. \etal, 2008,  Proceedings of the SPIE, 7021, 702116 %, Marisaldi, M., Fuschino, F.
\bibitem{lim06} Limongi, M. \& Chieffi, A., 2006, \apj, 647, 483
\bibitem{lim05} Limousin, O. \etal. 2005, IEEE Transactions on Nuclear Science, 52, 1595 %, Gevin, O., Lugiez, F.,
\bibitem{mar05} Marisaldi, M. \etal, 2005, IEEE Transactions on Nuclear Science, 52, 1842 %, Labanti, C., Soltau, H.
\bibitem{McClelland2015} McClelland, D., \etal, 2015, LIGO Scientific Collaboration, Instrument Science White Paper, LIGO Document T1500290-v2
\bibitem{mcc16} McConnell, M.~L.\ 2016, accepted for publication in New Astronomy Review, arXiv:1611.06579 
\bibitem{meyer} Meyer, M., \etal, 2017, Phys. Rev. Lett. 118, 011103 
\bibitem{moi07} Moiseev, A.A., \etal, 2007, Astroparticle Physics, 27, 339 % Hartman, R.~C., Ormes, J.~F.,
\bibitem{alexnew} Moiseev, A.A., \etal, 2007,  arXiv:1508.07349 
%%\bibitem{Murphy1987} Murphy, R. J. \etal, 1987, \apjs, 63, 721
\bibitem{or3} Moskalenko, I.V., Porter, T.A., and Digel, S.W., 2006, Astrophys. J. 652, L65
\bibitem{Nakar2007} Nakar, E., 2007, \physrep, 442, 166
%\bibitem{2010Sci...328...73N} Neronov, A., \& Vovk, I., 2010, Science, 328, 73 
\bibitem{nom84} Nomoto, K., Thielemann, F.-K. \& Yokoi, K., 1984, \apj, 286,  644
\bibitem{astroh} Odaka, H., \etal, 2012, Nucl. Instr. Methods A,  695, 179
\bibitem{pdg2015} Olive, K.A., \etal, 2014, Chin. Phys. C, 38, 090001  and 2015 update
\bibitem{dpol} Olsen, H., 1963, Phys. Rev., 131, 406
\bibitem{or1} Orlando, E., \& Strong, A., 2007, Astrophys. Space Sci. 309, 359
\bibitem{or2} Orlando, E., \& Strong, A., 2008, Astron. Astrophys. 480, 847
\bibitem{pac99} Paciesas, W.~S., \etal, 1999, \apjs, 122, 465 %, Meegan, C.~A., Pendleton, G.~N.,
\bibitem{paliya2016} Paliya, V. S. \etal, 2016, \apj 825, 74
\bibitem{Patricelli2016} Patricelli, B., \etal, 2016, arXiv:1606.06124
\bibitem{per06} Perotti, F., \etal, 2006, Nucl. Instr. Meth. Phys. Res. A, 556, 228 %Fiorini, M., Incorvaia, S., Mattaini, E., \& Sant'Ambrogio, E.
\bibitem{Petrovic14} Petrovi\'c, J., Pasquale, S.D., Zaharija\v{s}, G., 2014, \jcap, 10, 052
\bibitem{phi93} Phillips, M.~M., 1993, \apj, 413, L105 
%\bibitem{Pian2014} Pian, E. \etal, 2014, \aap, 570, A77 %, T{\"u}rler, M., Fiocchi, M.,
\bibitem{piano12} Piano, G., \etal, 2012, \aap, 545, A110
\bibitem{Prada:2004} Prada, F., \etal, 2004,
   %, A. Klypin, J. Flix, M.~Mart\'{\i}nez and E.~Simonneau 2004, 
   % Dark Matter Annihilation in the Milky Way Galaxy: Effects of Baryonic Compression,
	\prl, 93, 241301 
\bibitem{Punturo2010} Punturo, M.,  \etal, 2010, Classical and Quantum Gravity, 27, 194002
%\bibitem{rac14} Rachevski, A. \etal, 2014, Journal of Instrumentation, 9, P07014 %, Zampa, G., Zampa, N.
%\bibitem{Ramaty1979} Ramaty, R. \& Lingenfelter, R. E., 1979, Nature 278, 127
%\bibitem{Regimbau2015} Regimbau, T. \etal, 2015, \apj, 799, 69
\bibitem{Recchia2016} Recchia, S., \etal, 2016, \mnras, 462, L88
\bibitem{Recchia16wind} Recchia, S., Blasi, P. \& Morlino, G., 2016, \mnras, 462, 4227
\bibitem{pdg2016} Ringwald, A., Rosenberg, L.J. \& Rybka, G., 2016, ``Axions and other similar particles'', in 
Patrignani, C.,  \etal~(Particle Data Group), Chin. Phys. C, 40, 100001
\bibitem{Romero2014} Romero, G.E., Vieyro, F.L. \& Chaty, S., 2014, \aap, 562, L7 
\bibitem{roq03} Roques, J.P., \etal, 2003, \aap, 411, L91 %, Schanne, S., von Kienlin, A.,
\bibitem{roques15} Roques, J.P., \etal, 2015, \apjl,  813, 22
\bibitem{Rudaz:1986db} Rudaz, S., \etal, 1986, Phys. Rev. Lett., 56, 2128  
\bibitem{ruizl} Ruiz-Lapuente, P., \etal, 2016, \apj, 820, 142
%%\bibitem{Sba2016} Sbarrato, T., \etal, 2016, \mnras, 462, 1542
\bibitem{Schlick2014} Schlickeiser, R., \etal, 2014, \apj, 787, 35
\bibitem{scho96} Sch\"onfelder, V., \etal, 1996, \aap, 120
\bibitem{senno2016} Senno, N., \etal, 2016, \prd, 93, 083003 
%\bibitem{Sesana2016} Sesana, A. 2016, \prl, 116, 231102
\bibitem{Siegert2016} Siegert, T., \etal, 2016, Nature, 531, 341
\bibitem{diehl2016} Siegert, T., \etal, 2016, \aap,  595, 25
\bibitem{Skilling1976} Skilling, J., \& Strong, A.~W., 1976, \aap, 53, 253
%\bibitem{Strigari:2008ib}  L.~E.~Strigari \etal, 2008,
  %, J.~S.~Bullock, M.~Kaplinghat, J.~D.~Simon, M.~Geha, B.~Willman and M.~G.~Walker,
  %``A common mass scale for satellite galaxies of the Milky Way,''
	 Nature, 454, 1096
  %doi:10.1038/nature07222
  %[arXiv:0808.3772 [astro-ph]].
  %%CITATION = doi:10.1038/nature07222;%%
  %292 citations counted in INSPIRE as of 23 Sep 2016
\bibitem{tagliaferri2015} Tagliaferri, G., \etal, 2015, \apj, 807, 167
\bibitem{tak13} Takahashi, T., Uchiyama, Y., \& Stawarz, {\L}.\ 2013, Astroparticle Physics, 43, 142 
\bibitem{Takami2014} Takami, H., Kyutoku, K., \& Ioka, K., 2014, \prd, 89, 063006
\bibitem{masaaki}Tanaka, T., \etal, 2008 \apj, 685, 988-1004
\bibitem{tat07} Tatischeff, V. \& Hernanz, M., 2007, \apj, 663, L101 
\bibitem{tavani2009} Tavani, M., \etal, 2009, Nature, 462, 620
\bibitem{tav09} Tavani, M., \etal, 2009, \aap, 502, 995 % M., Barbiellini, G., Argan, A.,
\bibitem{tavani2011} Tavani, M., \etal, 2011, Science, 331, 736
\bibitem{tavani2013} Tavani, M., \etal, 2013, Nucl. Phys. (Proc. Suppl.), 243-244, 131.
\bibitem{the14} The, L.-S. \& Burrows, A. 2014, \apj, 786, 141
\bibitem{epol} Tsai, Y.S., 1974, Rev. Mod. Phys., 46, 815
\bibitem{tsy16} Tsygankov, S.~S., Krivonos, R.~A., Lutovinov, A.~A., et al.\ 2016, \mnras, 458, 3411 
\bibitem{Uchiyama2012} Uchiyama, Y., \etal, 2012, \apj, 749, 35
\bibitem{Veres2014} Veres, P. \& Meszaros, P. 2014, \apj, 787, 168 
\bibitem{volon11} Volonteri, M., \etal, 2011, \mnras, 416, 216.
\bibitem{vonbal14}von Ballmoos, P., 2014, Hyperfine Interact., 228, 1-3, 91.
  \bibitem{Walker:2009zp} 
  Walker, M.G.,  \etal, 2009,
  %, M.~Mateo, E.~W.~Olszewski, J.~Penarrubia, N.~W.~Evans and G.~Gilmore,
  %``A Universal Mass Profile for Dwarf Spheroidal Galaxies,''
  \apj, 704, 1274
  %Erratum: [Astrophys.\ J.\  {\bf 710}, 886 (2010)]
  %doi:10.1088/0004-637X/704/2/1274, 10.1088/0004-637X/710/1/886
  %[arXiv:0906.0341 [astro-ph.CO]].
  %%CITATION = doi:10.1088/0004-637X/704/2/1274, 10.1088/0004-637X/710/1/886;%%
  %290 citations counted in INSPIRE as of 23 Sep 2016
\bibitem{Wang2016} Wang, L.J., \etal, 2016, \apj, 823, 15
\bibitem{loeb2016} Wang, X. \& Loeb, A.,  2016, Nature Physics, 12, 1116
\bibitem{wick51} Wick, G.C., 1951, Phys. Rev., 81, 467
\bibitem{woo07} Woosley, S.E., \etal, \apj 662, 487 
\bibitem{wb2012} Wouters, D. \& Brun, P., 2012, \prd, 86, 043005
%\bibitem{zanin2016} Zanin, R. \etal, 2016, arXiv:1605.05914
\bibitem{yang50} Yang, C.N., 1950, Phys. Rev., 77, 722
\bibitem{Zdziarski2014} Zdziarski, A.A., Stawarz, {\L}., Pjanka, P. \& Sikora, M.\ 2014, \mnras, 440, 2238
\bibitem{ZhangBoettcher2013} Zhang, H. \& Boettcher, M., 2013, \apj, 774, 18
\bibitem{zog06} Zoglauer, A., Andritschke, R. \& Schopper, F., 2006, \nar, 50, 629
%\bibitem{zog13} Zoglauer, A., \& Boggs, S.~E., 2013, AAS/High Energy Astrophysics Division, 13, 117.06 


\end{thebibliography}
\end{document}